# Electric-current control of anomalous Hall effect

J. P. Guo[1,4], G. M. S. Brizolla[2], P. Rao[1], J. Shao[1,4], X. H. Huang[1,4], T. N. G. Meier[1,4], T. L. Xu[1], P. R. Ji[5], J. J. Finley[3,5], J. Fabian[2], J. Knolle[1], C. H. Back[1,3,4], and L. Chen[1,4]

[1]*Department of Physics, TUM School of Natural Sciences, Technical University of Munich, Munich, Germany*

[2]*Institute of Theoretical Physics, University of Regensburg, Regensburg, Germany*

[3]*Munich Center for Quantum Science and Technology (MCQST), Munich, Germany*

[4]*Center for Quantum Engineering (ZQE), Technical University of Munich, Munich, Germany*

[5]*Walter Schottky Institute and TUM School of Natural Science, Technical University of Munich, Munich, Germany*

**We demonstrate robust and reversible electric-current control of the anomalous Hall effect (AHE) in a two-dimensional $WTe_2/Fe_3GeTe_2$ (FGT) stack. Applying a current through $T_d$-$WTe_2$ leads to a giant modulation of the AHE of the adjacent FGT layer, with the relative change of the AHE conductivity exceeding 180%. Control experiments show that i) the observed effect is absent in pure FGT, ii) the modulation weakens in thicker FGT films, confirming its interfacial origin, and iii) the modulation peaks for bilayer $WTe_2$, indicating that the Berry-curvature dipole (BCD) plays the dominant role in the modulation. We propose that the charge current $I$ generates an out-of-plane magnetization $M_z^I$ via BCD in $WTe_2$**

**and $\mathbf{M}_z^I$ modifies the exchange splitting of FGT, thereby altering its Berry curvature and nontrivially influencing the AHE. This work reveals that the BCD in low-dimensional material systems is suitable for efficient spintronic applications. Moreover, the demonstrated method of AHE control offers new possibilities for control of static magnetism, i.e., for the study of AHE-transistors as well as electric-current control of quantum magnets, especially magnetic insulators.**

In 1880 Edwin H. Hall observed that when a current passes through ferromagnetic iron, a transverse voltage develops in the absence of an external magnetic field [1]. This phenomenon, known as the anomalous Hall effect (AHE), is now considered as one of the most fundamental effects in condensed matter physics [2]. The empirical relationship [2] of the anomalous Hall resistivity $\rho_{\mathrm{AHE}}$ is given by

$$\rho_{\mathrm{AHE}} = R_{\mathrm{S}}\mu_0 M_{\mathrm{z}} \quad (1)$$

Here, $\mu_0$ is the magnetic constant, $R_{\mathrm{S}}$ the anomalous Hall coefficient, and $M_{\mathrm{z}}$ the out-of-plane component of the magnetization **M**. This expression implies that $\rho_{\mathrm{AHE}}$ depends linearly on $M_{\mathrm{z}}$. Therefore, the AHE has frequently been employed to characterize diverse fundamental magnetic properties such as the Curie temperature, coercive field and magnetic anisotropies of ferromagnetic metals (FM), as well as the current-induced spin-orbit torques in heavy metal/FM bilayers [3-5]. On the other hand, it took decades to reach a general consensus on the underlying microscopic mechanisms responsible for AHE, which involve extrinsic and intrinsic contributions [2]. These are associated with scattering and the Berry curvature (BC), respectively, giving rise to scattering-dependent and scattering-independent anomalous Hall conductivities $\sigma_{\mathrm{AHE}}$. The Berry curvature describes the local curvature of the electronic wavefunction in momentum space, and has become an essential concept in modern condensed matter physics, explaining various physical phenomena including the AHE [6-8]. However, once a magnetic material has been prepared, its electronic band structure, and consequently its BC as well as $\sigma_{\mathrm{AHE}}$ remain fixed. From both fundamental and technological points of view, it would be highly desirable to be able to efficiently tune the BC by electrical means. Conventional approaches for tuning the AHE include applying a gate-voltage using capacitor-like structures or ionic gating, which adjusts the carrier concentration in magnetic materials [3,9-11]. However, these methods require relatively complex three-layer devices, and the degree of modulation is limited by the charging/discharging capacity. Therefore, it is important to develop new concepts and to unveil material

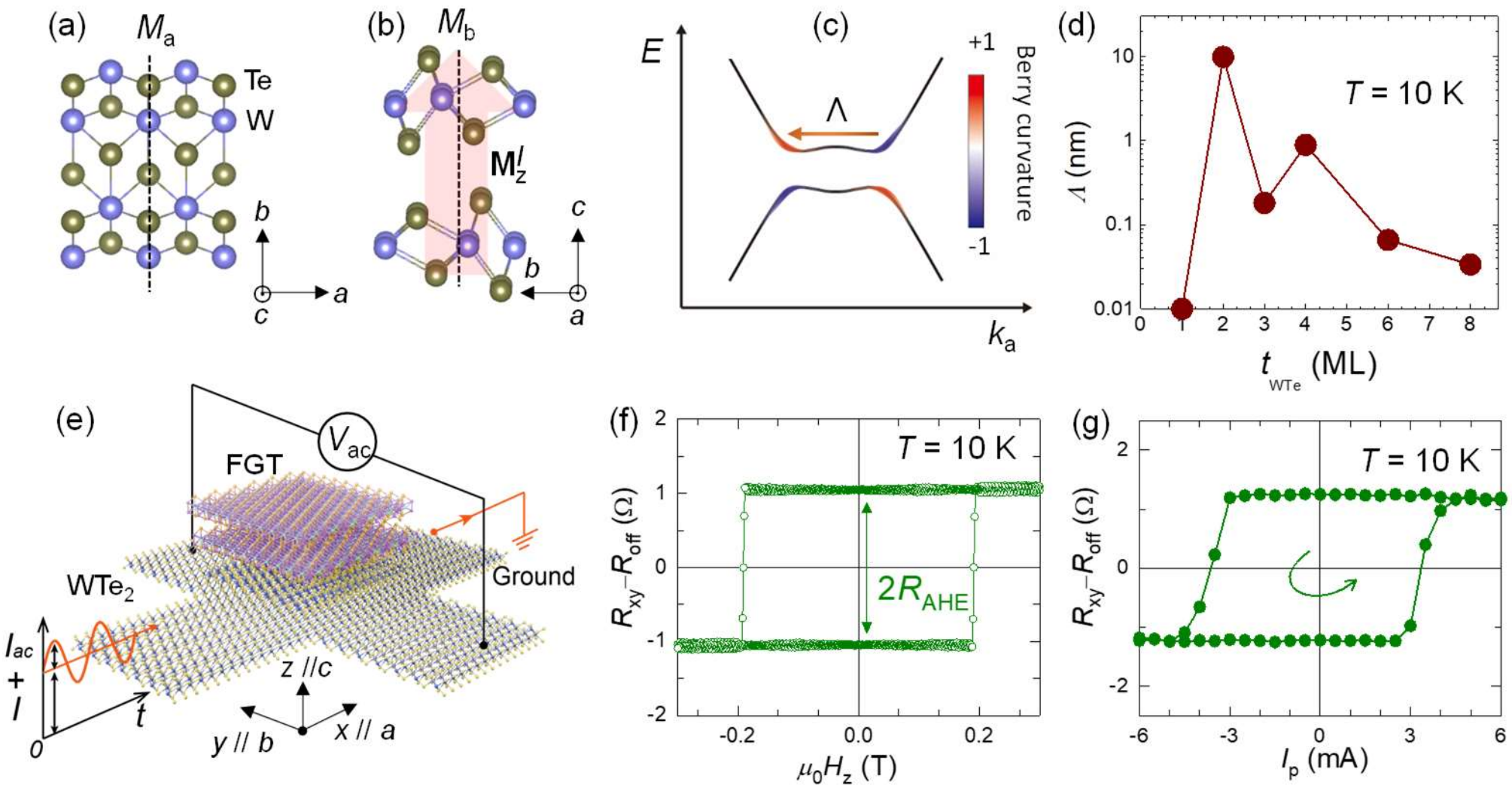

FIG. 1. (a) Top- and (b) side- view of the crystal structure of $WTe_2$, where **a**, **b** and **c** represent the crystal axes. $WTe_2$ has a mirror plane $M_a$ and a glide mirror plane $M_b$ as indicated by the dashed lines. Applying a charge current **I** along the low symmetry **a**-axis generates an out-of-plane magnetization $\mathbf{M}_z^{\mathrm{I}}$ (red arrow in b) due to BCD. (c) Schematic of the band structure of bilayer of $WTe_2$. The orange arrow represents the BCD formed by the layer-polarized Dirac fermions in bilayer $WTe_2$. (d) $t_{\mathrm{WTe}}$-dependence of the strength of the BCD (**Λ**) quantified by non-linear Hall measurements at $T$ = 10 K. Note the logarithmic scale. (e) Schematic of the experimental set up for current modulation of AHE. $I_{\mathrm{ac}}$ is the ac sense current, and $V_{\mathrm{ac}}$ is the transverse voltage detected by a lock-in amplifier, $I$ is the dc modulation current, and $I \gg I_{\mathrm{ac}}$ holds. (f) Anomalous Hall resistance $R_{\mathrm{xy}}$ (= $V_{\mathrm{ac}}/I_{ac}$) as a function of $H_z$ of the $WTe_2$/FGT stack at $T$ = 10 K. $R_{\mathrm{AHE}}$ is determined by the height of the AHE loop. (g) Pulsed current ($I_{\mathrm{p}}$) induced magnetization switching in the $WTe_2$/FGT stack without auxiliary magnetic field at $T$ = 10 K.

systems that also offer efficient tunability in a simpler structure.

To efficiently control the AHE using electrical currents, we have fabricated two-dimensional $WTe_2/Fe_3GeTe_2$ (FGT) stacks with $WTe_2$ thickness $t_{WTe}$ down to the monolayer limit (see Supplemental Material [12] for the full list of devices). $T_d$-$WTe_2$ is a transition-metal dichalcogenide with remarkable properties, e.g., non-saturating magnetoresistance [13], coexistence of quantum spin Hall insulator and superconducting states in the single monolayer regime [14-17], as well as the existence of a Berry curvature dipole (the dipole moment of the Berry curvature in momentum space represented by $\mathbf{\Lambda}$) [18-22] and quadrupole [23] along the low symmetry *a*-axis of few-layer $WTe_2$ (Figs. 1a, 1b and 1c). BCD leads to the observation of the in-plane circular photo-galvanic effect [21] and the nonlinear Hall effect (NLHE) in the absence of an external magnetic field [19,20,22]. NLHE can be understood as the current-induced AHE, in which an in-plane current generates an out-of-plane magnetization $\mathbf{M}_z^I$. $\mathbf{M}_z^I$ is given by the dot product of $\mathbf{I}$ and $\mathbf{\Lambda}$, and acts in the same way as the spontaneous magnetization for the AHE of ferromagnetic metals [18,24,25]. Folowing the recipe of Refs. 19 and 20, we have quantified the magnitude of $\mathbf{\Lambda}$ by using the NLHE for various $t_{WTe}$ at a temperature $T$ of 10 K [12]. As shown in Fig. 1d, $\Lambda = 0$ for monolayer $WTe_2$, it reaches a maximum of ~10 nm for the bilayer, and decreases as $t_{WTe}$ further increases. The enhanced dipole moment in bilayer $WTe_2$ arises from layer-polarized Dirac fermions where the symmetry along the *c*-axis is broken [19] (See Supplemental Material, Note 2 [12] for qualitative argument about the $WTe_2$ thickness dependence of $\Lambda$). Note that the generation of $\mathbf{M}_z^I$ by BCD is different from the spin accumulation induced by the interfacial spin-Rashba effect (SRE) of the normal two-dimensional states [26,27], because i) $\mathbf{M}_z^I$ points out-of-plane while SRE generates in-plane spin accumulation, and ii) the band structure of SRE has zero Berry curvature and thus no BCD [12].

The current-induced $\mathbf{M}_z^I$ can be a source of an out-of-plane anti-damping-torque [28] τ, τ ~ **M**×**M**×**z** acting on the adjacent FGT with perpendicular anisotropy [29,30] (evidenced by AHE as shown in Fig. 1f). Because τ directly counteracts the damping torque of FGT, field-free magnetization switching can be realised [31-34]. To confirm that this is also the case for our devices, we sweep a square shaped pulse current (width = 100 μs), and then detect the magnetic state by measuring AHE with a much smaller ac current of 10 μA for device DS [12]. As shown in Fig. 1g, deterministic field-free magnetization switching is achieved when the current is applied along the *a*-axis at $T = 10$ K. However, when the current is applied along the high symmetry *b*-axis, no switching is observed [12], indicating that $\mathbf{M}_z^I$ is generated less efficiently. Moreover, by simulating the switching curves with Mumax3, we estimate that the lower limit of effective spin-torque efficiency $\xi$ is 0.1 [12].

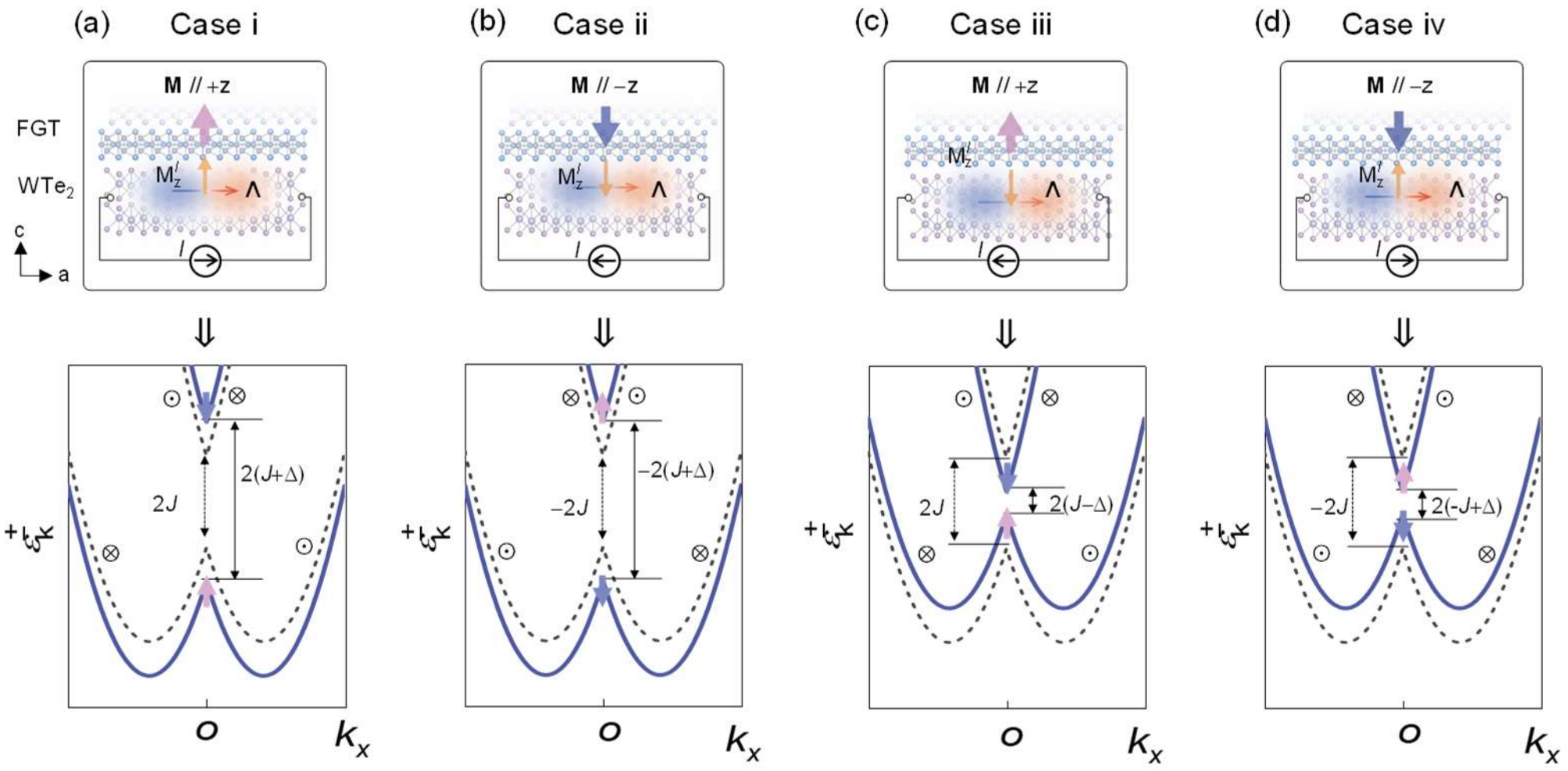

FIG. 2. (a) Upper panel: The application a positive charge current **I** generates a $\mathbf{M}_z^I$ (// +**z**) via BCD. If **M** of FM is also along +**z**, the effective field of $\mathbf{M}_z^I$ leads to an enhancement of the effective exchange coupling $E_{ex}$ from $J$ to $J$+Δ, where $J$ is the intrinsic exchange energy of FM and Δ the increase of the exchange energy. Lower panel: in a minimal two-band model, an avoided crossing between the two bands emerges, characterized by an energy splitting of 2($J$+Δ). The avoided crossing is accompanied by a non-zero BC, arising from the warped spin texture in **k**-space. At $k$ = 0, the spin directions of the lower and upper bands are represented by upward and downward arrows, respectively. For $k$ far from zero, the spin orientation is indicated by the symbol ⊗ (into the page) and ⊙ (out of the page). In this scenario, the enhanced exchange splitting leads to an enhanced BC, and consequently, an enhanced anomalous Hall conductivity $\sigma_{AHE}$. (b) For $\mathbf{M}_z^I$ // **M** // −**z**, an enhanced $\sigma_{AHE}$ is also expected. (c) and (d): If $\mathbf{M}_z^I$ and **M** are anti-parallel, $\mathbf{M}_z^I$ reduces the exchange splitting of FM. This leads to a reduced BC and $\sigma_{AHE}$. The dashed lines in the lower panels of (a), (b), (c) and (d) represent the band structure without modulation of the exchange interaction.

Besides the out-of-plane anti-damping-torque, the current-induced $\mathbf{M}_z^I$ couples with the magnetization of FGT via an effective exchange magnetic-field (upper panels of Fig. 2 and for the origin of the exchange field see Note 6 of Supplemental Material [12]). This exchange magnetic-field acts on FGT, and can modulate the effective exchange interaction $E_{ex}$ of FGT and thus static magnetic properties [35]. Depending on the relative direction of **M** and $\mathbf{M}_z^I$ (determined by the polarity of **I**), four cases arise (Fig. 2 and Table 1): parallel configuration i) and ii): When both **M** and $\mathbf{M}_z^I$ are aligned along the +z-direction and –z-direction, as respectively shown in Figs. 2a and 2b, $\mathbf{M}_z^I$ leads to an enhancement of $E_{ex}$. Specifically, $E_{ex} = J + \Delta$ for case i) and $E_{ex} = -J - \Delta$ for case ii), where $J$ is the inherent exchange interaction of FGT, and $\Delta$ the magnitude of the modulation by $\mathbf{M}_z^I$. Anti-parallel configuration iii) and iv): When $\mathbf{M}_z^I$ is aligned along –z (+z) and **M** is aligned along +z (–z) as shown in Fig. 2c (Fig. 2d), this leads to a reduction of $E_{ex}$, i.e., $E_{ex} = J - \Delta$ for case iii) and $E_{ex} = -J + \Delta$ for case iv). Note that the above senario, i.e., tuning of the exchange interaction by the inverse magnetic proximity effect, does not involve electron transfer and it must be distinguished from the modulation of exchange interaction by spin-polarised electron filling [35].

To give a perspective on how the modulation of $E_{ex}$ influences AHE, we consider a minimal two-band Hamiltonian with inversion and time reversal symmetry breaking [2,36],

$$\boldsymbol{H}(\mathbf{k}) = \frac{\hbar^2 \boldsymbol{k}^2}{2\boldsymbol{m}} + \hbar\boldsymbol{v}\mathbf{k}\cdot(\mathbf{z}\times\boldsymbol{\sigma}) - \boldsymbol{E}_{\mathbf{ex}}\boldsymbol{\sigma}_{\mathbf{z}} \quad (2)$$

where $\hbar$ is the reduced Planck constant, **k** the wavevector, $m$ the effective mass, $v$ the velocity, and **σ** the Pauli matrices. Although the band structure in real materials is more complex, leading to a more complex behaviour of $\sigma_{AHE}$ as a function of $E_{ex}$, Eq. (2) captures the essential physics that gives a major contribution to the AHE. If $\hbar vk \gg E_{ex}$ holds and one assumes that the Fermi level lies in the upper band, the magnitude of $\sigma_{AHE}$ for a certain $k$ is obtained as [12]

$$\sigma_{AHE} \approx \frac{e^2}{2h}\frac{E_{ex}}{\hbar vk}, \quad (3)$$

where $e$ is the electronic charge, and $h$ the Planck constant. Eq. (3) shows that $\sigma_{AHE}$ is linearly proportional to $E_{ex}$, and the magnitude of $\sigma_{AHE}$ is enhanced (reduced) by $\delta\left(=\frac{e^2}{2h}\frac{\Delta}{\hbar vk}\right)$ for case i and ii (case iii and iv). The modulation of $\sigma_{AHE}$ by $E_{ex}$ for each case is summarized in Table 1.

| Case | i | ii | iii | iv |
|---|---|---|---|---|
| **I** | +**x** | −**x** | −**x** | +**x** |
| $\mathrm{M}_z^I$ | +**z** | −**z** | −**z** | +**z** |
| **M** | +**z** | −**z** | +**z** | −**z** |
| $E_{ex}$ | $J+\Delta$ | $-J-\Delta$ | $J-\Delta$ | $-J+\Delta$ |
| $\sigma_{AHE}$ | $\sigma_{AHE}^0+\delta$ | $-\sigma_{AHE}^0-\delta$ | $\sigma_{AHE}^0-\delta$ | $-\sigma_{AHE}^0+\delta$ |

Table 1. Summary of the AHE modulation for cases i-iv, corresponding to the different scenarios schematically shown in Figs. 2a-2d, respectively. **I**: charge current, $\mathrm{M}_z^I$: current generated magnetization by BCD, **M**: magnetization of FM, $E_{ex}$: effective exchange splitting, $J$: intrinsic exchange splitting of FM, $\Delta$: modulation of the exchange splitting by $\mathrm{M}_z^I$, $\sigma_{AHE}$: anomalous Hall conductivity, $\sigma_{AHE}^0$: intrinsic anomalous Hall conductivity at $I$ = 0. The modulation of the anomalous Hall conductivity is denoted by $\delta$, where $\delta = \frac{e^2}{2h}\frac{\Delta}{\hbar vk}$.

To modulate AHE, the superposition of an ac detection current $I_{ac}$ and a dc modulation current $I$ was applied *simultaneously*, and the resulting ac transverse voltage $V_{ac}$ was measured (Fig. 1e and Ref. 12). Note that an out-of-plane magnetic field $H_z$ is applied in the measurements, and that the magnitude of $I$ is kept below the critical switching threshold, ensuring that no current-induced **M** switching occurs and thus spin-orbit torques don't play a role for the modulation. To match the symmetry of **I** and **M** for effective modulation (Table 1), the polarity of **I** (and thus $\mathbf{M}_z^I$) is synchronized with the polarity of **M** (upper panel of Fig. 3a). We define the modulation current $I_{mod}$ as, $I_{mod} = (I^{\uparrow} - I^{\downarrow})/2$, where $I^{\uparrow}$ ($I^{\downarrow}$) represents the charge current when the magnetization **M** points along the +**z** (–**z**) direction. The corresponding $\sigma_{xy}$ loop of device DM1 measured at 10 K is shown in the lower panel of Fig. 3a. The height of the AHE loop varies drastically when $I_{mod}$ is varied from +1 mA to −1 mA, which demonstrates a clear modulation of the AHE. The results can be understood from Table 1: case i (**I** // +**x** and **M** // +**z**) and case ii (**I** // −**x** and **M** // −**z**), respectively, lead to an enhancement of $\sigma_{AHE}$ to $\sigma^0_{AHE}+\delta$ and $-\sigma^0_{AHE}-\delta$. Therefore, the height of the loop increases by $2\delta$. In contrast, case iii (**I** // −**x** and **M** // +**z**) and case iv (**I** // +**x** and **M** // −**z**), respectively, lead to a decrease of $\sigma_{AHE}$ to $\sigma^0_{AHE}-\delta$ and $-\sigma^0_{AHE}+\delta$ , and thus the height of the loop reduces by $2\delta$. The magnitude of $\sigma_{AHE}$ for each $I_{mod}$ is summarized in Fig. 3b, and a drastic modulation of $\sigma_{AHE}$ is observed. Starting from $I_{mod}$ = 0 mA to 0.5 mA, $\sigma_{AHE}$ increases from 970 $\Omega^{-1}$m$^{-1}$ to 1800 $\Omega^{-1}$m$^{-1}$, and for $I_{mod}$ > 0.5 mA, $\sigma_{AHE}$ slightly decreases. For negative $I_{mod}$ from 0 mA to −0.5mA, $\sigma_{AHE}$ decreases from 970 $\Omega^{-1}$m$^{-1}$ to ~ 0 $\Omega^{-1}$m$^{-1}$, and for $I_{mod}$ < −0.5 mA, $\sigma_{AHE}$ slightly increases. Moreover, we demonstrate that another different modulation measurement, i.e., keeping the polarity of **I** (and thus $\mathbf{M}^I_z$) fixed while changing the direction of **M** by sweeping $H_z$, gives identical results as obtained in Fig. 3b [12].

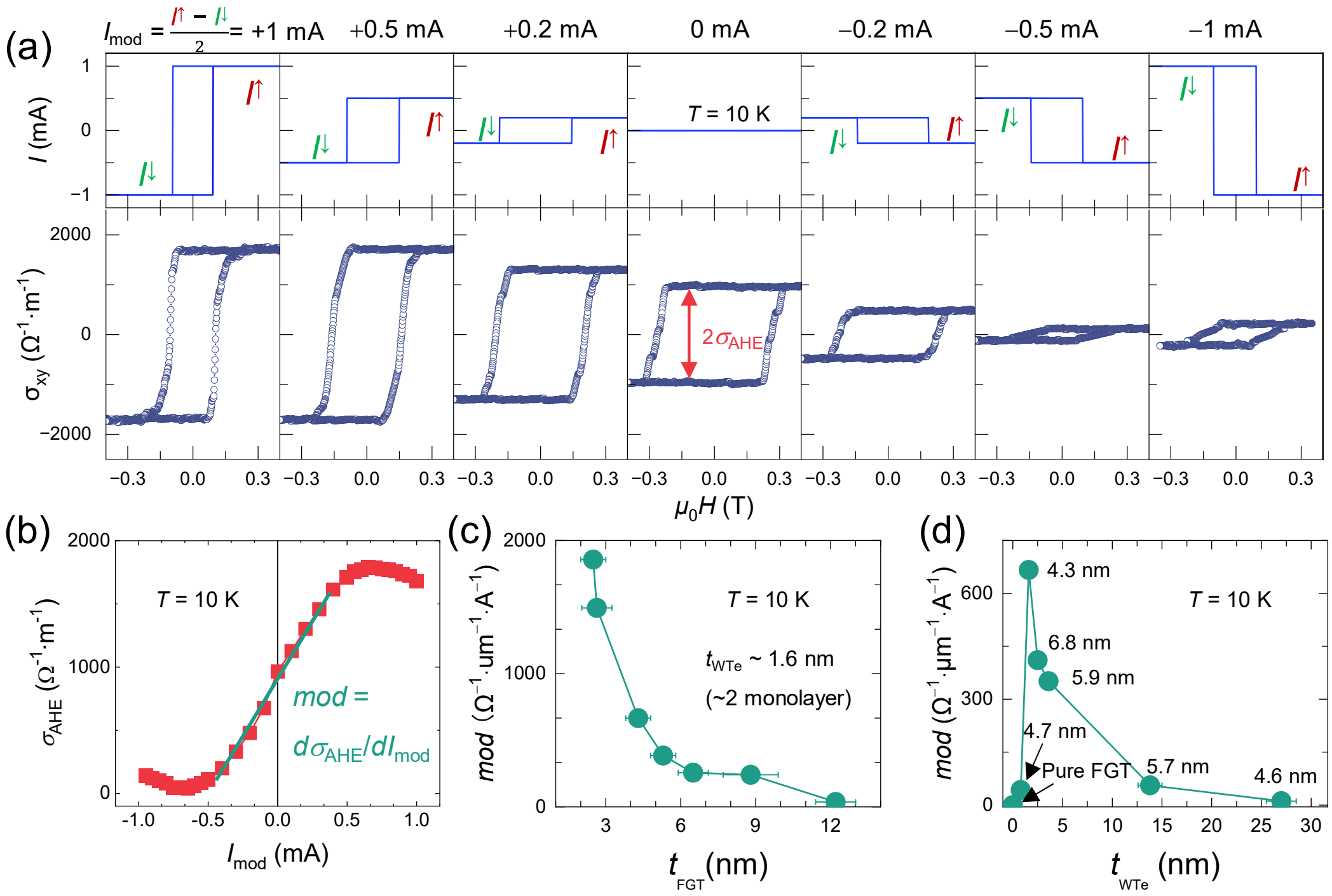

FIG. 3. (a) Upper panel: AHE modulation is performed by a **M**-dependent charge current. Here the modulation current $I_{mod}$ is defined as, $I_{mod} = (I^{\uparrow} - I^{\downarrow})/2$, where $I^{\uparrow}$ ($I^{\downarrow}$), as marked in the panel, represents the charge current when **M** points along the +z (–z) direction. From left to right: the sign of $I_{mod}$ changes from positive ($I_{mod} > 0$) to negative ($I_{mod} < 0$). Lower panel: the corresponding $\sigma_{xy}$ loops of device DM1 measured at 10 K. (b) $I_{mod}$-dependence of $\sigma_{AHE}$, and the modulation amplitude *mod* is obtained from the linear fit around $I = 0$. (c) *mod* as a function of $t_{FGT}$ for fixed $t_{WTe}$ of ~2 ML. See Supplementary Material how the modulation amplitude is corrected by the device width and the direction of **I**. (d) *mod* as a function of $t_{WTe}$ for $t_{FGT}$ approximately equals to 5 nm. The actual $t_{FGT}$ for each device, ranging from 4.3 nm to 5.9 nm, is indicated in the figure. The *x*-axis error bar of in (c) and (d) comes from the uncertainties of thickness measurements, while the y-axis error bar, obtained from the linear fit in (b), is smaller than the symbol size.

The relative modulation amplitude can be obtained by $\Delta\sigma_{AHE}/\sigma_{AHE}(0)$, where $\Delta\sigma_{AHE}$ is the change of AHE conductivity and $\sigma_{AHE}(0)$ is the AHE conductivity at $I_{mod}$ = 0. In our device, the ratio reaches a record value of ~180%, significantly exceeding values obtained by previous methods. For example, static gating [30] yields $\Delta\sigma_{AHE}/\sigma_{AHE}(0)$ ~ 40%, and high-pressure [37] achieves ~100%. Note that high pressure only suppresses the AHE, whereas our current-driven approach enables both enhancement and suppression, offering bidirectional control. Moreover, this drastic tunability indicates that our device functions analogously to a field-effect transistor, i.e., an anomalous Hall transistor, in which the charge current serves as the gate**,** and the transverse AHE signal functions as the source-drain channel.

We notice that the application of a dc current inevitably introduces Joule heating and changes $\sigma_{AHE}$ in a trivial way. However, a heating effect can be excluded because i) the $\sigma_{AHE}$-$I_{mod}$ relationship induced by Joule heating is an even function with respect to $I_{mod}$, which cannot explain the odd $\sigma_{AHE}$-$I_{mod}$ relationship observed experimentally. ii) The change of $R_{AHE}$ by Joule heating is quantified to be about 1.6 Ω, which is much smaller than the modulation of 13 Ω [12]. We have also studied several control devices [12], and the magnitude of the modulation *mod* can be approximately quantified by the slope of the $\sigma_{AHE}$-$I_{mod}$ trace (Fig. 3b) in the linear regime (i.e., $|I_{mod}|$ < 0.5 mA), i.e., $mod = d\sigma_{AHE}/dI_{mod}$. The results show that: i) For fixed $t_{WTe}$ ($t_{WTe}$ = 1.6 nm, ~2 monolayers), *mod* decreases as $t_{FGT}$ increases (Fig. 3c), which indicates that the modulation by $\mathbf{M}_z^I$ is an interfacial effect. ii) For pure FGT samples as well as for currents applied along the high symmetry *b*-axis of the $WTe_2$/FGT devices, negligible modulation is observed [12]; iii) *mod* depends on $t_{WTe}$ in a non-monotonic way (Fig. 3d): for monolayer $WTe_2$, the modulation is negligibly small; while for bilayer of $WTe_2$, the modulation is maximized; and when further increasing $t_{WTe}$, the modulation decreases. This trend is consistent with the $t_{WTe}$-dependence of Λ as shown in Fig. 1d, confirming that BCD plays the dominant role for the AHE modulation. Moreover, we have also measured the current-induced Hall signal in pure $WTe_2$, and the signal is only approximately 0.25 Ω at a current of 1 mA. This value is at least 13 times smaller than the modulation observed in the $WTe_2$/FGT bilayer. We therefore conclude that the observed modulation cannot be attributed to an additional current-induced Hall contribution originating from $WTe_2$.

To better understand the giant modulation of AHE, we study $\sigma_{AHE}$ by first-principle calculations. The results reveal that $\sigma_{AHE}$ of FGT depends strongly on both the Fermi level position and the spin-splitting. For example, shifting the Fermi level by just ~20 meV changes not only the magnitude but even the sign of $\sigma_{AHE}$ [12]. Moreover, as shown in Ref. [12], the calculated $\sigma_{AHE}$ varies strongly with Δ over an energy window of 100 meV — closely reflecting the key trends seen in the experimental data (Fig. 3b). From this, we infer that the modulation of AHE corresponds to a modulation of the exchange interaction by $\Delta \approx 50$ meV, which is about 3% of the intrinsic exchange of FGT ($J \sim 1.5$ eV according to Refs. 38 and 39). This result stands in sharp contrast to the oversimplified two-band model described by Eq. 2, which implies $\Delta = 0.93J$, significantly larger than that determined from a more realistic band structure.

Our results show that the anomalous Hall effect of the two-dimensional perpendicular magnetic metal $Fe_3GeTe_2$ can be drastically and reversibly modulated by a charge current in proximal few-layer $WTe_2$, where the current induced out-of-plane non-equilibrium magnetization by Berry curvature dipole plays the dominating role. The non-linear dependence of $\sigma_{AHE}$ on the charge current, along with the giant modulation, has not been achieved by any other existing method. The giant modulation suggests the feasibility of an anomalous Hall transistor. The proposed mechanism for such modulation could be oversimplified, and further theoretical studies are necessary to fully understand the current-generated torques and effective exchange magnetic-fields. Even larger modulation of AHE could be expected with improved devices capable of sustaining substantially larger dc current, or exploring other materials hosting stronger BCD [25,40-44]. Besides the AHE control, this approach can be used for the control of other static magnetic properties of magnetic materials, e.g., the phase transition temperature, the magnetic anisotropies, Curie temperature, etc. Since the control of AHE is independent of electron filling/transfer, this unique method opens up new avenues for controlling quantum magnetic insulators. For example, this approach could be extended to magnetic topological insulators to raise the observation temperature of the quantum anomalous Hall effect, which is currently limited by the relatively small exchange gap of a few tens of meV [11]. It could also be ultilised to drive a topological insulator into a magnetic Weyl semimetal state and to tune the separation between Weyl points [45], providing a novel method for manipulating emerging magnetic correlated states.

## Acknowledgements

This work was funded by the Deutsche Forschungsgemeinschaft by TRR 360–492547816 and SFB1277-314695032, by the excellence cluster MCQST under Germany's Excellence Strategy EXC-2111 (Project no. 390814868), and by FLAG-ERA JTC 2021-2DSOTECH. J. Guo, J. Shao and P. Ji were partially supported by the China Scholarship Council (CSC).

## Author contributions

L.C. and C.H.B planned the study. L.C. had the general idea for AHE control. J.G. J.S, P.J, X.H, J. F. and T.N.G.M fabricated the devices. J.G collected the data which have been analysed by J.G and L.C. G.B. R.P. J. K. and J.F. did the theoretical calculations. L.C. wrote the manuscript with input from all other co-authors. All authors discussed the results.

## Data availability

The data that support the findings of this article are not publicly available. The data are available from the authors upon reasonable request.

# Figure captions

FIG. 1. (a) Top- and (b) side- view of the crystal structure of $WTe_2$, where **a**, **b** and **c** represent the crystal axes. $WTe_2$ has a mirror plane $M_a$ and a glide mirror plane $M_b$ as indicated by the dashed lines. Applying a charge current **I** along the low symmetry **a**-axis generates an out-of-plane magnetization $\mathbf{M}_z^{\mathrm{I}}$ (red arrow in b) due to BCD. (c) Schematic of the band structure of bilayer of $WTe_2$. The orange arrow represents the BCD formed by the layer-polarized Dirac fermions in bilayer $WTe_2$. (d) $t_{WTe}$-dependence of the strength of the BCD (**Λ**) quantified by non-linear Hall measurements at $T$ = 10 K. Note the logarithmic scale. (e) Schematic of the experimental set up for current modulation of AHE. $I_{ac}$ is the ac sense current, and $V_{ac}$ is the transverse voltage detected by a lock-in amplifier, $I$ is the dc modulation current, and $I \gg I_{ac}$ holds. (f) Anomalous Hall resistance $R_{xy}$ (= $V_{ac}/I_{ac}$) as a function of $H_z$ of the $WTe_2$/FGT stack at $T$ = 10 K. $R_{AHE}$ is determined by the height of the AHE loop. (g) Pulsed current ($I_p$) induced magnetization switching in the $WTe_2$/FGT stack without auxiliary magnetic field at $T$ = 10 K.

FIG. 2. (a) Upper panel: The application a positive charge current **I** generates a $\mathbf{M}_z^{\mathrm{I}}$ (// +**z**) via BCD. If **M** of FM is also along +**z**, the effective field of $\mathbf{M}_z^{\mathrm{I}}$ leads to an enhancement of the effective exchange coupling $E_{ex}$ from $J$ to $J$+Δ, where $J$ is the intrinsic exchange energy of FM and Δ the increase of the exchange energy. Lower panel: in a minimal two-band model, an avoided crossing between the two bands emerges, characterized by an energy splitting of 2($J$+Δ). The avoided crossing is accompanied by a non-zero BC, arising from the warped spin texture in **k**-space. At $k$ = 0, the spin directions of the lower and upper bands are represented by upward and downward arrows, respectively. For $k$ far from zero, the spin orientation is indicated by the symbol ⊗ (into the page) and ⊙ (out of the page). In this scenario, the enhanced exchange splitting leads to an enhanced BC, and consequently, an enhanced anomalous Hall conductivity $\sigma_{AHE}$. (b) For $\mathbf{M}_z^{\mathrm{I}}$ // **M** // –**z**, an enhanced $\sigma_{AHE}$ is also expected. (c) and (d): If $\mathbf{M}_z^{\mathrm{I}}$ and **M** are anti-parallel, $\mathbf{M}_z^{\mathrm{I}}$ reduces the exchange splitting of FM. This leads to a reduced BC and $\sigma_{AHE}$. The dashed lines in the lower panels of (a), (b), (c) and (d) represent the band structure without modulation of the exchange interaction.

FIG. 3. (a) Upper panel: AHE modulation is performed by a **M**-dependent charge current. Here the modulation current $I_{mod}$ is defined as, $I_{mod}$ = ($I^{\uparrow}$ – $I^{\downarrow}$)/2, where $I^{\uparrow}$ ($I^{\downarrow}$), as marked in the panel, represents the charge current when **M** points along the +z (–z) direction. From left to right: the sign of $I_{mod}$ changes from positive ($I_{mod} > 0$) to negative ($I_{mod} < 0$). Lower panel: the corresponding $\sigma_{xy}$ loops of device DM1 measured at 10 K. (b) $I_{mod}$-dependence of $\sigma_{AHE}$, and the modulation amplitude *mod* is obtained from the linear fit around $I = 0$. (c) *mod* as a function of $t_{FGT}$ for fixed $t_{WTe}$ of ~2 ML. See Supplementary Material how the modulation amplitude is corrected by the device width and the direction of **I**. (d) *mod* as a function of $t_{WTe}$ for $t_{FGT}$ approximately equals to 5 nm. The actual $t_{FGT}$ for each device, ranging from 4.3 nm to 5.9 nm, is indicated in the figure. The *x*-axis error bar of in (c) and (d) comes from the uncertainties of thickness measurements, while the y-axis error bar, obtained from the linear fit in (b), is smaller than the symbol size.

Table 1. Summary of the AHE modulation for cases i-iv, corresponding to the different scenarios schematically shown in Figs. 2a-2d, respectively. **I**: charge current, $\mathrm{M}_z^I$: current generated magnetization by BCD, **M**: magnetization of FM, $E_{ex}$: effective exchange splitting, *J*: intrinsic exchange splitting of FM, $\Delta$: modulation of the exchange splitting by $\mathrm{M}_z^I$, $\sigma_{AHE}$: anomalous Hall conductivity, $\sigma_{AHE}^0$: intrinsic anomalous Hall conductivity at $I = 0$. The modulation of the anomalous Hall conductivity is denoted by $\delta$, where $\delta = \frac{e^2}{2h}\frac{\Delta}{\hbar v k}$.

Supplementary Materials for

# Electric-current control of anomalous Hall effect

J. P. Guo[1,4], G. M. S. Brizolla[2], P. Rao[1], J. Shao[1,4], X. H. Huang[1,4], T. N. G. Meier[1,4], P. R. Ji[5], J. J. Finley[3,5], J. Fabian[2], J. Knolle[1], C. H. Back[1,3,4], and L. Chen[1,4]

[1]*Department of Physics, Technical University of Munich, Munich, Germany*
[2]*Institute of Theoretical Physics, University of Regensburg, Regensburg, Germany*
[3]*Munich Center for Quantum Science and Technology (MCQST), Munich, Germany*
[4]*Center for Quantum Engineering (ZQE), Technical University of Munich, Munich, Germany*
[5]*Walter Schottky Institute and TUM School of Natural Science, Technical University of Munich, Munich, Germany*

## Table of Contents:

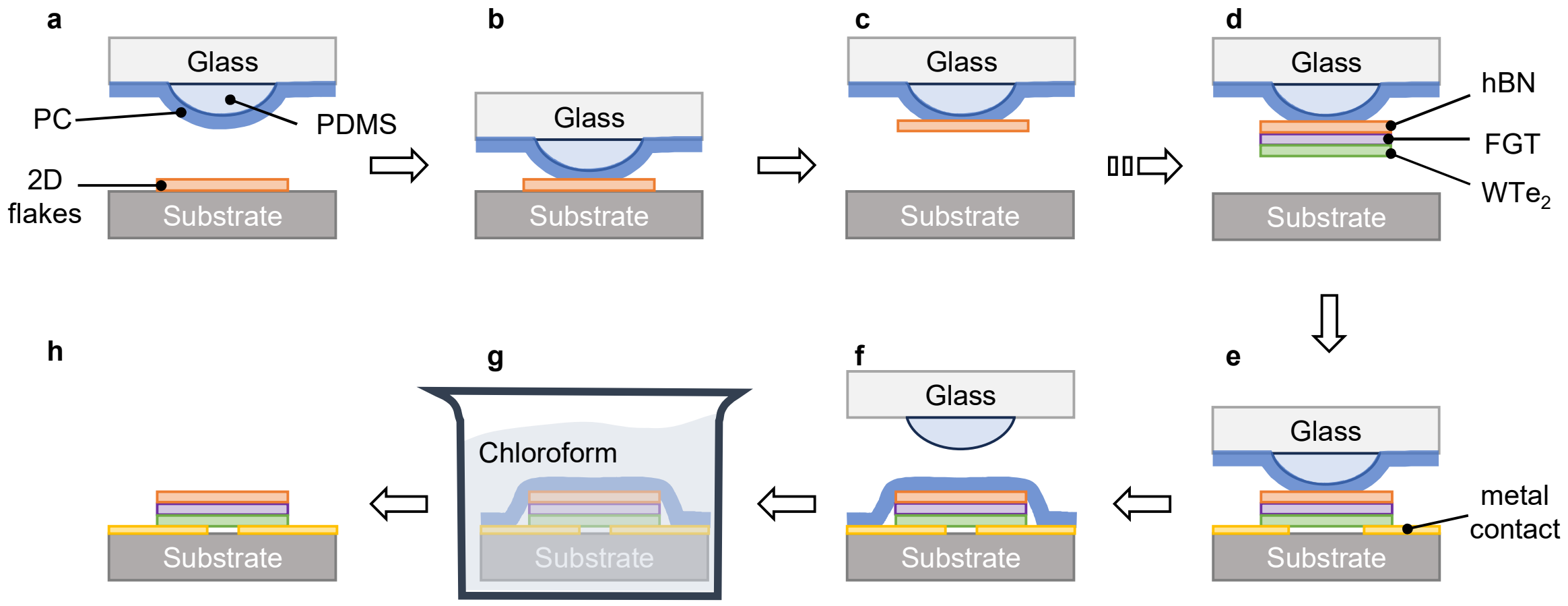

**Figure S1 | Schematic illustration of the stacking process for the fabrication of the $WTe_2$/FGT devices.** (**a**) A stamp, consisting of dome-shaped polydimethylsiloxane (PDMS) with a thin polycarbonate (PC) layer on top, is prepared to pick up the 2D materials. (**b**) The PDMS/PC stamp brings the PC layer into contact with the target flake at 110 °C. (**c**) The stamp lifts off from the substrate, and the flake is picked up. (**d**) By repeating the processes (**b**) and (**c**), the h-BN/FGT/$WTe_2$ stack is formed. (**e**) The stamp is brought into contact with the pre-patterned electrodes on the Si/$SiO_x$ substrate. (**f**) By heating up the Si/$SiO_x$ substrate to 170 °C, the 2D stack is released onto the substrate together with the melted PC layer. (**g**) The sample is immersed in chloroform to dissolve the PC layer. (**h**) Completed devices with different kinds of electrodes have been fabricated.

## 1. Experimental Methods

### 1.1 Materials and device fabrication

Bulk $WTe_2$, $Fe_3GeTe_2$, and hBN crystals were purchased from HQ Graphene. We fabricated $WTe_2$/$Fe_3GeTe_2$/hBN devices using a previously reported layer-by-layer dry transfer technique [S1]. As shown in Fig. S1, thin flakes were prepared by mechanical exfoliation using tape, with hBN flakes exfoliated onto Si/$SiO_2$ (285 nm) substrates, while $WTe_2$ and FGT flakes were exfoliated onto sapphire substrates. The stacking process was carried out using a PDMS/PC stamp. We sequentially picked up the hBN, FGT, and $WTe_2$ layers, and then transferred them onto the pre-patterned bottom Au-electrodes. The bottom Au-electrodes, with cross-, Hall-bar- and circular-shapes, were defined via electron beam lithography (EBL) on a $Si/SiO_2$ (285 nm) substrate, and consist of evaporated Ti (3 nm)/Au (7 nm) layers. All the exfoliation and the stacking procedures were conducted inside a nitrogen-filled glovebox to prevent degradation, and both $O_2$ and $H_2O$ concentration are less than 0.1 ppm. After device encapsulation, atomic force microscopy (AFM) was employed to identify the thickness of each layer.

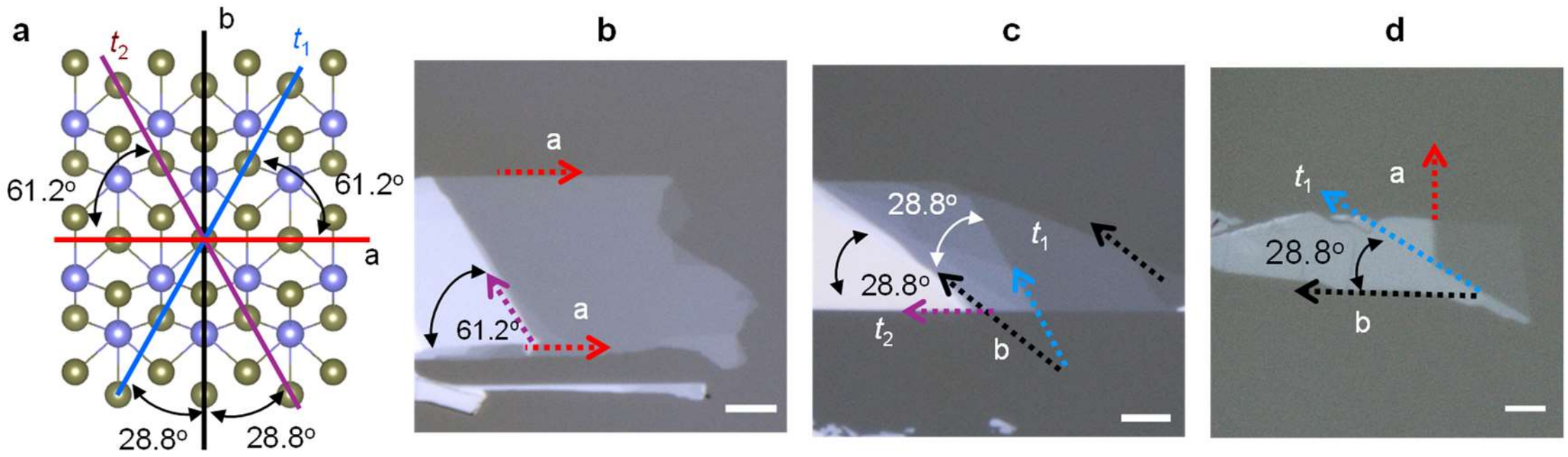

**Figure S2 | Determining the crystallographic axis of $WTe_2$.** (**a**) Top view of the $WTe_2$ crystal structure, where the four possible cleaving directions are labeled by solid lines. Here **a** is the low symmetry *a*-axis, **b** the high symmetry *b*-axis, $t_1$ and $t_2$ the direction of Te chains. $t_1$ ($t_2$) forms an angle of 61.2° (28.8°) with respect to the *a*- (*b*-)axis. (**b**) An example of identifying the *a*-axis of $WTe_2$ by two straight edges forming an angle of 61.2°. In this case the long straight edge is the *a*-axis. (**c**) Second example of identifying the *b*-axis by three straight edges. Here the middle edge forms an angle of 28.8° with the other two. In this case, the middle edge is the *b*-axis. (**d**) Third example of identifying the *a*- and *b*-axis by three edges. Here the two perpendicular edges are a- and b-axis. By measuring the angle between orthogonal edges with a third straight edge, one can determine which orthogonal edge is the *a*- and *b*-axis. The scale bar in (**b**), (**c**), and (**d**) is 5 μm.

During device fabrication, the a- and b-axes of $WTe_2$ need to be approximately aligned with the Hall bar geometry. As shown in Fig. S2a, $WTe_2$ tends to cleave preferentially along the a- and b-axes, as well as along the Te chain directions (labeled as $t_1$ and $t_2$). $t_1$ and $t_2$ form an angle of 61.2°(28.8°) with respect to the *a*- (*b*)-axis. The crystal orientation can thus be determined from the relative angles of the long edges of the exfoliated flakes. We first identify long, straight edges on selected flakes, and then determine the *a*- and *b*-axis by the angles between the edges. Because monolayer and bilayer flakes often exhibit irregular geometries, we determine their axis by neighboring thicker flakes. Three typical examples have been shown in Fig. S2c-d.

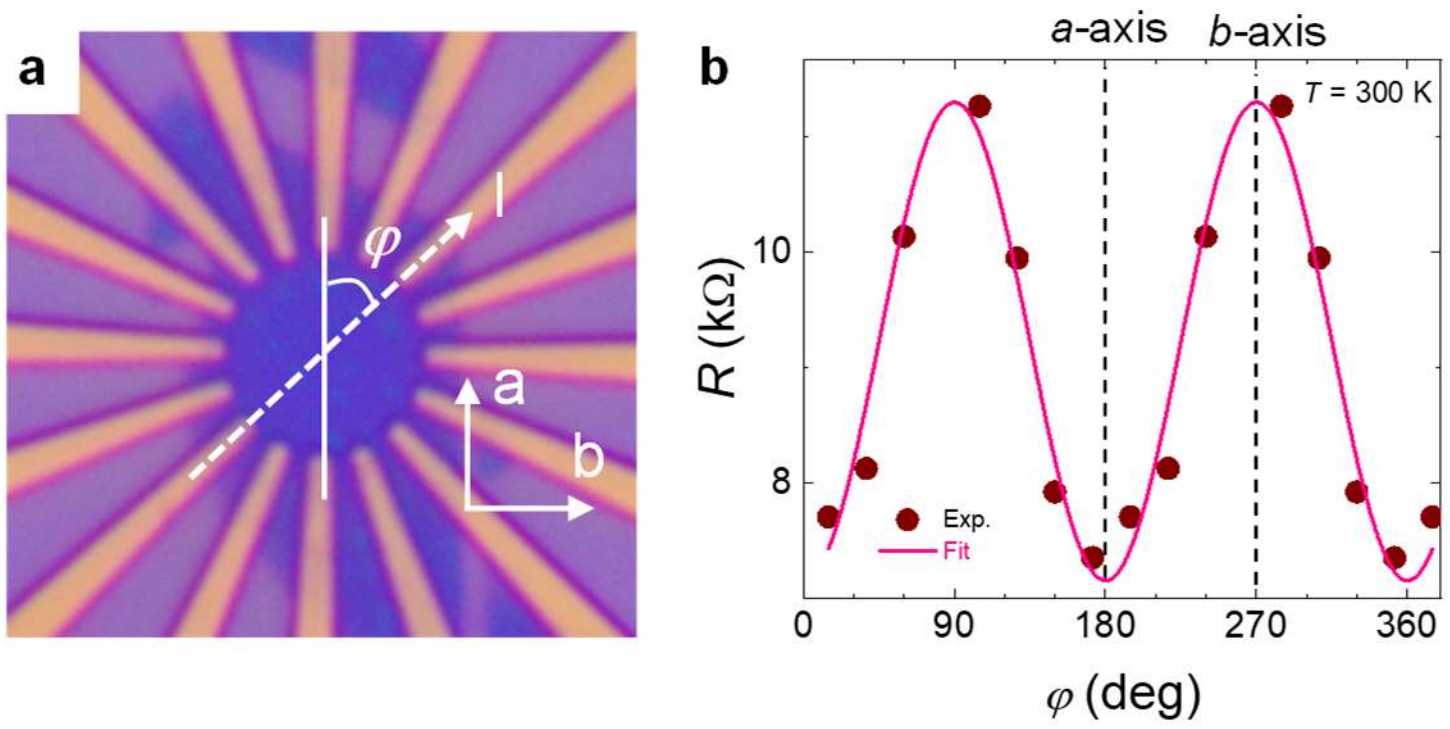

**Figure S3 | Angular dependence of the resistance of $WTe_2$.** (**a**) Optical image of the circular disk used to measure the resistance anisotropy of $WTe_2$ ($t_{WTe}$ ~ 7 nm). **a** (**b**) is the low (high) symmetry axis, and $\varphi$ is the angle between the *a*-axis and the current **I**. (**b**) $\varphi$-dependence of the resistance measured at 300 K.

The orientation of the $WTe_2$ flakes is further confirmed by resistance measurements. A disk device (Fig. S3a) is used to measure the resistance anisotropy of $WTe_2$. A charge current **I** is applied along each of the 16 electrodes and the two-terminal resistance $R$ is measured. The $\varphi$-dependence of $R$ is summarized in Fig. S3b, where $\varphi$ is defined as the angle between **I** and the *a*-axis. The resistance shows two-fold symmetry, which is consistent with the crystal symmetry of few layers of $WTe_2$. The data can be well fitted by [24]

$$R = R_b \cos^2\varphi + R_a \sin^2\varphi, \qquad \text{(S1)}$$

where $R_b$ ($R_a$) is the resistance along the *a*-(*b*-)axis. The fitting gives a resistance anisotropy $R_a/R_b$ of 0.64, which is consistent with previous reports [24].

| Device | ∠ (deg) | $t_{WTe}$ (nm) | $t_{FGT}$ (nm) | $l$ (μm) | $w$ (μm) |
|---|---|---|---|---|---|
| DS | 0 | 7.0±1.0 | 2±0.9 | 5.5 | 2 |

**Table S1. Parameters for magnetization switching devices**. ∠ is the angle between the dc current **I** and the low symmetry *a*-axis of $WTe_2$ (as defined in Fig. S3a), $t_{WTe}$ the thickness of $WTe_2$, $t_{FGT}$ the thickness of FGT, and $l$ ($w$) the length (width) of the Hall bar.

| Device | ∠ (deg) | $t_{WTe}$ (nm) | $t_{FGT}$ (nm) | $l$ (μm) | $w$ (μm) | $R_{AHE}(0)$ (Ω) | $R^{max}_{AHE}(+I)$ (Ω) | $R^{min}_{AHE}(-I)$ (Ω) | $\Delta R_{AHE}$ (Ω) | $\Delta R_{AHE}/R_{AHE}(0)$ (%) | $dR_{AHE}/dI$ (Ω/mA) | *mod* ($\Omega^{-1}\cdot\mu m^{-1}\cdot A^{-1}$) |
|---|---|---|---|---|---|---|---|---|---|---|---|---|
| DM1 | 0 | 1.5±0.2 | 2.5±0.5 | 3 | 3 | 7.1 | 13.5 | 0.3 | 13.2 | 185.9 | 13.7 | 1856.3 |
| DM2 | 29 | 1.8±0.3 | 2.4±0.6 | 2 | 2 | 15.7 | 23.5 | 13.0 | 10.5 | 66.9 | 9.8 | 1241.5 |
| DM3 | 29 | 1.6±0.2 | 4.3±0.5 | 2 | 2 | 6.5 | 9.0 | 6.4 | 2.6 | 40.0 | 2.5 | 559.0 |
| DM4 | 0 | 1.5±0.3 | 5.3±0.5 | 5 | 5 | 8.8 | 10.1 | 7.6 | 2.5 | 28.4 | 1.7 | 385.3 |
| DM5 | 0 | 1.3±0.3 | 6.5±0.6 | 2 | 2 | 1.7 | 1.9 | 1.4 | 0.5 | 29.4 | 0.5 | 256.6 |
| DM6 | 29 | 1.5±0.8 | 8.8±1.1 | 3 | 3 | 3.5 | 4.1 | 2.8 | 1.3 | 37.1 | 0.7 | 241.0 |
| DM7 | 0 | 1.6±0.3 | 12.2±0.8 | 2 | 2 | 2.9 | 2.9 | 2.7 | 0.2 | 6.8 | 0.1 | 36.6 |

**Table S2. Device parameters for devices with various FGT thicknesses**. $R_{AHE}(0)$ is the anomalous Hall resistance at $I$ = 0, $R^{max}_{AHE}(+I)$ is the maximum $R_{AHE}$ for positive current, $R^{max}_{AHE}(-I)$ is the minimum $R_{AHE}$ for negative current, $\Delta R_{AHE} = R^{max}_{AHE}(+I) - R^{max}_{AHE}(-I)$, and the ratio $\Delta R_{AHE}/R_{AHE}(0)$ shows the relative change of the AHE. $dR_{AHE}/dI$ is the slope obtained from the $R_{AHE}$-$I$ curve around $I$ = 0. *mod* is the corrected modulation magnitude (see text). The $t_{FGT}$-dependence of *mod* is summarized in Fig. 3c.

| Device | ∠ (deg) | $t_{WTe}$ (nm) | $t_{FGT}$ (nm) | $l$ (μm) | $w$ (μm) | $R_{AHE}(0)$ (Ω) | $R^{max}_{AHE}(+I)$ (Ω) | $R^{min}_{AHE}(-I)$ (Ω) | $\Delta R_{AHE}$ (Ω) | $\Delta R_{AHE}/R_{AHE}(0)$ (%) | $dR_{AHE}/dI$ (Ω/mA) | *mod* ($\Omega^{-1}\cdot\mu m^{-1}\cdot A^{-1}$) |
|---|---|---|---|---|---|---|---|---|---|---|---|---|
| DM8 | n.a | 0 | 4.6 | 3 | 3 | 36.7 | 23.0 | 23.0 | 0.5 | 0.01 | 0 | 0 |
| DM9 | 0 | 0.8±0.3 | 4.7±0.7 | 1 | 1 | 11.6 | 8.2 | 7.5 | 0.6 | 5.2 | 0.57 | 42.7 |
| DM3 | 29 | 1.6±0.2 | 4.3±0.5 | 2 | 2 | 6.5 | 9.0 | 6.4 | 2.6 | 40.0 | 2.5 | 559.0 |
| DM10 | 0 | 2.5±0.3 | 6.8±0.7 | 2 | 2 | 6.9 | 10.4 | 7.1 | 3.3 | 47.0 | 2.06 | 410.7 |
| DM11 | 0 | 3.6±0.4 | 5.9±0.9 | 3 | 3 | 3.0 | 5.5 | 3.8 | 1.7 | 56 | 1.06 | 351.0 |
| DM12 | 0 | 13.8±1.2 | 5.7±0.8 | 2 | 2 | 2 | 4.4 | 4.03 | 0.37 | 18.5 | 0.14 | 56.2 |
| DM13 | 0 | 27.0±1.5 | 4.6±0.4 | 5 | 5 | 1.2 | 1.2 | 1.3 | 0.1 | 8.3 | 0.01 | 11.3 |

**Table S3. Device parameters for devices with various $WTe_2$ thicknesses**. The $t_{WTe}$-dependence of modulation is summarized in Fig. 3d.

| Device | ∠ (deg) | $t_{WTe}$ (nm) | $t_{FGT}$ (nm) | $l$ (μm) | $w$ (μm) | $R_{AHE}(0)$ (Ω) | $R^{max}_{AHE}(+I)$ (Ω) | $R^{min}_{AHE}(-I)$ (Ω) | $\Delta R_{AHE}$ (Ω) | $\Delta R_{AHE}/R_{AHE}(0)$ (%) | *mod* ($\Omega^{-1}\cdot\mu m^{-1}\cdot A^{-1}$) |
|---|---|---|---|---|---|---|---|---|---|---|---|
| DM4 | 0 | 1.5±0.3 | 5.3±0.5 | 5 | 5 | 8.8 | 10.1 | 7.6 | 2.5 | 28.4 | 385.3 |
| DM4 | 90 | 1.5±0.3 | 5.3±0.5 | 5 | 5 | 9.2 | 9.3 | 9.2 | 0.1 | 1 | 65.8 |
| DM3 | 29 | 1.6±0.2 | 4.3±0.5 | 2 | 2 | 6.5 | 9.0 | 6.4 | 2.6 | 40.0 | 559.0 |
| DM3 | 61 | 1.6±0.2 | 4.3±0.5 | 2 | 2 | 6.2 | 7.7 | 6.0 | 1.7 | 27.4 | 362.5 |
| DM2 | 29 | 1.8±0.3 | 2.4±0.6 | 2 | 2 | 15.7 | 23.5 | 13.0 | 10.5 | 66.9 | 1241.5 |
| DM2 | 61 | 1.8±0.3 | 2.4±0.6 | 2 | 2 | 16.5 | 23.4 | 14.7 | 8.7 | 52.7 | 1077.7 |
| DM10 | 0 | 2.5±0.3 | 6.8±0.7 | 2 | 2 | 6.9 | 10.4 | 7.1 | 3.3 | 47.0 | 410.7 |
| DM10 | 90 | 2.5±0.3 | 6.8±0.7 | 2 | 2 | 6.6 | 8.1 | 7.5 | 0.6 | 9 | 55.9 |

**Table S4. Device parameters for devices with different current directions**. Negligible modulation is observed when the charge current is along the high-symmetry axis (∠=90°). When the charge current is applied neither along the low-symmetry axis nor along the high-symmetry axis, sizeable modulation is observed.

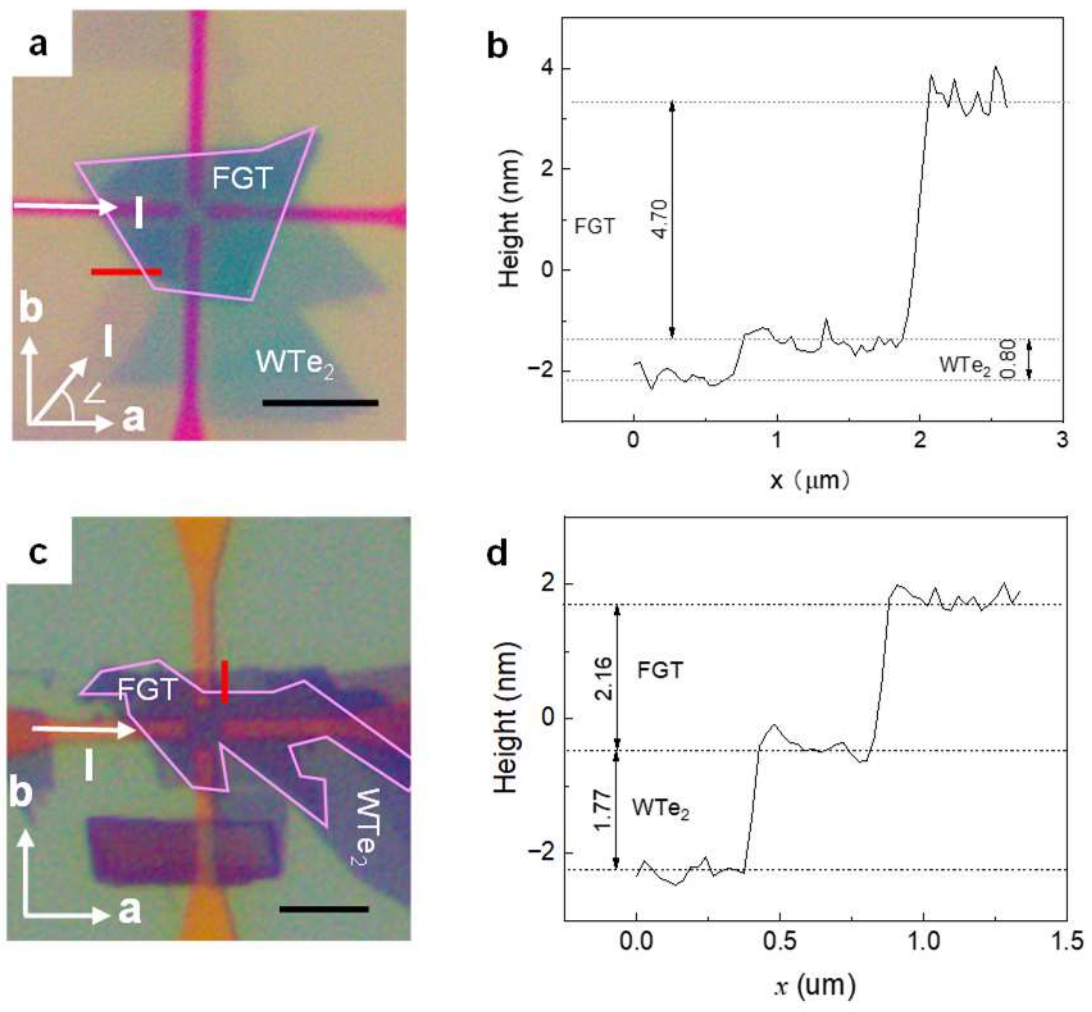

**Figure S4 | Optical images and AFM measurements of the $WTe_2$/FGT devices.** (**a**) Optical image and (**b**) atomic force microscopy measurements of a monolayer $WTe_2$ (0.8 nm)/FGT (4.7 nm) device. (**c**) and (**d**): a bilayer of $WTe_2$ (1.8 nm)/FGT (2.2 nm) device. The dashed lines in (**a**) and (**c**) show the scanning direction for the thickness measurements. The scale bar in (**a**) and (**b**) is 5 μm.

A summary of all the studied devices is listed in Tables S1-S4, where the magnetization switching device is named by DS, and the AHE modulation devices are named by DM. To study the dependence of modulation on the thickness of FGT ($t_{FGT}$), we keep the thickness of $WTe_2$ ($t_{WTe}$) at about two monolayers (1 monolayer ~ 0.8 nm) while changing $t_{FGT}$ from 2.5 nm to 12.2 nm (Table S2). To study the dependence of modulation on $t_{WTe}$, we kept $t_{FGT}$ around 5 nm while changing $t_{WTe}$ from 0.8 nm to 27 nm (Table S3). Devices with different current directions have also been fabricated (Table S4). Figs. S4a and S4b, respectively, show the optical images of the Hall devices with a monolayer and a bilayer of $WTe_2$, where the thickness is determined by AFM measurements and by contrast of optical images (not shown).

Because some devices are not perfectly aligned along the a-axis ($\angle = 29^{\circ}$ for some devices), a correction factor of $\frac{1}{\cos\angle}$ should be considered. The areal current density is given by $I/w$, where $w$ is the width (ranging from 1 μm and 5 μm) of the device. Therefore, one should also consider the width as a correction factor. The corrected modulation amplitude of AHE conductivity *mod* is obtained by $mod = \frac{d\sigma_{\mathrm{AHE}}}{dI} = \frac{w}{\rho_{xx}^2}\frac{dR_{\mathrm{AHE}}}{dI}\frac{1}{\cos\angle}$, where $\sigma_{\mathrm{AHE}}$ is the anomalous Hall conductivity, $\rho_{xx}$ ($= R_{xx}wt/l$) the longitudinal resistivity, and $R_{xx}$ the resistance. The dependencies of *mod* on $t_{FGT}$ and $t_{WTe}$ are plotted in Figs. 3c and 3d.

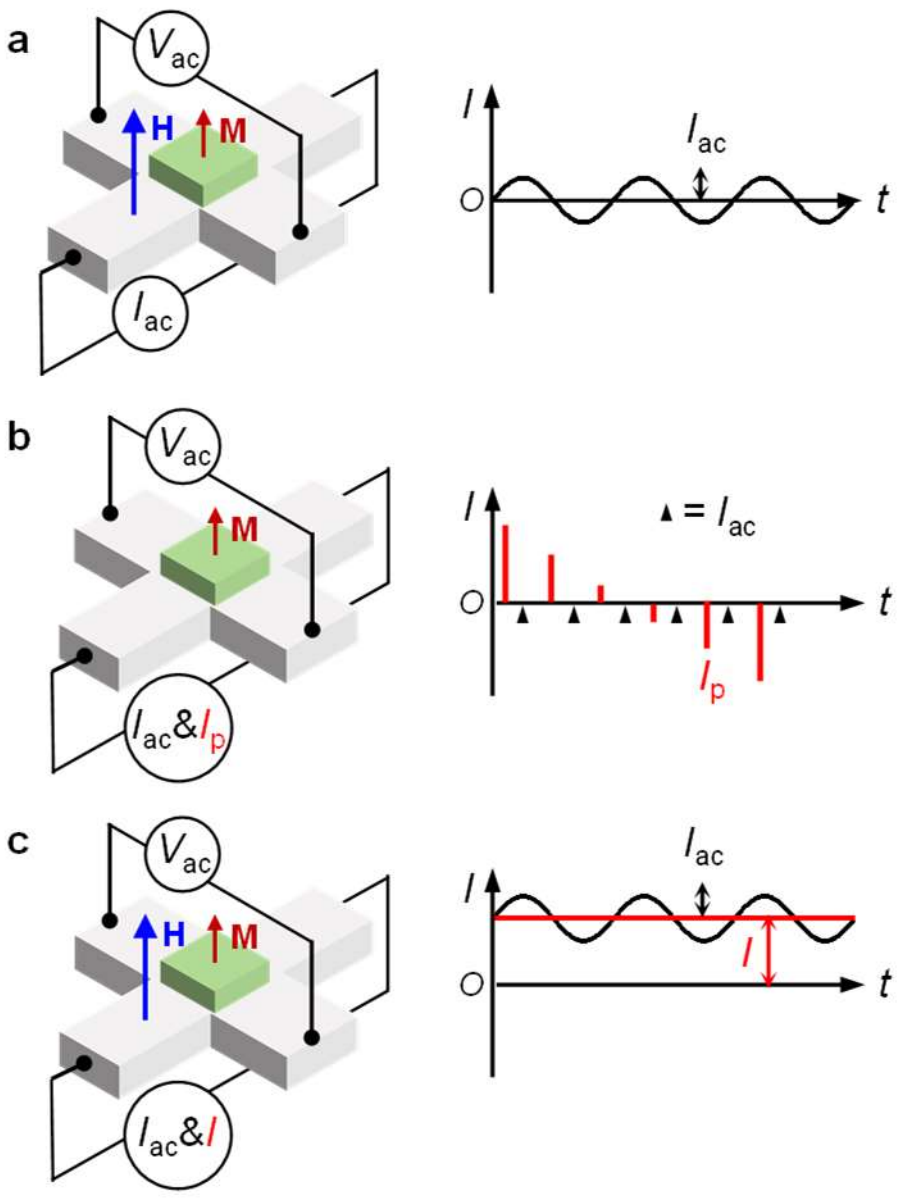

**Figure S5 | The three measurement configurations used in this work.** (**a**) By sending a low frequency ac current $I_{ac}$ from the source to the drain contacts, the transverse voltage $V_{ac}$ is detected by a lock-in amplifier. The AHE-loop is measured by sweeping the external magnetic-field $H_z$. (**b**) Magnetization switching measurements are realized by first sending a pulsed current $I_p$ and then detecting the magnetization state by $V_{ac}$. (**c**) For the modulation of AHE by a charge current $I$, $I$ and $I_{ac}$ are simultaneously applied, and the modulated magnetic state is detected by measuring $V_{ac}$. Note that the magnitude of $I$ is much smaller than the critical current for magnetization switching.

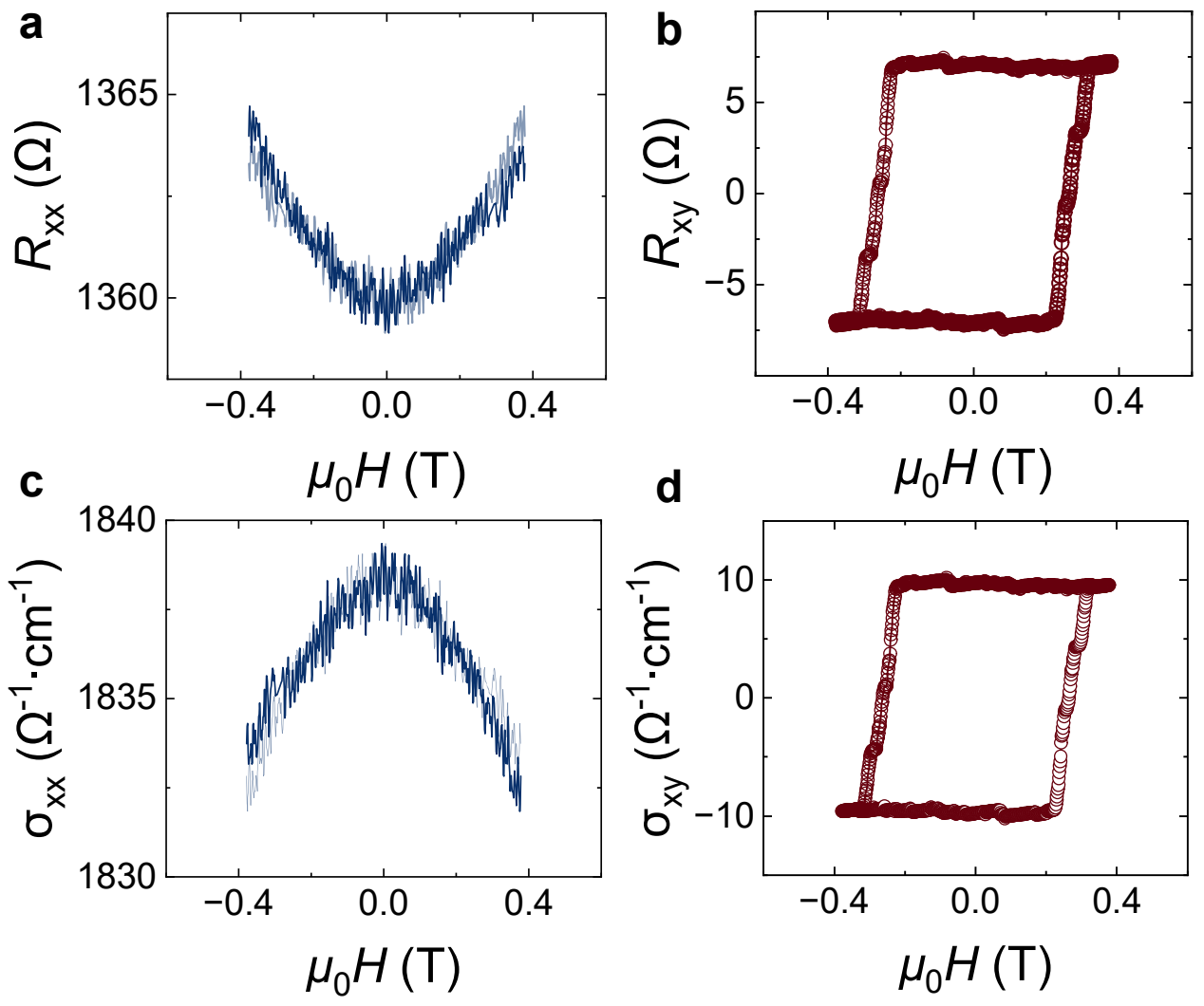

**Figure S6 | Out-of-plane magnetic-field dependence of resistance and conductivities.** (**a**) and (**b**): longitudinal and transverse resistance as a function of out-of-plane magnetic-field $H$. (**c**) and (**d**): calculated longitudinal and transverse conductivity as a function of $H$.

## 1.2 Magneto-transport measurements

All the electrical measurements were performed in a cryostat equipped with a 3D vector magnet. As shown in Fig. S5a, to measure the AHE hysteresis loop, a low frequency ac current $I_{ac}$ (rms amplitude of 10 μA with a frequency of 31 Hz) is applied between the source and drain contacts by a Keithley 6221 current source, and the transverse voltage $V_{ac}$ is detected by a SR830 lock-in amplifier. The AHE hysteresis loop is measured by sweeping the out-of-plane magnetic field $H_z$, and the transverse resistance $R_{xy}$ is obtained by $R_{xy} = V_{ac}/I_{ac}$.

For magnetization switching measurements, a combination of the Keithley 6221 current source and a Keithley 2182A nanovoltmeter is used to send the pulsed current $I_p$ (pulse width 100 μs). After the pulse has passed, the transverse voltage $V_{ac}$ is detected using the same condition as depicted in Fig. S5a to determine the magnetization state of FGT. Before applying $I_p$, the initial magnetic state is prepared by sweeping the out-of-plane magnetic-field to 0.4 T (larger than the coercive field) and by decreasing the magnetic-field to zero in no-overshoot mode.

For modulation of AHE measurements, a superposition of $I_{ac}$ and the dc current $I$ is generated using a Keithley 6221 current source (i.e., in wave mode with an offset). The transverse voltage $V_{ac}$ was measured by a lock-in amplifier by sweeping $H_z$. We sweep the magnetic field while keeping a constant current $I$ or changing polarity of $I$ as the magnetization $\mathbf{M}$ reverses. These two cases are schematically shown in Fig. S12.

Misalignment of the transverse contacts cannot be completely avoided and thus $R_{xy}$ also contains a signal from $R_{xx}$. A standard symmetrization procedure is implemented to separate the Hall effects from magnetoresistance, which is based on the fact that $R_{xy}$ is antisymmetric whereas $R_{xx}$ is symmetric with respect to $H_z$. Finally, the normal Hall effect of $WTe_2$ and FGT, which is linear with $H_z$, has also been subtracted from $R_{xy}$. As an example, Figs. S6a and S6b respectively show the symmetrized $R_{xx}$ and $R_{xy}$ as a function of out-of-plane magnetic-field. The transverse conductivity $\sigma_{xy}$ is obtained by $\sigma_{xy} = \rho_{xy}/(\rho_{xx}^2 + \rho_{xy}^2)$, where $\rho_{xy}$ (= $R_{xy}t$) is the transverse resistivity and $t$ (= $t_{FGT}$ + $t_{WTe}$) the total thickness.

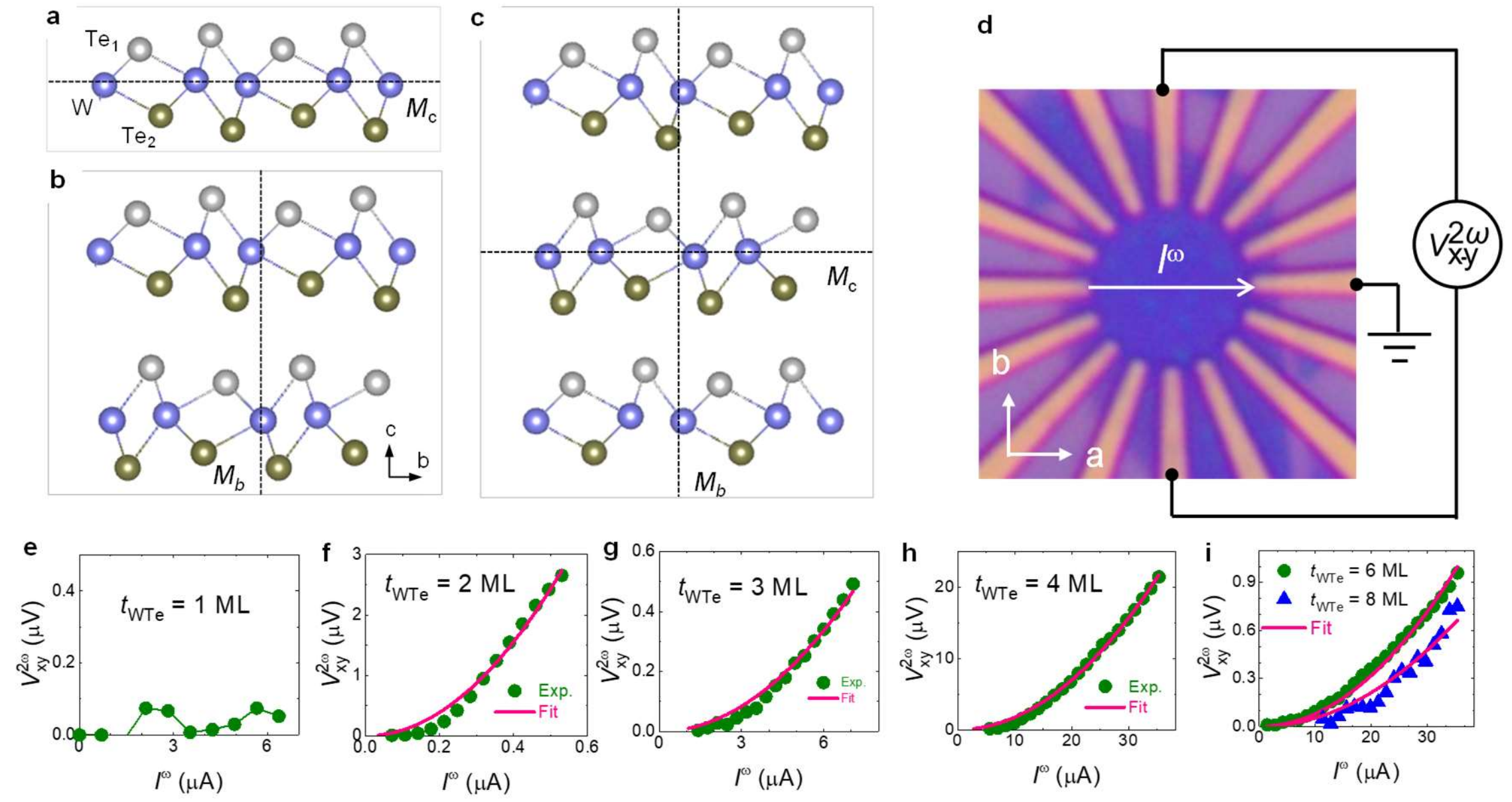

**Figure S7 | Nonlinear Hall effect induced by the Berry curvature dipole in $WTe_2$.** (**a**), (**b**) and (**c**): Crystal structure of monolayer, bilayer and trilayer of $WTe_2$ viewed in the *b*-*c* plane. The monolayer and trilayer samples have an approximate inversion plane, which are indicated by the dashed lines. The vertical dash lines in **b** and **c** are the glide mirror planes.(**d**) Schematic of the nonlinear Hall effect measurements. (**e**)-(**i**): The second harmonic Hall voltage $V_{xy}^{2\omega}$ as a function of $I^{\omega}$ for various $t_{WTe}$.

## 2. Non-linear Hall effect and quantification of the Berry curvature dipole in $WTe_2$

Previous photo-galvanic [21] and non-linear Hall effect (NLHE) [19,20,22] measurements have respectively revealed the existence of a Berry curvature dipole (BCD) $\mathbf{\Lambda}$ in $WTe_2$. The formation of BCD obeys a rather stringent symmetry constraint, and it needs inversion-symmetry breaking along the *c*-direction and uneven Berry curvature near the Fermi level. In the monolayer regime (Fig. S7a), inversion symmetry is approximately (but not exactly) satisfied, and the presence of inversion symmetry forces the BCD to be zero. In bilayer of $WTe_2$ (Fig. S7b), the stacking breaks the inversion symmetry between the layers and leads to an enhanced BCD along the a-axis. In trilayer and thicker samples (Fig. S7c), the inversion symmetry is preserved again, which also forbids the formation of a BCD. Note that because $M_b$ involves a half-cell translation along the *c*-axis and hence the inversion is broken on the sample surface, nonzero BCD should exist on the surface. The magnitude of BCD is expected to decrease with $t_{WTe}$ since the surface/bulk ratio decrease as $t_{WTe}$ increases.

To quantify the magnitude of Λ, we performed NLHE measurements of our $WTe_2$ flakes with different thicknesses. We found that in the monolayer regime, the resistance of the device (~ 200 kΩ along the *a*-axis) is too large to detect NLHE (Fig. S7e). For bilayer of $WTe_2$, a maximized NLHE signal is detected (Fig. S7f). As $t_{WTe}$ further increases, NLHE signal decreases (Figs. S7g-S7i). All these results qualitatively match with the symmetry analysis as shown in Figs. S7a-S7c. According to Ref. 19, $V_{xy}^{2\omega}$ is given by

$$V_{xy}^{2\omega} = \frac{e^3\tau}{2\hbar^2}\Lambda\frac{I_\omega^2}{w\sigma_b\sigma_a^2}, \tag{S2}$$

where $\tau$ is the momentum relaxation time, and $\sigma_a$ ($\sigma_b$) the zero-field conductivity in 2D (in units of $\Omega^{-1}$) along *a*- (*b*-)axis. The magnitude of $\tau$ can be determined by $\sigma_a = \frac{ne^2\tau}{m^*}$, where *n* is the charge density, and $m^*$ the effective mass. Therefore, equation S2 is rewritten as

$$V_{xy}^{2\omega} = \frac{em^*}{2\hbar^2 nw\ {}_b\sigma_a}\Lambda I_\omega^2 \tag{S3}$$

By using equation S3 to fit the experimental data, the magnitude of $\Lambda$ is determined by using $m^* = 0.3\ m_0$ ($m_0$ is the free electron mass), $n = 10^{17}\ m^{-2}$, $w = 10\ \mu m$ and the corresponding $\sigma_a$ values for each $t_{WTe}$. The $t_{WTe}$-dependence of $\Lambda$ is plotted in Fig. 1d, which reaches a maximized value of ~ 10 nm for bilayer of $WTe_2$.

The current-induced magnetization $\mathbf{M}_z^I$ via BCD can be regarded as an effective magnetic field $B_{eff}$ in $WTe_2$. $B_{eff}$ can be approximately evaluated as follows: By using $V_{xy}^{2\omega} = \frac{w}{\sigma_b} j_{xy}^{2\omega}$, and $I_\omega = \sigma_a w E$, equation S3 is expressed in terms of the NLHE current $j_{xy}^{2\omega}$, i.e.,

$$j_{xy}^{2\omega} = \frac{e^3\tau}{2\hbar^2}\Lambda E^2 \tag{S4}$$

where $E$ is the applied electric-field. On the other hand, $j_{xy}^{2\omega}$ is given by [S2]

$$j_{xy}^{2\omega} = \sigma_{xy}E = \frac{\sigma_a \omega_c \tau}{1+\omega_c^2\tau^2}E \tag{S5}$$

where $\omega_c$ (= $eB_{eff}/m^*$) is the cyclotron frequency. By comparing equation S4 with equation S5, and assuming that $\omega_c t \ll 1$, we have

$$B_{eff} = \frac{e^2 m^*}{2\hbar^2\sigma_a}\Lambda E. \tag{S6}$$

By using $\Lambda$ = 10 nm, $\sigma_a = 1.0\times10^{-3}\ \Omega^{-1}$ (for the modulation device DM1), and $E = 0.5\times10^6$ V/m (corresponding to $|I|$ = 1 mA), $B_{eff}$ = 1.5 T is obtained, which is much larger than effective fields obtained in spin Hall and spin Rashba systems [S3]. Further experimental and theoretical studies are needed to understand the mechanism underlying the generation of such enormous effective fields.

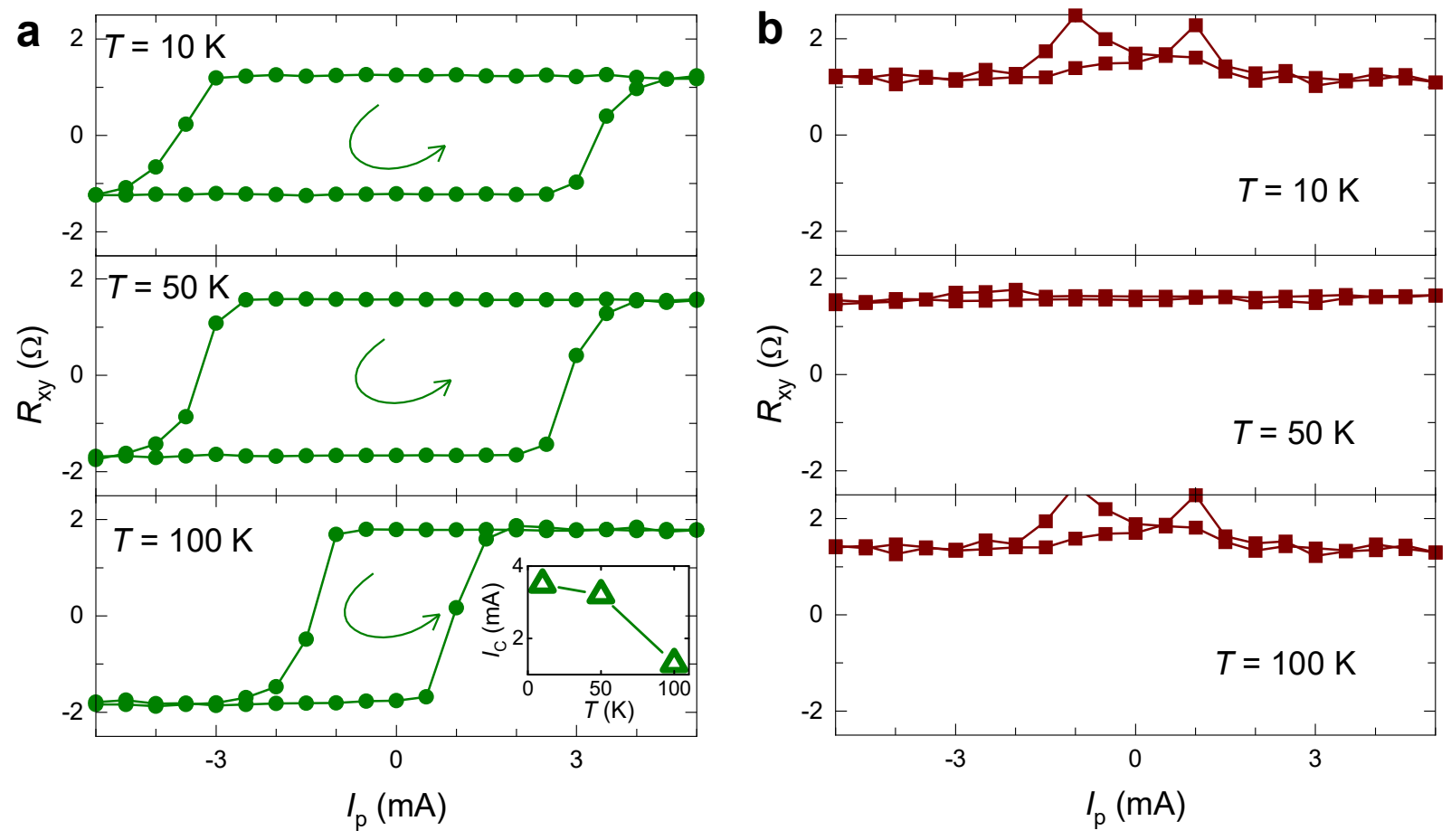

**Figure S8 | *I*-dependence of $R_{xy}$ measured for the pulsed current $I_p$ // a and $I_p$ // b.** (**a**) Current (**$I_p$ // a**) induced magnetization switching at zero magnetic-field measured at $T$ = 10 K, 50 K and 100 K. The curved arrows show the sweeping direction of the charge current. The inset shows the critical current as a function of temperature. (**b**) However, when the pulsed current is applied along the high symmetry *b*-axis, no switching is observed for all the measured temperatures.

## 3. No magnetization switching when I // b

As shown in Fig. S8a, field-free magnetization switching is achieved for $T$ = 10 K, 50 K and 100 K when the pulsed current is applied along the *a*-axis of $WTe_2$, and the switching current decreases as $T$ increases, which is due to the reduction of the effective perpendicular anisotropy. However, if the current is applied along the high symmetry *b*-axis and no in-plane magnetic-field is applied along the current direction, magnetization switching does not occur (Fig. S8b). This indicates that no or not sufficient $M_z^I$ is generated.

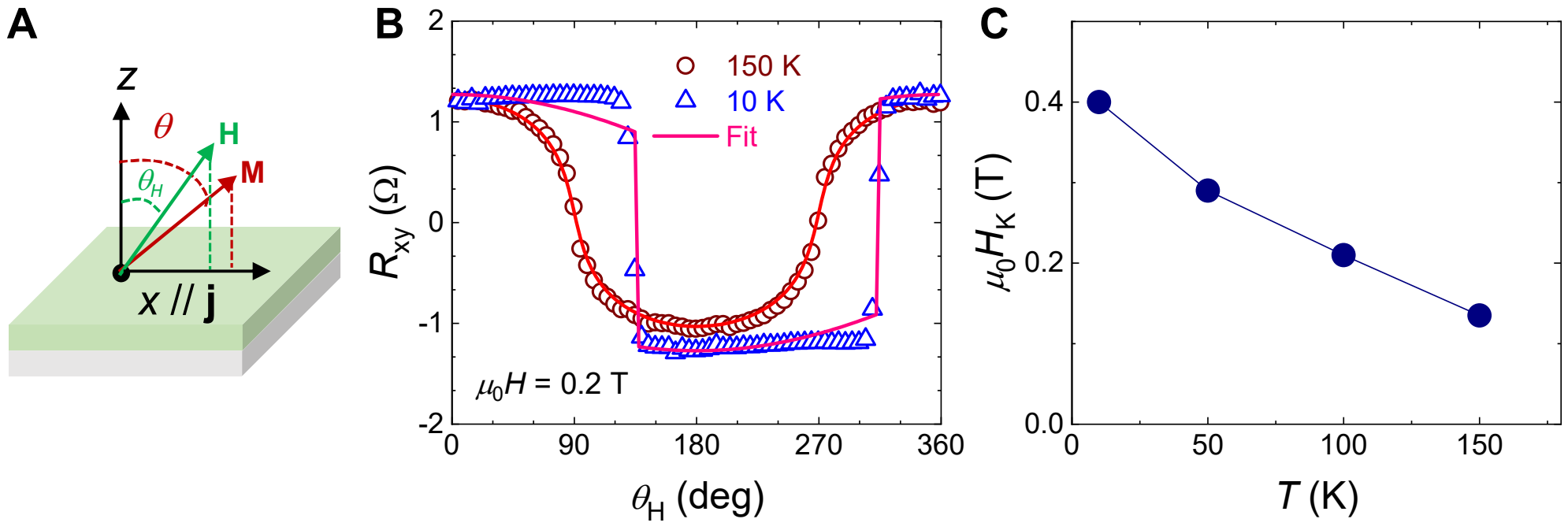

**Figure S9. Quantification of the perpendicular anisotropy of the magnetization switching device.** (**A**) Schematic of the coordinate system used in the analysis where $\theta$ and $\theta_H$, respectively, represent the polar angles of **M** and **H**. (**B**) $\theta_H$-dependence of $R_{xy}$ measured under a fixed magnetic field $\mu_0 H$ = 0.2 T at $T$ = 10 K and 150 K for Device DS (Table 2). (**C**) $T$-dependence of the effective perpendicular anisotropy field $H_K$.

## 4. Quantification of the perpendicular anisotropy of the $WTe_2$/FGT stack

To quantify the spin-torque efficiency of the magnetization switching device, the magnitude of the perpendicular anisotropy needs to be determined. The magneto-static energy $F$ can be written as [S3]

$$F = \frac{M}{2}\{-2H[\sin\theta \sin\theta_H \cos(\varphi - \varphi_H) + \cos\theta \cos\theta_H] - H_K \cos^2\theta\}. \qquad \text{(S7)}$$

Here $H_K$ is the effective perpendicular anisotropy field including the demagnetizing field along the ±**z**-direction (positive $H_K$ represents the easy-axis pointing out-of-plane), $\theta_H$ ($\theta$) the polar angle of **H** (**M**) as defined in Fig. S9a. $H_K$ originates from the competition between the demagnetizing field $H_d$ ($H_d = M$) of the flake and the perpendicular anisotropy field $H_\perp$ ($H_\perp = \frac{2K_\perp}{\mu_0 M}$) of FGT, i.e., $H_K = H_\perp - M$, where $K_\perp$ is the perpendicular anisotropy energy density, and $\mu_0$ the magnetic constant. To quantify $H_K$, we measured $R_{xy}$ by rotating **M** with a fixed $H$. Fig. S9b shows the $\theta_H$-dependence of $R_{xy}$ for $\mu_0 H$ = 0.2 T. The magnitude of $H_K$ is determined by fitting the $\theta_H$-$R_{xy}$ trace, i.e, $R_{xy} \sim \cos\theta$, where $\theta$ is determined from equation S7 by imposing the energy minimum condition $\partial F/\partial\theta = 0$ and $\partial^2 F/\partial\theta^2 > 0$. As shown in Fig. S9c, $\mu_0 H_K$ is ~ 0.4 T at 10 K and $H_K$ decreases as $T$ increases. In comparison with thicker FGT films, the significant reduction of $H_K$ in our samples is probably due to the reduction of the FGT thickness.

## 5. Quantification of the effective spin-torque efficiency by micro-magnetic simulations

The spatial- and time-dependent magnetization can be obtained by solving the Landau-Lifshitz-Gilbert equation in the presence of the spin-orbit torque [S4] term $\boldsymbol{\tau}_{SOT}$

$$\frac{d\mathbf{m}}{dt} = -\frac{1}{1+\alpha^2}\gamma\mathbf{m}\times\mathbf{H}_{eff} - \frac{\alpha}{1+\alpha^2}\gamma\mathbf{m}\times\left(\mathbf{m}\times\mathbf{H}_{eff}\right) + \boldsymbol{\tau}_{\mathrm{SOT}} \tag{S8}$$

where $\mathbf{m}$ is the unit magnetization, $t$ is time, $\alpha$ the damping constant, $\gamma$ the gyromagnetic ratio, and $\mathbf{H}_{eff}$ the effective magnetic-field originating from exchange interaction, perpendicular magnetic anisotropy, demagnetization field, and Dzyaloshinskii-Moriya interaction. The spin-orbit torque term $\boldsymbol{\tau}_{SOT}$ is given by

$$\boldsymbol{\tau}_{\mathrm{SOT}} = -a_J\frac{\gamma}{1+\alpha^2}[(1+\alpha\eta)\mathbf{m}\times(\mathbf{m}\times\mathbf{p}) + (\eta-\alpha)\mathbf{m}\times\mathbf{p}] \tag{S9}$$

Here $\eta\left(=\frac{b_J}{a_J}\right)$ is the ratio of the field-like toque $b_J$ and the damping-like torque $a_J$, and $\mathbf{p}$ the spin polarization unit vector. The magnitude of $a_J$ is given by

$$a_J = \frac{\hbar\xi j}{2Met_{\mathrm{FGT}}} \tag{S10}$$

where $\xi$ is the effective spin-torque efficiency, $\hbar$ the reduced Plank constant, $e$ the electronic charge and $j$ the current density flowing in $WTe_2$. In Ref. S4, it is shown that one can use the Slonczewski spin-transfer torque $\boldsymbol{\tau}_{SL}$ implemented in MuMax3 to simulate $\boldsymbol{\tau}_{SOT}$. In a three-layer structure with a fixed magnetization layer, a spacer layer and a free magnetization layer, $\boldsymbol{\tau}_{SL}$ is given by

$$\boldsymbol{\tau}_{\mathrm{SL}} = \gamma\beta\frac{\epsilon+\alpha\varepsilon'}{1+\alpha^2}\mathbf{m}\times(\mathbf{p}\times\mathbf{m}) - \gamma\beta\frac{\varepsilon'-\alpha\epsilon}{1+\alpha^2}\mathbf{m}\times\mathbf{p} \tag{S11}$$

with

$$\beta = \frac{\hbar j}{M_s e t_{\mathrm{FGT}}}$$

$$\epsilon = \frac{P\,\Lambda^2}{(\Lambda^2 + 1) + (\Lambda^2 - 1)(\mathbf{p}\cdot\mathbf{m})}$$

Here the parameters $\varepsilon'$ and $\Lambda$ characterize the properties of the magnetic trilayers, $P$ is the degree of spin polarization of the fixed layer, and $\mathbf{p}$ the direction of the spin-polarization. By letting $\Lambda = 1$ and $\varepsilon' = \eta\epsilon$, equation S11 is rewritten as

$$\boldsymbol{\tau}_{\mathrm{SL}} = -\beta \frac{P}{2}\frac{\gamma}{1+\alpha^2}[(1+\alpha\eta)\mathbf{m}\times(\mathbf{m}\times\mathbf{p}) + (\eta-\alpha)\mathbf{m}\times\mathbf{p}] \quad \text{(S12)}$$

By comparing equations S9 and S12, one notices that $P = \frac{2a_J}{\beta} = \xi$ and $\varepsilon' = \frac{\eta\xi}{2}$ hold, and this explains how to simulate $\tau_{\mathrm{SOT}}$ by $\boldsymbol{\tau}_{\mathrm{SL}}$.

In the Mumax3 simulation, we set a grid size of $N_x \times N_y \times N_z$ = 512×512×2, and the dimension of $Dim_x \times Dim_y \times Dim_z$ = 1.0 μm×1.0 μm × $t$ nm, where $t$ (= 2 nm) is the thickness of FGT. Periodic boundary conditions are set along the $x$ and $y$-directions. Since magnetization switching is governed by the damping-like torque $a_J$, the field-like torque $b_J$ (and thus $\eta$) is set to 0. We initialize the magnetization along the (0, 0, 1) direction and the spin polarization unit vector is **p** = (0, 0, −1). To take thermal effects into account, reduced values $M$ and $K_\perp$ are used. The current density $j$ at 10 K is determined by $\frac{I_{cri}}{wt_{WTe}} = 3.5\times10^{11}$ A/m$^2$, where $w$ (= 2 μm) is the width of the device. $A_{ex}$ is fixed to be $0.95\times10^{-12}$ J/m [S5]. In the literature, one can find few reports concerning the DMI constant and the damping value for $WTe_2$/FGT bilayers [S6-S8]. Therefore, we determine the magnitude of $\xi$ as a function of $\alpha$ for various DMI values. As shown in Fig. S10, $\xi$ increases as the DMI increases. For DMI values > $4\times10^{-4}$ Jm$^{-2}$, $\xi$ is almost independent of $\alpha$. However, for DMI < $4\times10^{-4}$ Jm$^{-2}$, the dependencies of $\xi$ on $\alpha$ show a complicated behavior, i.e., for $\alpha < 0.01$, $\xi$ is almost independent of $\alpha$; but for $\alpha > 0.01$, $\xi$ increases linearly with $\alpha$. This behavior is in sharp contrast to the macro-spin model [S9], i.e, a single magnetic domain model with $A_{ex} = 0$ and DMI = 0, which shows that $\xi \propto \alpha H_K$. We believe that the Mumax3 simulations give more realistic $\xi$ values. Since both the DMI and $\alpha$ values of the $WTe_2$/FGT bilayers are unknown to us, we roughly estimate from Fig. S10 that the lower limit of $\xi$ is 0.1. The large $\xi$ value is probably due to the fact that both $WTe_2$ and FGT layers are terminated by Te (van der Waals gap), and thus $WTe_2$/FGT bilayers have a Te-Te interface. This unique interface is barrier free for spin transport and thus the spin transparency is expected to be enhanced. Further experimental and theoretical studies are needed for a more accurate measurement of $\xi$ as well as for the microscopic understanding of $\xi$.

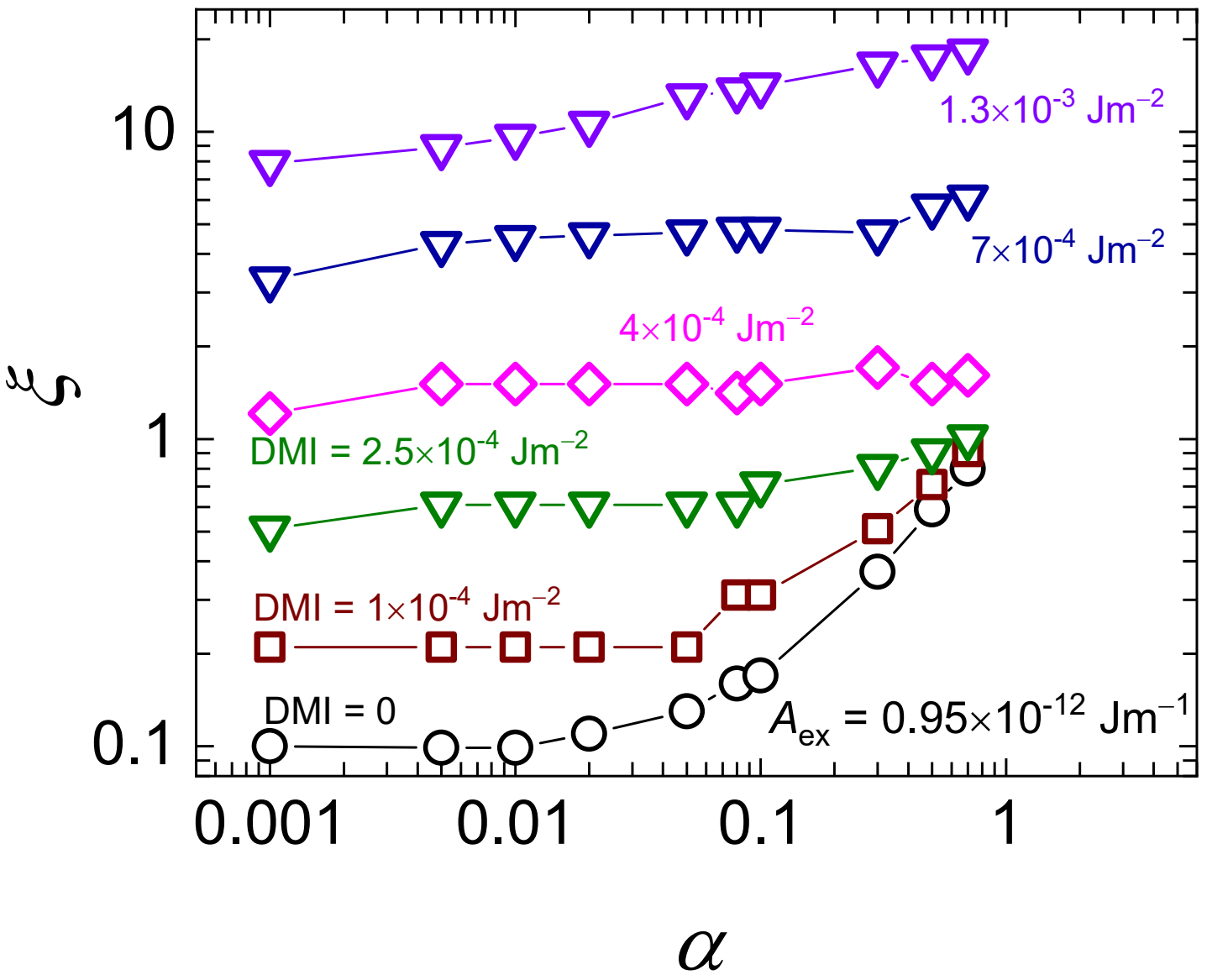

**Figure S10. Dependence of the effective spin-torque efficiency $\xi$ on damping $\alpha$ for various DMI values.**

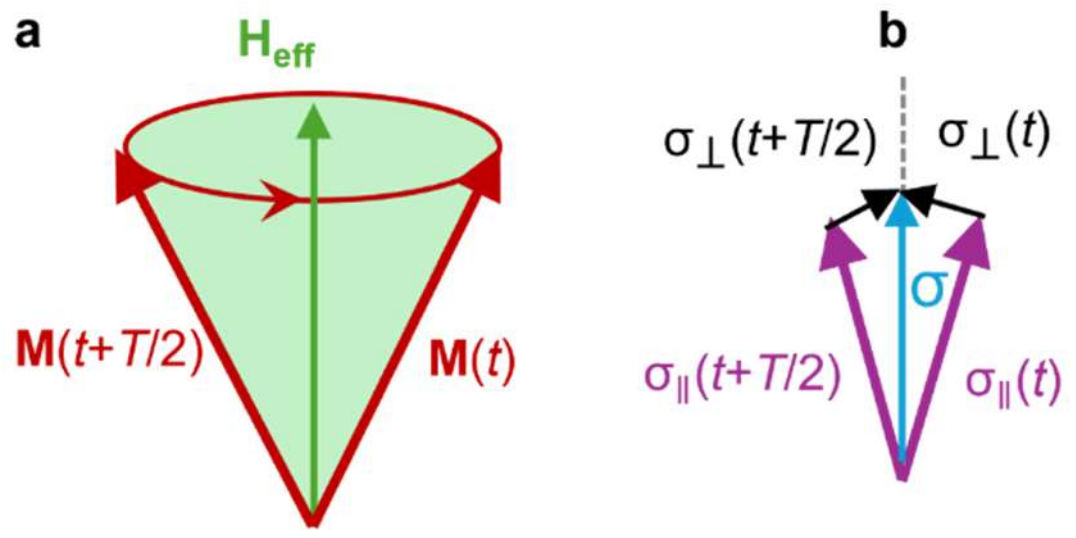

**Fig. S11.** (**a**) The magnetization precesses around the effective magnetic field $\mathbf{H}_{\text{eff}}$, which consists of both the external magnetic field and the perpendicular magnetic anisotropy field. On average, the magnetization is aligned along the $z$-direction. $\mathbf{M}(t)[\mathbf{M}(t+T/2)]$ denotes the magnetization at time $t$($t$+$T$/2), where $T$ is the precession period. (**b**) The out-of-plane spin accumulation $\sigma$ is decomposed into components parallel and perpendicular to the directions of $\mathbf{M}(t)$ and [$\mathbf{M}(t+T/2)$], respectively, which are denoted as $\boldsymbol{\sigma}_{\parallel}(t)$, $\boldsymbol{\sigma}_{\parallel}(t+T/2)$, $\boldsymbol{\sigma}_{\perp}(t)$ and $\boldsymbol{\sigma}_{\perp}(t+T/2)$.

## 6. Consideration of modulation of exchange interaction by z-polarized spin accumulation

In the AHE modulation experiments, the modulation current is much smaller than the critical magnetization switching current and an external magnetic-field has been applied, the magnetization stays along z-direction. Since the direction of the magnetization doesn't change, there are no SOTs-related effects.

We can also consider it more quantitatively: at finite temperature, the magnetization precesses around the effective magnetic field $\mathbf{H}_{eff}$ as shown in Figure S11. The current induced out-of-plane spin accumulation $\boldsymbol{\sigma}$ in $WTe_2$ is decomposed into components parallel and perpendicular to the directions of $\mathbf{M}(t)$ and $[\mathbf{M}(t+T/2)]$, respectively. The transverse components $\boldsymbol{\sigma}_{\perp}(t)$ and $\boldsymbol{\sigma}_{\perp}(t+T/2)$ contribute to the damping-like torque, which is respectively given by $\boldsymbol{\tau}(t) = \mathbf{M}(t) \times \mathbf{M}(t) \times \boldsymbol{\sigma}_{\perp}(t)$ and $\boldsymbol{\tau}(t+T/2) = \mathbf{M}(t+T/2) \times \mathbf{M}(t+T/2) \times \boldsymbol{\sigma}_{\perp}(t+T/2)$. Obviously, $\boldsymbol{\tau}(t) + \boldsymbol{\tau}(t+T/2) = 0$. Therefore, $\boldsymbol{\sigma}$ generates no antidamping torques. However, the parallel components $\boldsymbol{\sigma}_{\parallel}(t)$ and $\boldsymbol{\sigma}_{\parallel}(t+T/2)$ give rise to an out-of-plane effective magnetic field acting on the magnetization, which modulates the exchange interaction of FGT.

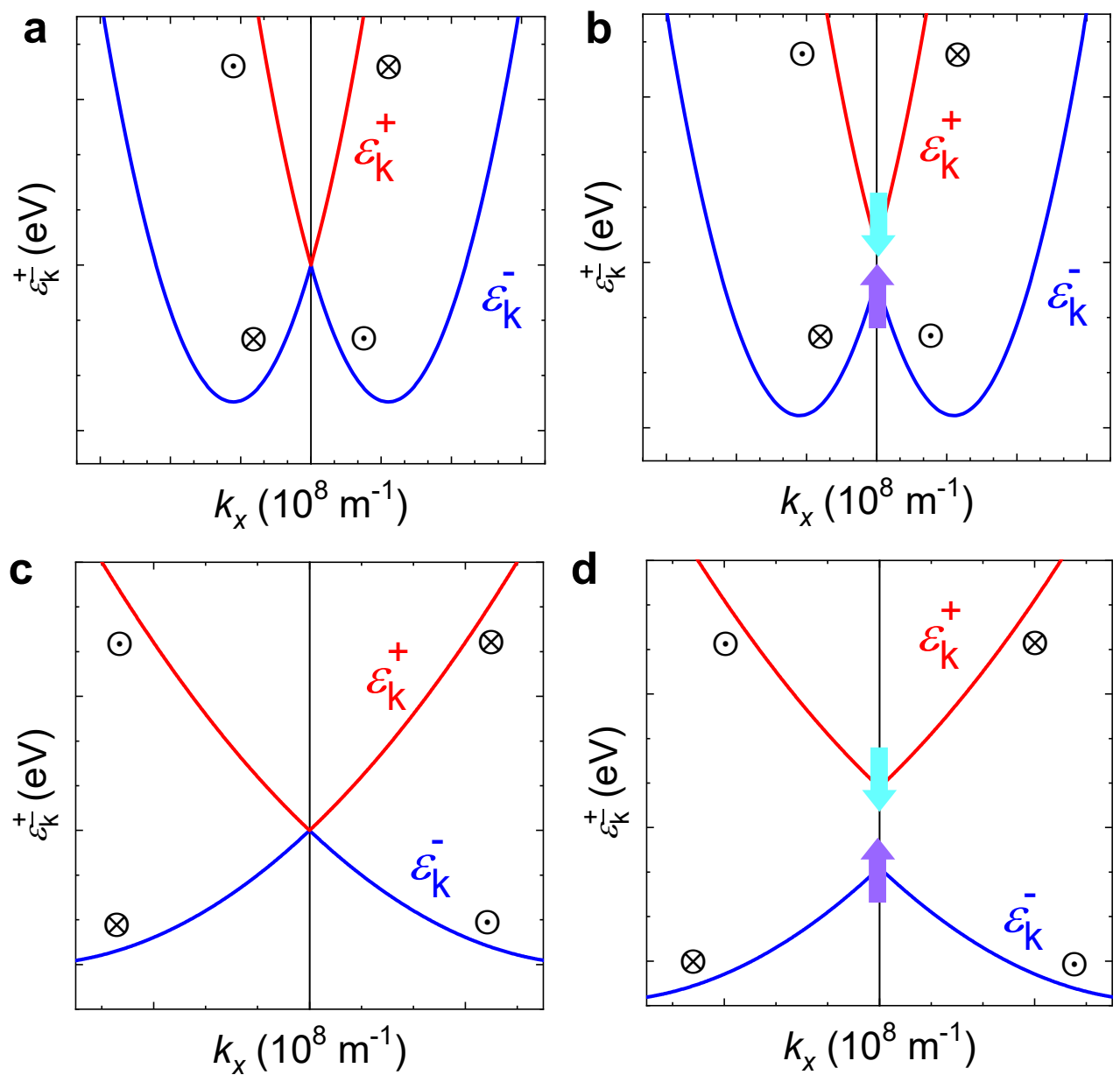

**Figure S12 | Electronic band dispersion for the two-band model given by equation S13.** Electronic band dispersion for $E_{ex} = 0$ (**a**) and $E_{ex} \neq 0$ (**b**). (**c**) and (**d**): the same as (**a**) and (**b**) but for $\hbar v k \gg \frac{\hbar^2 k^2}{2m}$.

## 7. The two band model for the AHE

To give a perspective on how exchange splitting influences the AHE, we consider a mimimal two-band model proposed by Onda, Sugimoto and Nagaosa [34], which offers a simple theoretical receipe for understanding the AHE. This model fully takes into account the parity anomaly associated with the avoided crossing of the band dispersion, and the Hamiltonian is

$$h(k) = \frac{\hbar^2 k^2}{2m} + \hbar v(\mathbf{k} \times \boldsymbol{\sigma}) \cdot \mathbf{z} - E_{ex}\sigma_z, \tag{S13}$$

where $E_{ex}$ is the effective exchange splitting ($E_{ex} > 0$ corresponds to **M** // +**z** and $E_{ex} < 0$ corresponds to **M** // −**z**), **σ** the Pauli matrices, *v* the velocity, *k* the wavevector, and *m* the effective mass. The dispersion relation $\varepsilon^{\pm}$and the wave function are obtained as

$$\varepsilon_k^{\pm} = \frac{\hbar^2 k^2}{2m} \pm \sqrt{E_{ex}^2 + \hbar^2 v^2 k^2}, \tag{S14}$$

$$|u_-(\mathbf{k})\rangle = \begin{pmatrix} \cos\frac{\theta}{2} \\ -ie^{i\varphi}\sin\frac{\theta}{2} \end{pmatrix}, |u_+(\mathbf{k})\rangle = \begin{pmatrix} \sin\frac{\theta}{2} \\ ie^{i\varphi}\cos\frac{\theta}{2} \end{pmatrix} \tag{S15}$$

where +(–), resprectively, represents the upper (lower) band, $\cos\theta = \frac{E_{ex}}{\sqrt{E_{ex}^2+\hbar^2 v^2 k^2}}$, and $\varphi = \tan^{-1}\frac{k_y}{k_x}$.

From the dispersion relation one notices that: i) The second term breaks the inversion symmetry, which leads to the linear dispersion with velocity *v*. ii) The third term breaks time reversal symmetry, and leads to the energy splitting of the upper (+) and the lower (–) band at the avoided-crossing point ($k = 0$) by $2E_{ex}$. Note that the magnetization plays the role of the spin-orbit interaction for AHE. When $E_{ex} = 0$, equation S13 is reduced to the two-dimensional Rashba band model (Fig. S12a), which is frequently used in semiconductor quantum wells, and in this case *v* is regarded as the spin-orbit interaction.

The **k**-dependent spin expectation $\langle \boldsymbol{\sigma}^{\pm} \rangle$ is obtained as

$$\langle \boldsymbol{\sigma}^{\pm} \rangle = \begin{pmatrix} \mp \frac{\hbar v k_y}{\sqrt{\hbar^2 v^2 k^2 + E_{ex}^2}} \\ \pm \frac{-\hbar v k_x}{\sqrt{\hbar^2 v^2 k^2 + E_{ex}^2}} \\ \mp \frac{E_{ex}}{\sqrt{\hbar^2 v^2 k^2 + E_{ex}^2}} \end{pmatrix}. \tag{S16}$$

When $E_{ex} = 0$, the Rashba band posseses a magnetic monopole at $\mathbf{k} = 0$ since the spin expectation value is not well defined, and for $k \neq 0$, the spin expectation value lies in-plane forming a typical helical spin texture. For ferromagnetic metals with $E_{ex} \neq 0$, the spin alignment changes significantly. For $k = 0$, the spin expectation is along the $+\mathbf{z}(-\mathbf{z})$ direction for the lower (upper) band, and for $k \to \pm\infty$, the spins keep the same alignment as the case of two-dimensional Rashba band model ($E_{ex} = 0$). Therefore, when $k$ changes from $+\infty$ to $-\infty$, the spin texture forms a half skyrmion configuration, leading to a winding number of ±1/2 for the upper and lower bands. The Berry curvature $b_z$ of these two bands for certain $k$ can be obtained as

$$b_z^{\pm}(k) = \nabla_{\mathbf{k}} \times \left[ i \langle u_{\pm}(\mathbf{k}) | \nabla_{\mathbf{k}} | u_{\pm}(\mathbf{k}) \rangle \right] = \pm \frac{v^2 E_{ex}}{2(\hbar^2 v^2 k^2 + E_{ex}^2)^{\frac{3}{2}}} \tag{S17}$$

The anomalous Hall conductivity can then be calculated using the Kubo formula

$$\sigma_{xy}^{\pm} = \frac{e^2}{\hbar} \int \frac{d^2\mathbf{k}}{(2\pi)^2} f\left[\varepsilon_k^{\pm}(k)\right] b_z^{\pm}(k) = \pm \frac{e^2}{2h} \frac{E_{ex}}{\sqrt{\hbar^2 v^2 k^2 + E_{ex}^2}} \tag{S18}$$

where $e$ is the electronic charge, $h$ the Planck constant ($h = 2\pi\hbar$). Equations S17 and S18 show that: i) The magnetization (i.e., $E_{ex} \neq 0$) is necessary for a sizeable $b_z$ and thus AHE. ii) Both $b_z$ and $\sigma_{xy}$ are of opposite signs for the upper and lower bands. iii) Both $|b_z|$ and $|\sigma_{xy}|$ are maximised when $k$ approaches 0, and for large enough $k$, both values apprach zero. iv) $\sigma_{xy}$ is linearly proportional to the effective exchange interaction and thus the direction of $\mathbf{M}$, which is known as the parity anomaly. v) For the model given by equation S13, the half Hall conductance ($\sigma_{xy} = e^2/2h$) cannot be reached because the state with $k = 0$ can not be realised. However, for Dirac fermions (e.g., in a three-dimensional topological insulator [S10]), $\hbar vk \gg \frac{\hbar^2 k^2}{2m}$ holds, and the dispersion relation for $E_{ex} = 0$ and $E_{ex} \neq 0$ is respectively plotted in Figs. 12c and 12d. If the Fermi energy lies in the exchange gap and thus $k = 0$ holds, both the top and bottom surfaces, give a maximum value of $e^2/2h$, which leads to the observation of the quantum AHE with $\sigma_{xy} = e^2/h$.

In ferromagnetic metals, if we assume that $\hbar vk >> E_{ex}$, Equations S17 and S18 can be simplified to

$$b_z^{\pm}(k) \approx \pm \frac{E_{ex}}{2vk^3} \tag{S19}$$

$$\sigma_{xy} \approx \pm \frac{e^2}{2h} \frac{E_{ex}}{\hbar vk} \tag{S20}$$

The above equations show that both the Berry curvature and the anomalous Hall conductivity are linearly proportional to $E_{ex}$, which provides the theoretical basis for tuning the AHE via the modulation of the exchange interaction.

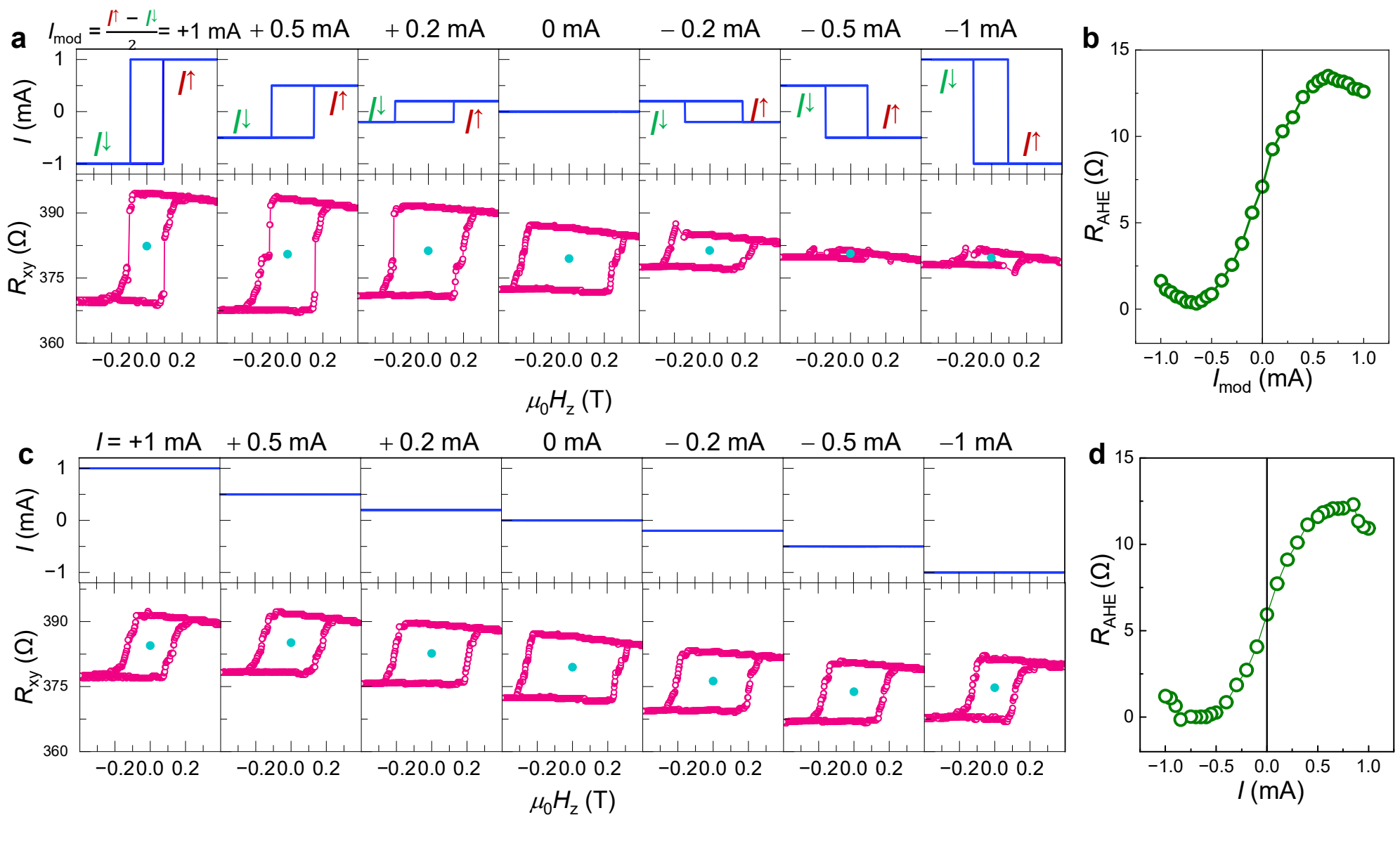

**Figure S13 | Comparing the modulation by M-independent I and by M-dependent I.** (**a**) Upper panel: the

modulation is performed by a **M**-dependent charge current. Here the modulation current $I_{mod}$ is defined as, $I_{mod}$ = ($I^{\uparrow}$ – $I^{\downarrow}$)/2, where $I^{\uparrow}$ ($I^{\downarrow}$), as marked in the panel, represents the charge current when **M** points along the +z (–z) direction. From left to right: the sign of $I_{mod}$ changes from positive ($I_{mod}$ > 0) to negative ($I_{mod}$ < 0). Lower panel: the corresponding $R_{xy}$ loops. In this case, the center of each loop remains nearly constant with respect to $I_{mod}$ while the height of loop exhibits a strong dependence on $I_{mod}$. (**b**) $I_{mod}$-dependence of $R_{AHE}$. (**c**) Upper panel: a **M**-independent dc current $I$ is varied from +1 mA (left panel) to −1 mA (right panel). Lower panel: the corresponding $R_{xy}$ loops for device DM1 measured at 10 K, and the center of each loop $R^{c}_{xy}$ is marked by a solid point. The variation of $R^{c}_{xy}$ indicates a modulation of the anomalous Hall resistance $R_{AHE}$ by $I$ (for details see text). (**d**) $I$-dependence of $R_{AHE}$ obtained from (**c**), showing consistent results with (**b**).

## 8. Consistent results by M-independent modulation current

Here, we show another measurement to demonstrate that the AHE modulation is a genuine effect. For clarity, the raw $R_{xy}$ loops without subtracting the offset resistance $R_{off}$, originating from the misalignment of the detection contacts, are presented.

Figure S13a shows the modulation of AHE by a **M**-dependent $I_{mod}$, which is same result as shown in Fig. 3 of the main text. Here the center of loops $R^{c}_{xy}$, marked by the solid symbol, does not change while the height of the loop drastically varies for different $I_{mod}$, as summarized in Figure S13b.

For an alternative measurement, the polarity of **I** (and thus $\mathbf{M}_z^I$) is fixed while changing the direction of **M** by sweeping $H_z$. As shown in Fig. S13c, when $I$ is varied from +1 mA to −1 mA, the height of the AHE loop remains almost unchanged, while the center of the loop shifts drastically. This result can be understood as follows: Assume that **I** // +**x** and **I** generates $\mathbf{M}_z^I$ // +**z**. $\mathbf{M}_z^I$ reduces $\rho_{xy}$ from $-\rho^{0}_{xy}$ to $-\rho^{0}_{xy}+\varepsilon$ for **M** // +**z** (case i), but enhances $\rho_{xy}$ for **M** // −**z** from $\rho^{0}_{xy}$ to $\rho^{0}_{xy}+\varepsilon$ (case iv), where $\varepsilon = \delta\left(\rho^{0}_{xy}\right)^2$ and $\rho^{0}_{xy}$ ($=1/\sigma^{0}_{xy}$) the intrinsic anomalous Hall resistivity at $I = 0$. Accordingly, $R^{c}_{xy}$ shifts upwards by $+\varepsilon$ but the height ($= 2\rho^{0}_{xy}$) of the loop remains unchanged. Similarly, if **I** // −**x** and **M** switches from +**z** to −**z** (cases ii and iii), $R^{c}_{xy}$ shifts downwards by $-\varepsilon$, which is opposite to the case of **I** //+**x**. Therefore, one can determine the modulation amplitude of the AHE at each $I$, quantified by $\varepsilon/t$ ($t$ is the thickness of FM), from the vertical shift of the loop, i.e., $\varepsilon/t = R^{c}_{xy}(I) - R^{c}_{xy}(I=0)$. Therefore, the magnitude of the AHE resistance at each $I$, $R_{AHE}(I)$, is obtained by summing $\varepsilon/t$ and $R_{AHE}(0)$, $R_{AHE}(I) = R_{AHE}(0) + \varepsilon/t$, where $R_{AHE}(0)$ is the magnitude of the AHE resistance at $I = 0$. The obtained $I$-dependence of $R_{AHE}$ is summarized in Fig. S13d, showing consistent results as Fig. S13b.

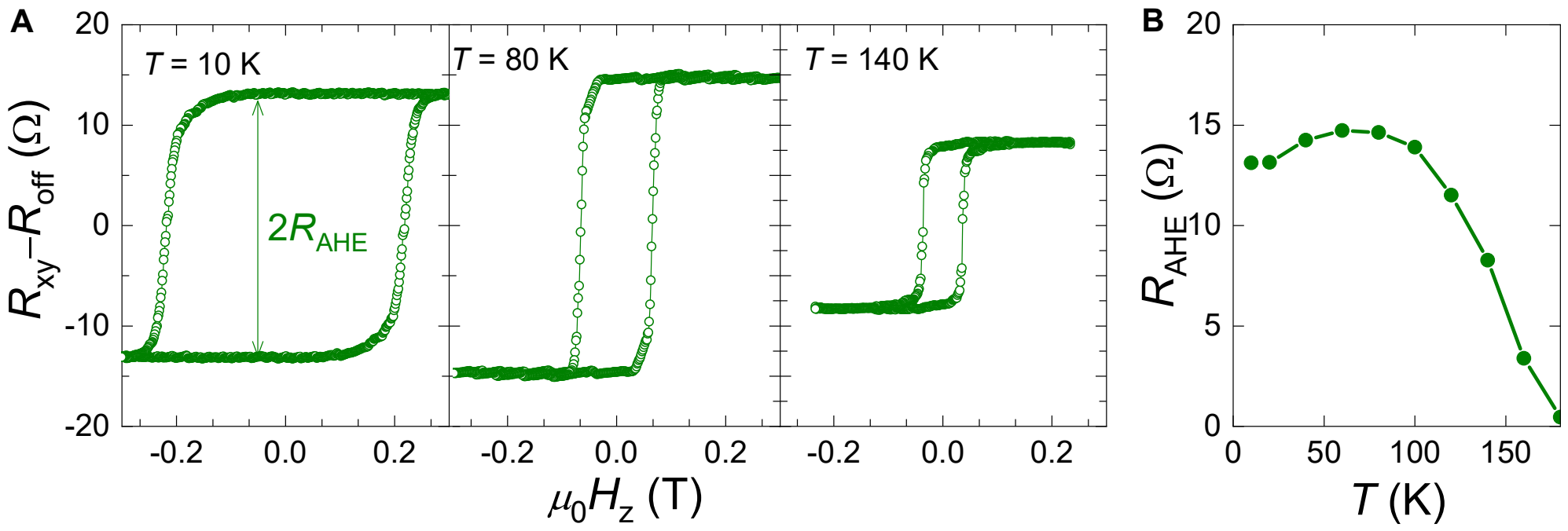

**Figure S14. Temperature dependence of the anomalous Hall resistance of the $WTe_2$/FGT stack (Device: DM2).** (**A**) Anomalous Hall loops measured at 10 K, 80 K and 140 K. The anomalous Hall resistance $R_{AHE}$ is determined by the height of the loop (height = $2R_{AHE}$). (**B**) Temperature dependence of $R_{AHE}$.

## 9. Temperature dependence of $R_{AHE}$ of the $WTe_2$/FGT stack

Fig. S14a shows the hysteresis loops of the transverse resistance $R_{xy}$ measured at different temperatures $T$ for Device DM2, and the offset voltage $V_{off}$, originating from misalignment of the transverse electrodes, has been subtracted for each loop. Each loop shows a square shape. The coercive field reaches ~ 0.25 T at 10 K and decreases monotonously as $T$ increases, which is a typical behaviour for ultrathin FGT with perpendicular magnetic anisotropy. The magnitude of $R_{AHE}$ at each $T$ is obtained from the height of the loop (height = $2R_{AHE}$), and a non-monotonic dependence of $R_{AHE}$ on $T$ is observed (Fig. S14b). This non-monotonic behaviour, is not supported by the conventional expression for AHE, $\rho_{AH} \sim M_z$. Here, $M_z$ is the out-of-plane component of the magnetization. Similiar bahavior has also been obseved in the perovskite $SrRuO_3$ (Ref. 8) and the magnetic semiconductor (Ga,Mn)As [S11], in which the intrinsic Berry curvature is responsible for the AHE.

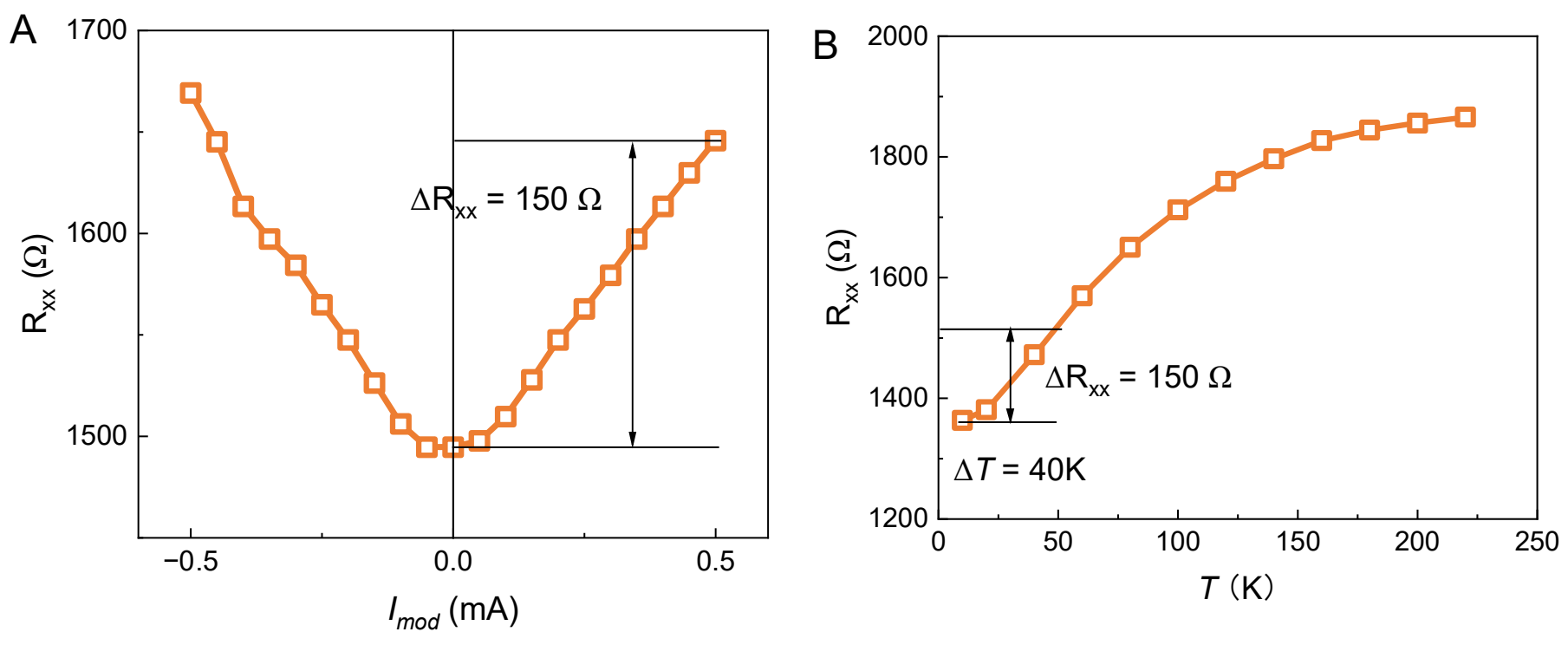

**Figure S15. Quantification of Joule heating induced by the application of a dc current for Device DM2.** (**A**) *I*-dependence of the longitudinal resistance $R_{xx}$. $R_{xx}$ increases by 150 Ω after applying |*I*| = 0.5 mA. (**B**) *T*-dependence of $R_{xx}$ of the same device. By comparing (**A**) and (**B**), we find that a resistance increase of 150 Ω corresponds to a temperature increase Δ*T* of 40 K.

## 10. Quantification of current-induced Joule heating

The application of a charge current in magnetization switching and in modulation experiments inevitably introduces Joule heating. To quantify the effect of Joule heating, we measure both the dependencies of the longitudinal resistance $R_{xx}$ on *I* and *T*. As shown in Fig. S15a, the *I*-dependence of $R_{xx}$ shows an even behavior with respect to *I* and the application of |*I*| = 0.5 mA leads to an increase of $R_{xx}$ by 150 Ω. The *T*-dependence of $R_{xx}$ (Fig. S15b) shows that the device is metallic with a negative temperature coefficient. By comparing Fig. S15a and S15b, we determine that the application of |*I*| = 0.5 mA leads to an increase of the sample temperature by ~40 K, which in turn leads to an increase of $R_{AHE}$ by 1.3 Ω according to the temperature dependence of $R_{AHE}$ measured for the same device (Fig. S14b). This is much smaller than the magnitude of $R_{AHE}$ modulation (~10.5 Ω, Table S2). Moreover, it is expected that Joule heating results in an even $R_{AHE}$-*I* relationship; while the modulation shows mainly an odd $R_{AHE}$-*I* behavior (Fig. 3b and Fig.S13). Therefore, we confirm that the modulation of the $R_{AHE}$ does not mainly originate from the Joule heating effect.

## 11. First-principle calculations

First-principles calculations were carried out within the framework of density functional theory (DFT) [S12, S13], as implemented in the Quantum ESPRESSO package [S14-S16]. The Brillouin zone was sampled using a 14 × 14 × 1 Monkhorst–Pack grid [S17], and a plane-wave energy cut-off of 160 Ry was adopted. Structural relaxations were performed until the forces on all atoms were smaller than $10^{-5}$ Ry/Bohr, with total energy changes between ionic steps below $10^{-5}$ Ry.

Spin-orbit coupling (SOC) effects were fully taken into account via fully relativistic pseudopotentials generated using the optimized norm-conserving Vanderbilt pseudopotential (ONCVPSP) method [S18-S20], as distributed in the PseudoDojo library [S21]. The exchange–correlation energy was treated within the local density approximation (LDA) using the Perdew–Wang parametrization [S22, S23], which has been shown to accurately capture the magnetic properties of $Fe_3GeTe_2$.

The relaxed lattice vectors were fixed at $\mathbf{a}_1$ = (3.9463, 0, 0) Å, $\mathbf{a}_2$ = (−1.9732, 3.4176, 0) Å, and $\mathbf{a}_3$ = (0, 0, 23.2200) Å, introducing a large vacuum spacing along the out-of-plane direction to minimize spurious interactions between periodic images.

To efficiently interpolate the electronic structure, we constructed Wannier functions (WFs) based on initial projections onto Fe-centered *d*-orbitals and Te/Ge-centered *p*-orbitals at the corresponding crystallographic sites. Rather than performing maximal localization, we employed a disentanglement procedure [S24-S26] to isolate the low-energy manifold, followed by a symmetry-based post-processing of the resulting Wannier Hamiltonian to restore the underlying crystal symmetries [S27].

The real-space Hamiltonian matrix elements are defined as

$$H_{mn}(\mathbf{R}) = \langle \mathbf{0}m|\hat{H}|\mathbf{R}n\rangle \quad \text{(S21)}$$

Where **R** connects unit cells, and *m*, *n* label Wannier orbitals. Fourier transformation yields the momentum-space Hamiltonian

$$H_{mn}(\mathbf{k}) = \sum_{\mathbf{R}} e^{i\mathbf{k}\cdot\mathbf{R}} H_{mn}(\mathbf{R}), \quad \text{(S22)}$$

which forms the basis for subsequent Berry curvature and transport calculations.

Velocity operators are evaluated by differentiating $H(\mathbf{k})$ with respect to the crystal momentum:

$$v_\alpha(\mathbf{k}) = \frac{i}{\hbar}\frac{\partial H(\mathbf{k})}{\partial k_\alpha} \quad \text{(S23)}$$

which, in atomic units ($\hbar = 1$), reduces to

$$[v_\alpha(\mathbf{k})]_{mn} = \sum_R iR_\alpha e^{i\mathbf{k}\cdot\mathbf{R}} H_{mn}(\mathbf{R}). \quad \text{(S24)}$$

Spin expectation values are computed using the spin operator $\hat{S}_\alpha = I_{orb} \otimes s_\alpha$, where $s_\alpha = \frac{1}{2}\sigma_\alpha$ acts on the spinor degrees of freedom ($\hbar = 1$). In the spinor-interlaced basis, this operator enables the evaluation of the spin expectation value for each eigenstate $|u_{n\mathbf{k}}\rangle$ as

$$\langle S_z\rangle_{n\mathbf{k}} = \langle u_{n\mathbf{k}}|\hat{S}_z|u_{n\mathbf{k}}\rangle = \sum_{i,j} C_{in}^*(\mathbf{k})[\hat{S}_z]_{ij} C_{jn}(\mathbf{k}) \quad \text{(S25)}$$

where $C(\mathbf{k})$ denotes the matrix of eigenvectors of the Hamiltonian $H(\mathbf{k})$.

Additionally, spin-resolved weights (or spin projections) for each state are defined via projectors $P_{\uparrow/\downarrow}$ onto the spin-up and spin-down sectors:

$$s_{n,\uparrow/_\downarrow}(\mathbf{k}) = \left\langle u_{n\mathbf{k}} \middle| P_{\uparrow/_\downarrow} \middle| u_{n\mathbf{k}} \right\rangle = \sum_{i\in\uparrow/_\downarrow} |C_{in}(\mathbf{k})|^2, \tag{S26}$$

where the summation is over basis indices associated with each spin channel.

The intrinsic anomalous Hall conductivity (AHC), specifically its transverse component $\sigma_{xy}$, is obtained as the Brillouin zone (BZ) integral of the Berry curvature $\Omega_{xy}(\mathbf{k})$ summed over occupied states:

$$\sigma_{xy} = \int \frac{d^2k}{(2\pi)^2} \sum_n \Omega_{xy}^{(n)}(\mathbf{k})\, \Theta(E_F - E_{n\mathbf{k}}) \tag{S27}$$

where the Berry curvature contribution from each band $n$ reads

$$\Omega_{xy}^{(n)}(\mathbf{k}) = -2\mathrm{Im} \sum_{m\neq n} \frac{\langle u_{n\mathbf{k}}|\hat{v}_x|u_{m\mathbf{k}}\rangle \langle u_{m\mathbf{k}}|\hat{v}_y|u_{n\mathbf{k}}\rangle}{(E_{m\mathbf{k}} - E_{n\mathbf{k}})^2} \tag{S28}$$

Here, $E_F$ is the Fermi level and $\Theta$ is the Heaviside step function selecting the occupied states. The calculation is performed in the clean limit and at zero temperature.

The convergence of the calculated $\sigma_{xy}$ was carefully analyzed as a function of the **k**-mesh density. In this work, a uniform 700 × 700 k-mesh was used, ensuring numerical precision.

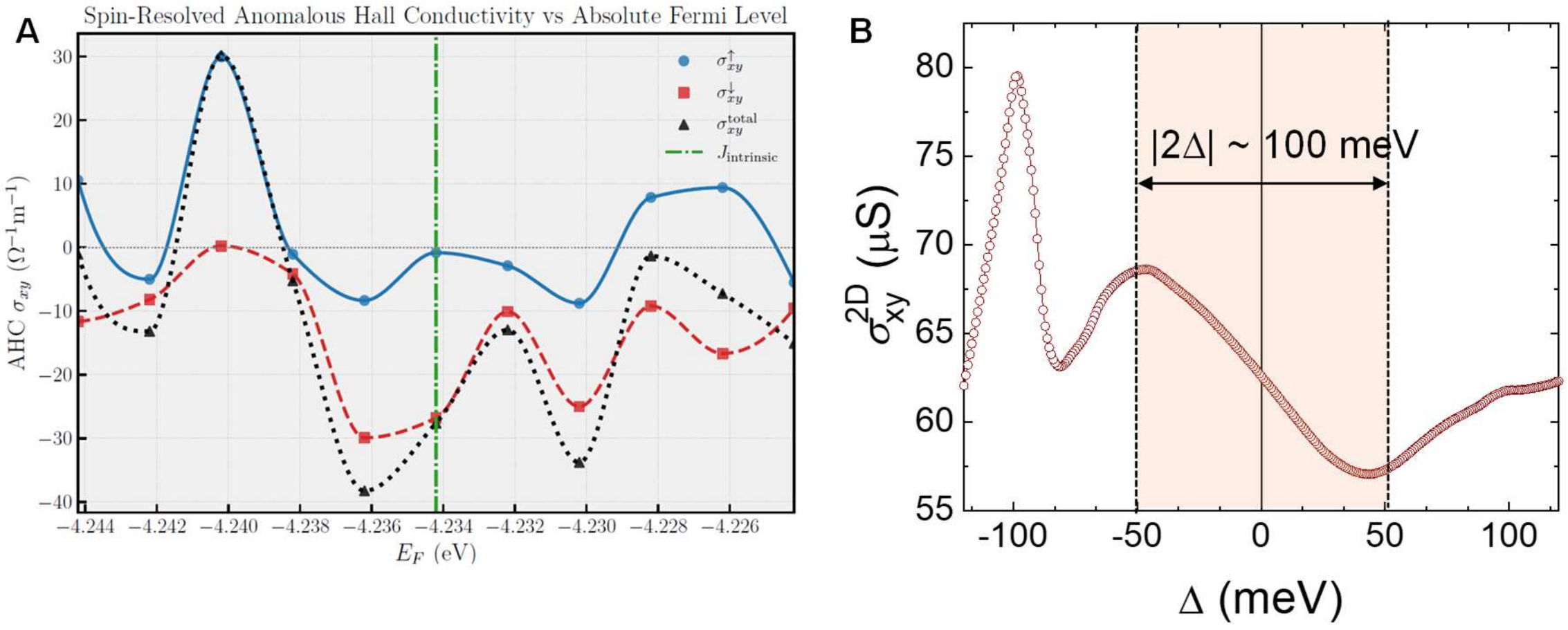

**Figure S16. Theoretical calculations.** (**A**) $\sigma_{xy}$ as a function of Fermi-level position. The Fermi level is shifted rigidly with respect to the zero point (as indicated by the green line). (**B**) Calculated two-dimensional AHE conductivity as a function of Δ in monolayer FGT.

To systematically investigate the influence of electronic filling on the AHC and the spin magnetization, we introduced a rigid shift $\Delta E_{\mathrm{F}}$ applied to the global Fermi level $E_{\mathrm{F,global}}$. This defines a single effective Fermi level at each scan point:

$$E_{\mathrm{F,eff}} = E_{\mathrm{F,global}} + \Delta E_{\mathrm{F}} \tag{S29}$$

The eigenvalues $E_{n\mathbf{k}}$, eigenvectors $|u_{n\mathbf{k}}\rangle$, and velocity matrices $v_{x/y}(\mathbf{k})$ of the original unperturbed Hamiltonian were computed once on the dense mesh. For each value of $\Delta E_{\mathrm{F}}$, the occupations were updated according to $E_{\mathrm{F,eff}}$ , and the AHC and spin magnetization were recalculated accordingly.

The anomalous Hall conductivity (AHC) is exquisitely sensitive to the fine details of the electronic structure in the vicinity of the Fermi level. To elucidate this sensitivity, we monitor the evolution of both the spin-resolved and total AHC under a rigid shift of the Fermi energy, an approach that mimics carrier doping or electrostatic gating while keeping the underlying band structure unchanged. Although such a shift constitutes a purely equilibrium modification of the occupation numbers, it nevertheless provides a stringent probe of the intrinsic Berry-curvature landscape: a pronounced variation of the AHC with energy signals the presence of sharp Berry curvature "hot spots"—often generated by spin–orbit-coupled avoided crossings—that lie close to the Fermi surface.

Our results (Fig. S16a) reveal multiple sign reversals and peak-valley structures within a narrow window of ~ 20 meV around the reference Fermi level. This behavior is strikingly reminiscent of the seminal observations by Fang *et al.* on $SrRuO_3$ (Ref. 8), where non-monotonic and sign changing AHC trends were traced back to momentum-space magnetic monopoles associated with near-degenerate bands. The close analogy underscores that $Fe_3GeTe_2$ hosts similarly localized, energy-dependent Berry curvature features that dominantly control its Hall response.

Beyond the equilibrium perspective, the steep energy-dependence of $\sigma_{xy}$ carries direct implications for non-equilibrium spin phenomena. In the presence of an external electric field, the induced spin accumulation (Edelstein and spin-Hall-like effects) is governed by the energy derivative $\partial\sigma_{xy}/\partial E_F$; hence the large slopes and sign changes obtained here imply a sizeable intrinsic spin–orbit-torque susceptibility. Consequently, modest electrostatic gating, chemical substitution, or ionic-liquid control that move $E_F$ by only a few meV should enable efficient tuning— polarity reversal—of both charge and spin currents in this material.

On the other hand, our single-layer model omits several proximity-induced perturbations supplied by the adjacent $WTe_2$ layer: (i) Rashba SOC generated by the interfacial electric field, (ii) Te–Te orbital hybridization that reshapes the low-energy band dispersion, and (iii) current-driven orbit-to-spin torque that dynamically renormalizes the effective exchange splitting $J$. These effects shift and amplify Berry-curvature "hot rings" near $E_F$, thereby enhancing the AHC response, features that are absent or strongly suppressed in the isolated-FGT calculation.

More broadly, the rigid-shift analysis highlights the extreme sensitivity of $\sigma_{xy}$ to the fine details of the electronic structure. Pronounced peak-valley structures and multiple sign reversals within a ~ 100 meV window trace directly to Berry-curvature hot spots associated with spin–orbit-coupled avoided crossings (Fig. S16b). In a non-equilibrium context, the same steep energy dependence implies a large intrinsic spin-orbit-torque susceptibility, because the Edelstein-induced spin accumulation is proportional to $\partial\sigma_{xy}/\partial E_F$. Hence, modest tuning of $E_F$ by gating, ionic-liquid control, or small dc currents—should enable efficient, even polarity-reversing, control of both charge and spin currents in this van-der-Waals magnet.

Although the rigid-band model captures the qualitative interplay between exchange splitting and Berry curvature, quantitative agreement with experiment demands inclusion of interfacial Rashba SOC, orbital hybridization, and current-induced exchange renormalization. Mapping those proximity-enhanced Berry-curvature hot spots therefore provides practical design rules for optimizing anomalous Hall, spin Hall, and Edelstein responses in $WTe_2/Fe_3GeTe_2$ heterostructures.

In summary, the rigid-shift analysis, despite its simplicity, emerges as a powerful predictor of spin-orbit-driven transport functionality. By mapping the Berry-curvature hot spots and their energetic proximity to $E_F$, it offers practical design guidelines for optimizing anomalous Hall, spin Hall, and Edelstein responses in $Fe_3GeTe_2$ and related van-der-Waals magnets.

## 12. Estimation of the modulation of exchange interaction by the two-band model

According to equation S20 and Fig. 3b, the AHE conductivity at $I = 0, \pm 1$ mA is respectively given by

$$\frac{e^2}{2h}\frac{J}{\hbar vk} = 970 \tag{S30}$$

$$\frac{e^2}{2h}\frac{J+\Delta}{\hbar vk} = 1800 \tag{S31}$$

$$\frac{e^2}{2h}\frac{J-\Delta}{\hbar vk} = 0 \tag{S32}$$

By subtracting equation S31 and equation S32, we obtain

$$2\Delta\frac{e^2}{2h}\frac{1}{\hbar vk} = 1800 \tag{S33}$$

By comparing equation S33 and equation S30, $\Delta = 0.93J$ is determined. This result stands in sharp contrast to the more realistic band structure calculation, which implies $\Delta = 0.03J$.

## 13. No modulation of AHE in pure FGT

To confirm that the modulation of the AHE is a genuine effect originating from $WTe_2$, we perform control experiments on pure FGT. As shown in Fig. S26, $R_{AHE}$ shows an even function with respect to $I$ and decreases as $|I|$ increases, which is purely caused by Joule heating. No detectable $I$-odd modulation is observed.

## 14. FGT and $WTe_2$ thickness dependence of the modulation

Figs. S17, S19, S21, S22, S23, S25, S27, S29, S31, S32 and S34 show the modulation of the AHE in two different ways (the same Fig. 3 in main text) for various FGT and $WTe_2$ thicknesses. The results have been respectively summarized in Tables S2 and S3.

## 15. Current-orientation dependence of the modulation

Similar to magnetization switching measurements, we have also performed modulation measurements along the high-symmetry axis for the $WTe_2$/FGT stack. As shown in Figs. S21c, S28, S30, S33 and S35, no detectable modulation is observed, confirming that current-induced out-of-plane magnetization is generated when **I** // b-axis. We have also measured two more devices for **I** aligned along 29° and 61° (i.e., neither along the a-axis nor along the b-axis), and sizable modulation is observed for both current directions (Figs. S17-S20, Figs. S23 and S24). The results have been summarized in Table 5.

## 16. Current induced Hall response in pure $WTe_2$

Figs. S36 shows the dc current induced Hall response in pure $WTe_2$. A signal of 0.25 Ω is observed at $I$ = 1 mA. This value is much smaller than the modulation observed in the $WTe_2$/FGT bilayer (Tables S3 and S4). We therefore conclude that the observed modulation cannot be attributed to an additional current-induced Hall contribution originating from $WTe_2$.

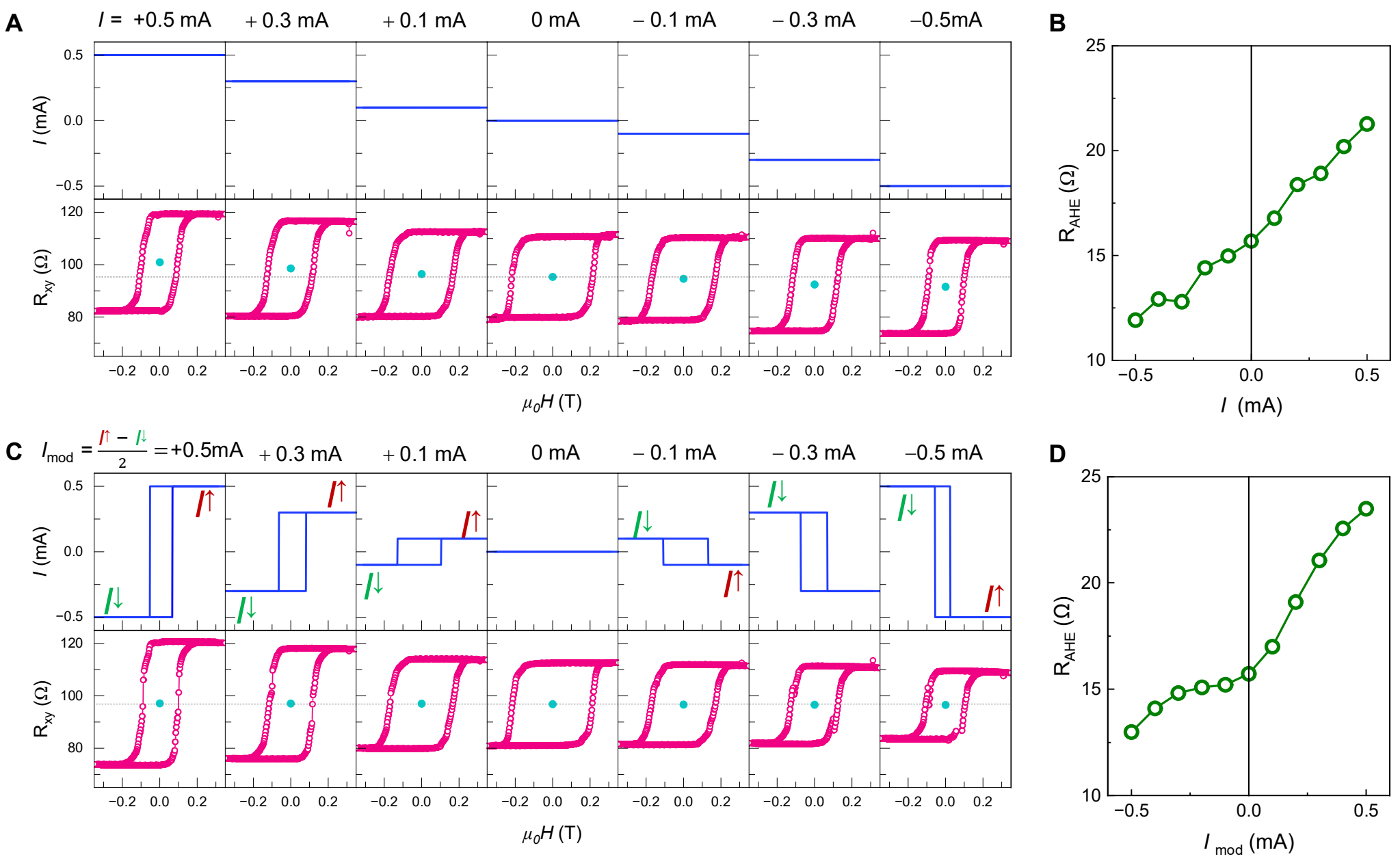

**Figure S17. Device DM2: modulation of the AHE for ∠ = 29º.** (**A**) and (**C**): $R_{xy}$ loops at different $I$ and $I_{mod}$. (**B**) and (**D**): $I$ and $I_{mod}$ dependence of $R_{AHE}$.

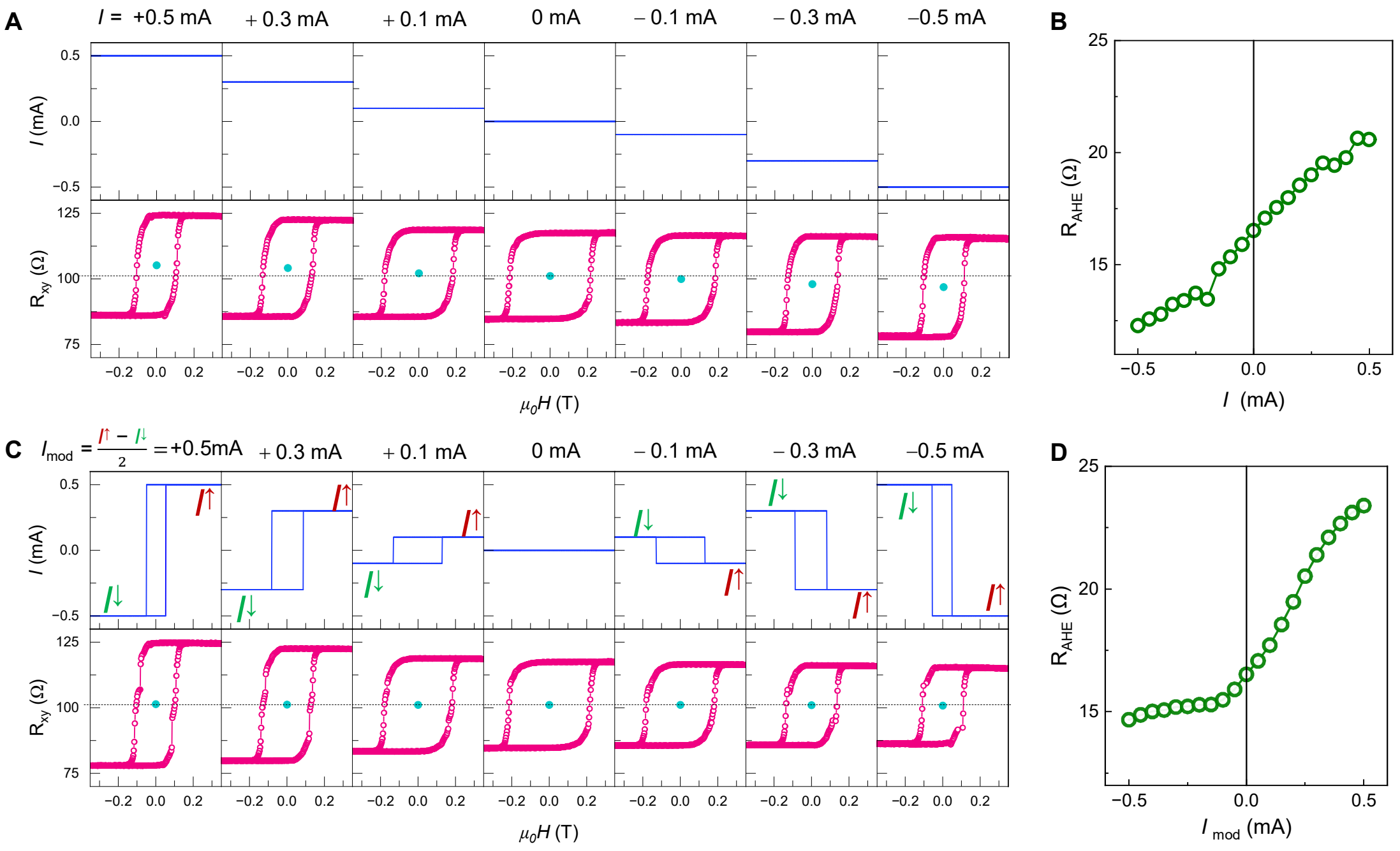

**Figure S18. Device DM2: modulation of the AHE for ∠ = 61º.** (**A**) and (**C**): $R_{xy}$ loops at different $I$ and $I_{mod}$. (**B**) and (**D**): $I$ and $I_{mod}$ dependence of $R_{AHE}$.

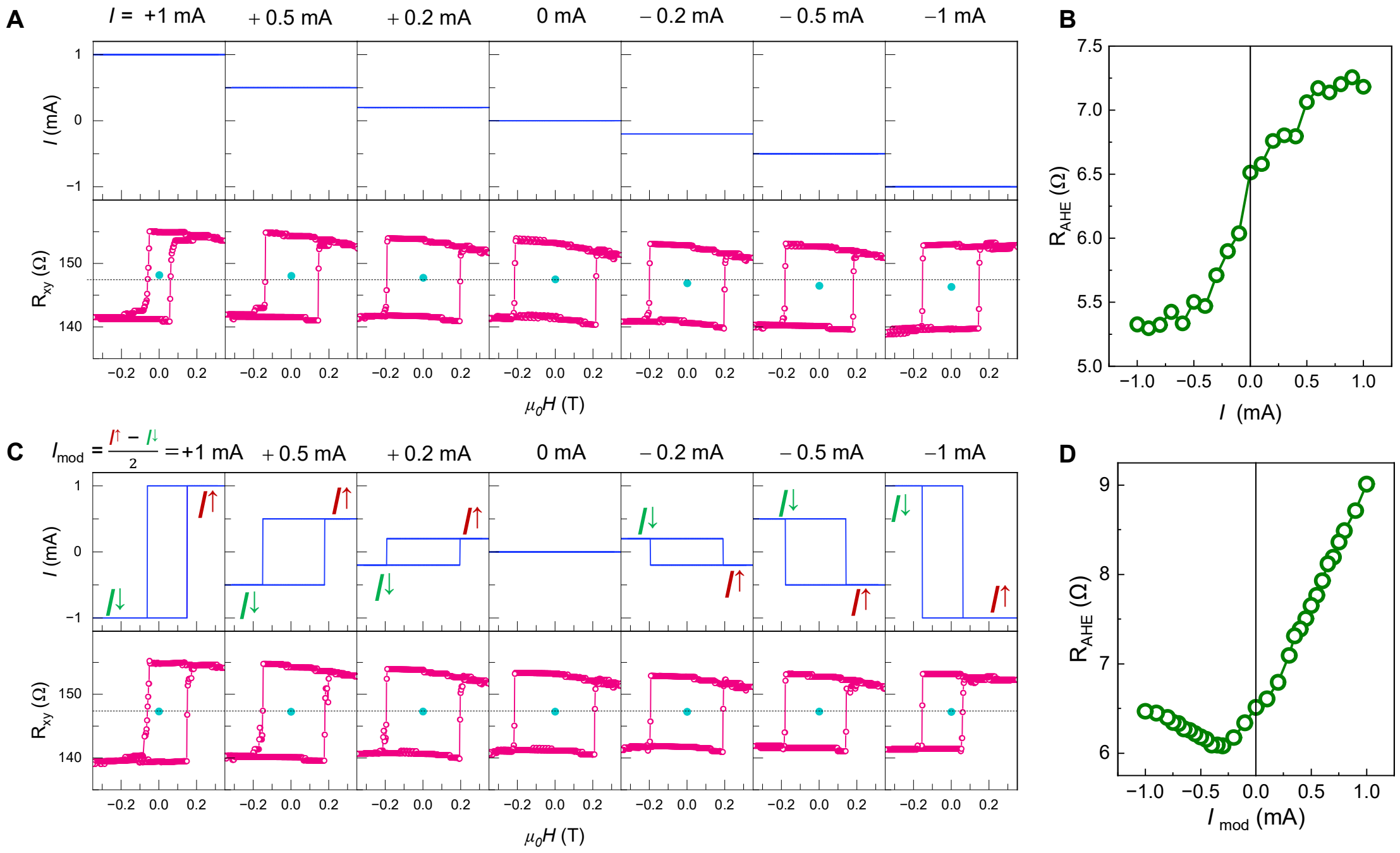

**Figure S19. Device DM3: modulation of the AHE for ∠ = 29°.** (**A**) and (**C**): $R_{xy}$ loops at different $I$ and $I_{mod}$. (**B**) and (**D**): $I$ and $I_{mod}$ dependence of $R_{AHE}$.

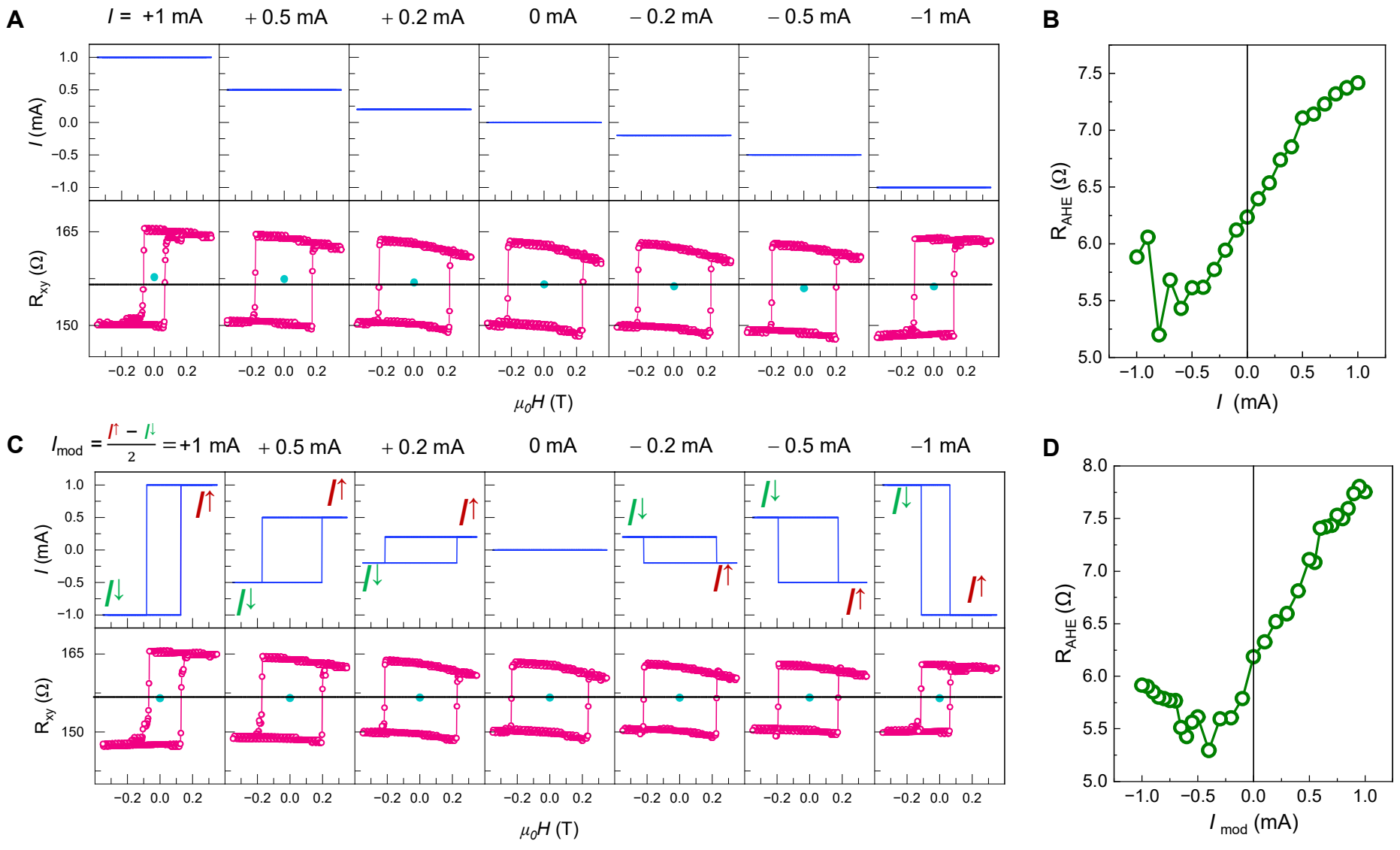

**Figure S20. Device DM3: modulation of the AHE for ∠ = 61°.** (**A**) and (**C**): $R_{xy}$ loops at different $I$ and $I_{mod}$. (**B**) and (**D**): $I$ and $I_{mod}$ dependence of $R_{AHE}$.

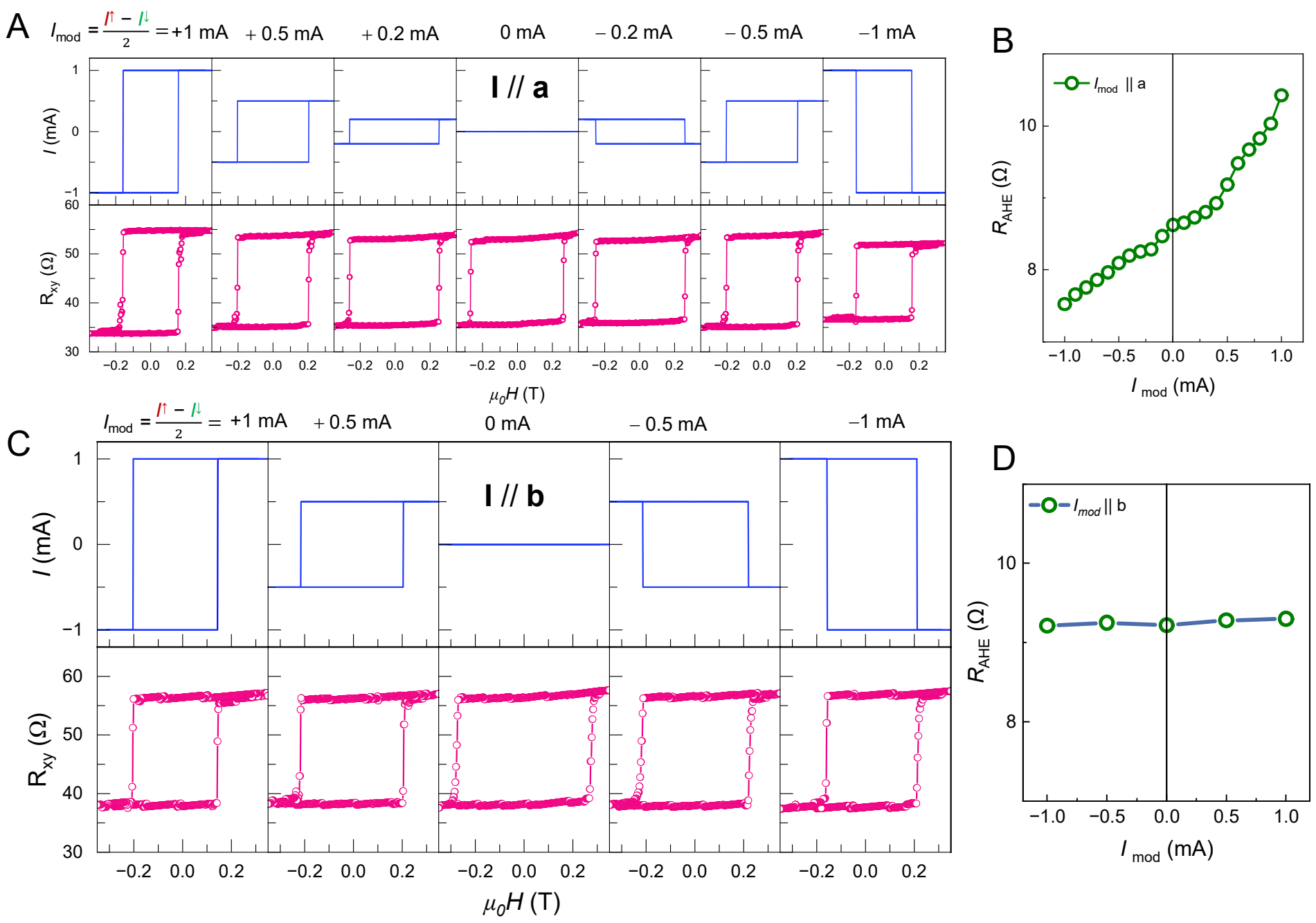

**Figure S21. Device DM4: modulation of the AHE for ∠ = 0º and 90º.** (**A**) and (**C**): $R_{xy}$ loops at different $I_{mod}$ for **I** // **a** and **I** // **b**. (**B**) and (**D**): $I_{mod}$ dependence of $R_{AHE}$ **I** // **a** and **I** // **b**.

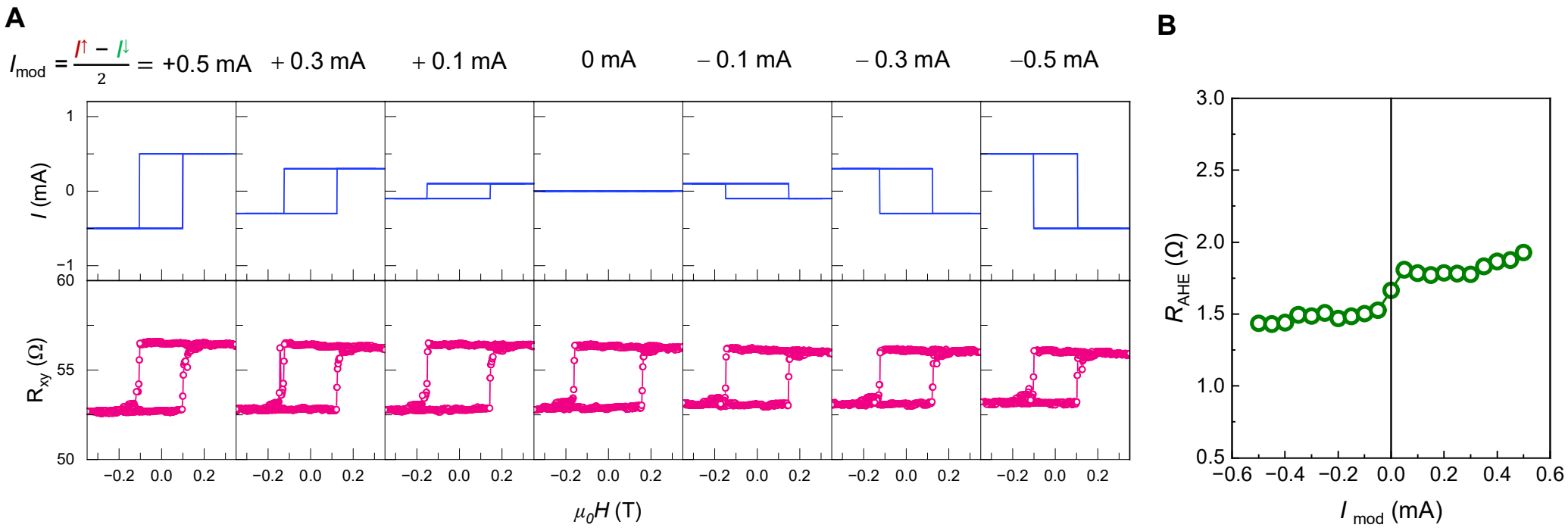

**Figure S22. Device DM5: modulation of the AHE for ∠ = 0º.** (**A**): $R_{xy}$ loops at different $I_{mod}$. (**B**) $I_{mod}$ dependence of $R_{AHE}$.

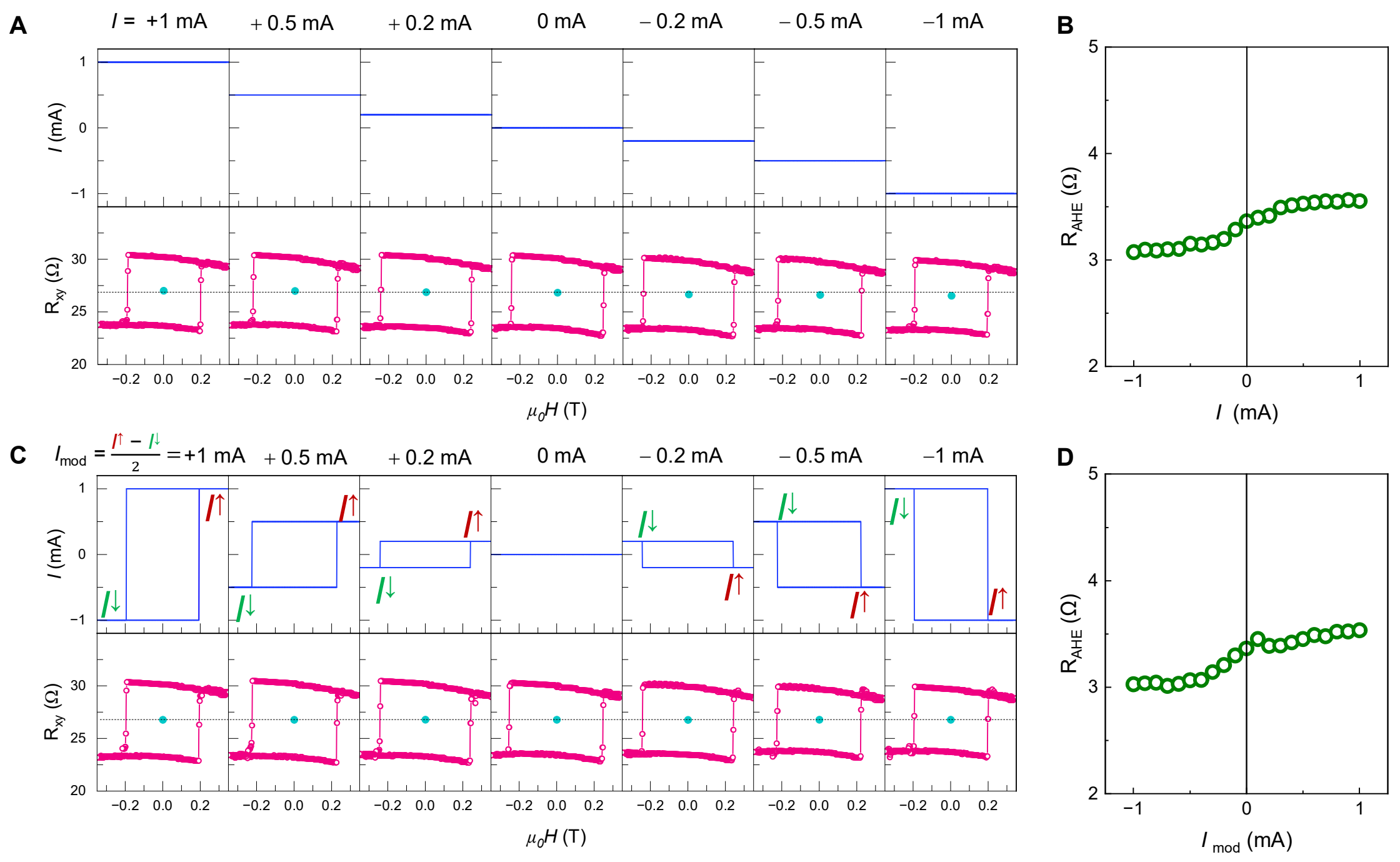

**Figure S23. Device DM6: modulation of the AHE for ∠ = 29º.** (**A**) and (**C**): $R_{xy}$ loops at different $I$ and $I_{mod}$. (**B**) and (**D**): $I$ and $I_{mod}$ dependence of $R_{AHE}$.

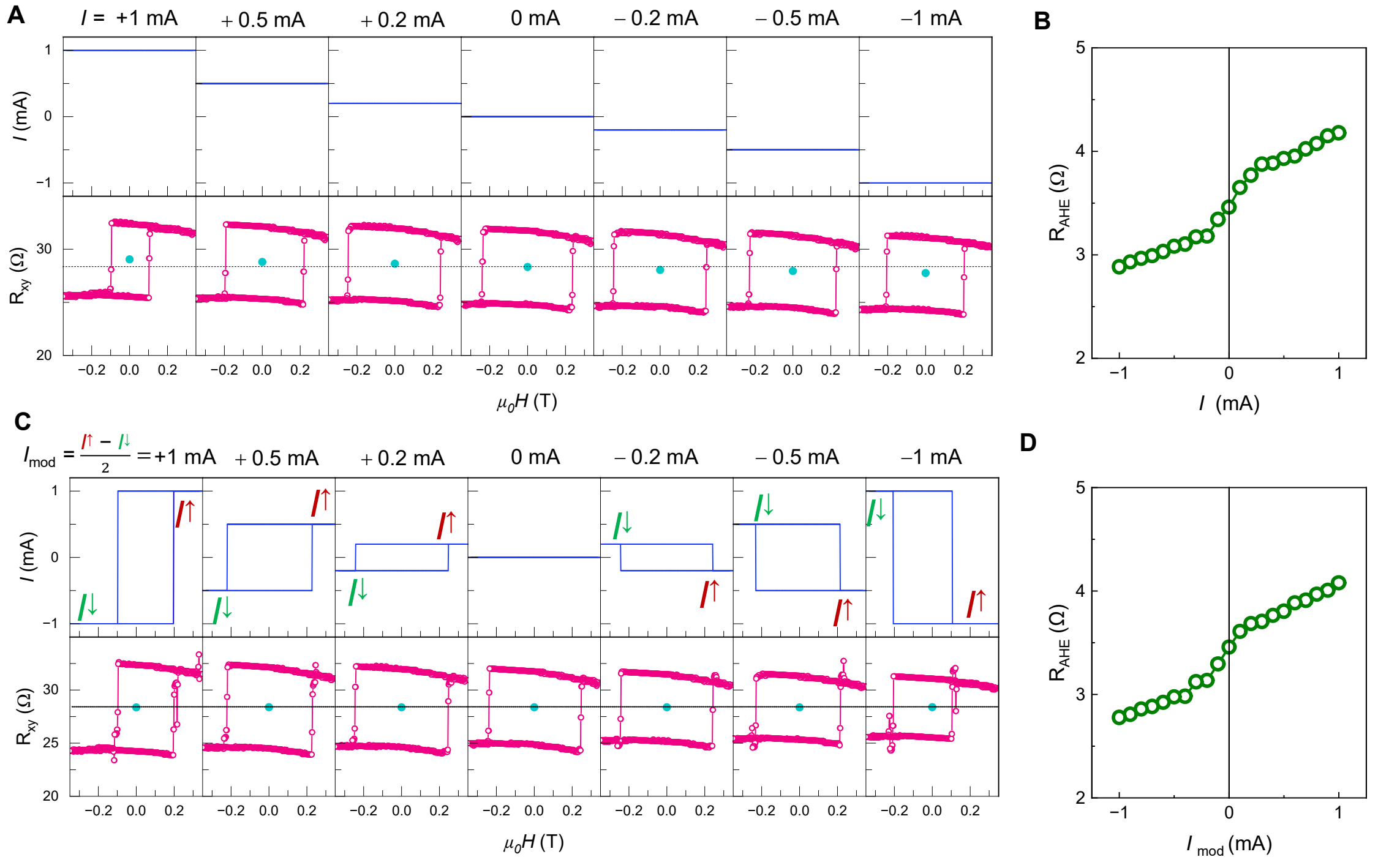

**Figure S24. Device DM6: modulation of the AHE for ∠ = 61º.** (**A**) and (**C**): $R_{xy}$ loops at different $I$ and $I_{mod}$. (**B**) and (**D**): $I$ and $I_{mod}$ dependence of $R_{AHE}$.

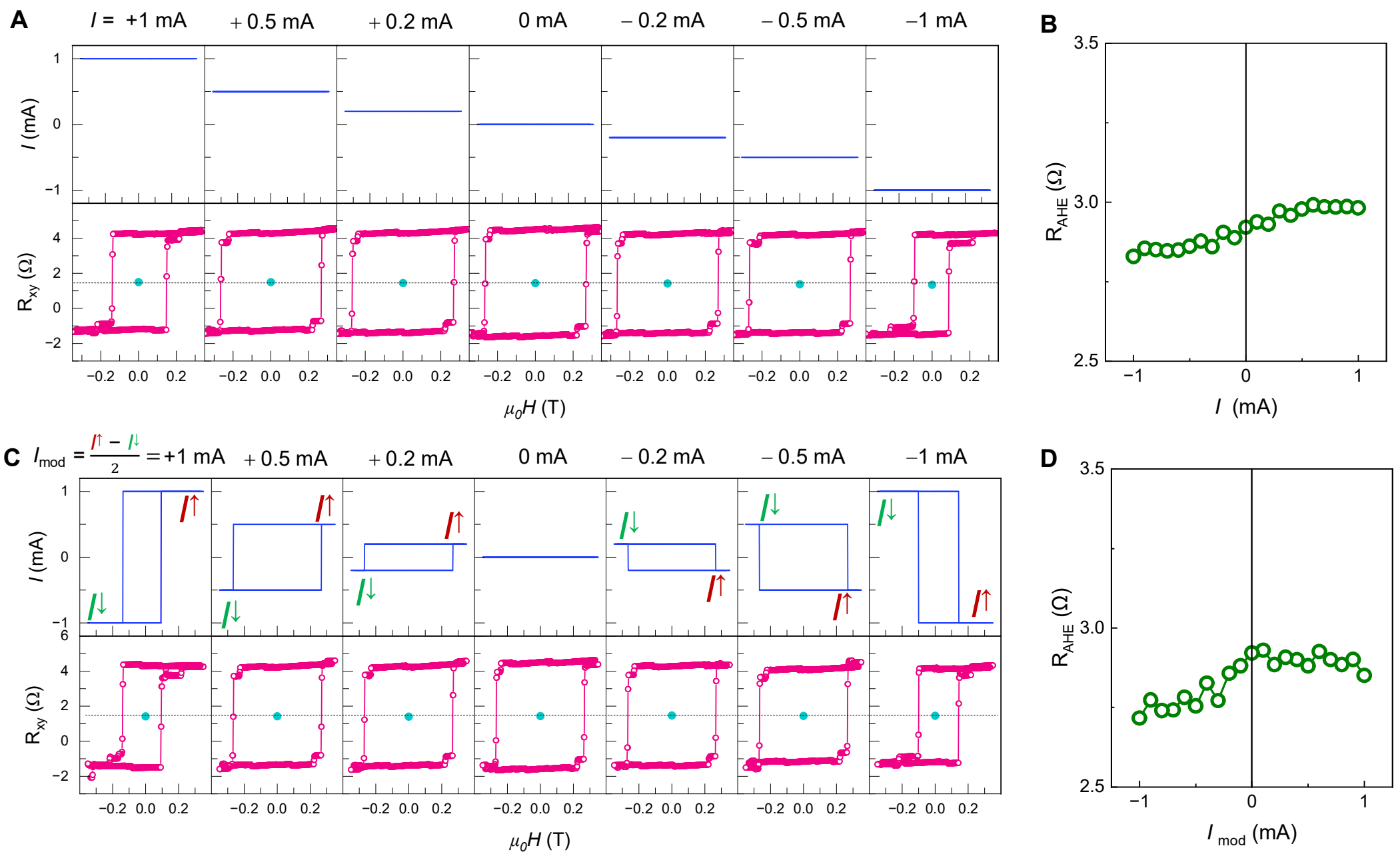

**Figure S25. Device DM7: modulation of the AHE for ∠ = 0°.** (**A**) and (**C**): $R_{xy}$ loops at different $I$ and $I_{mod}$. (**B**) and (**D**): $I$ and $I_{mod}$ dependence of $R_{AHE}$.

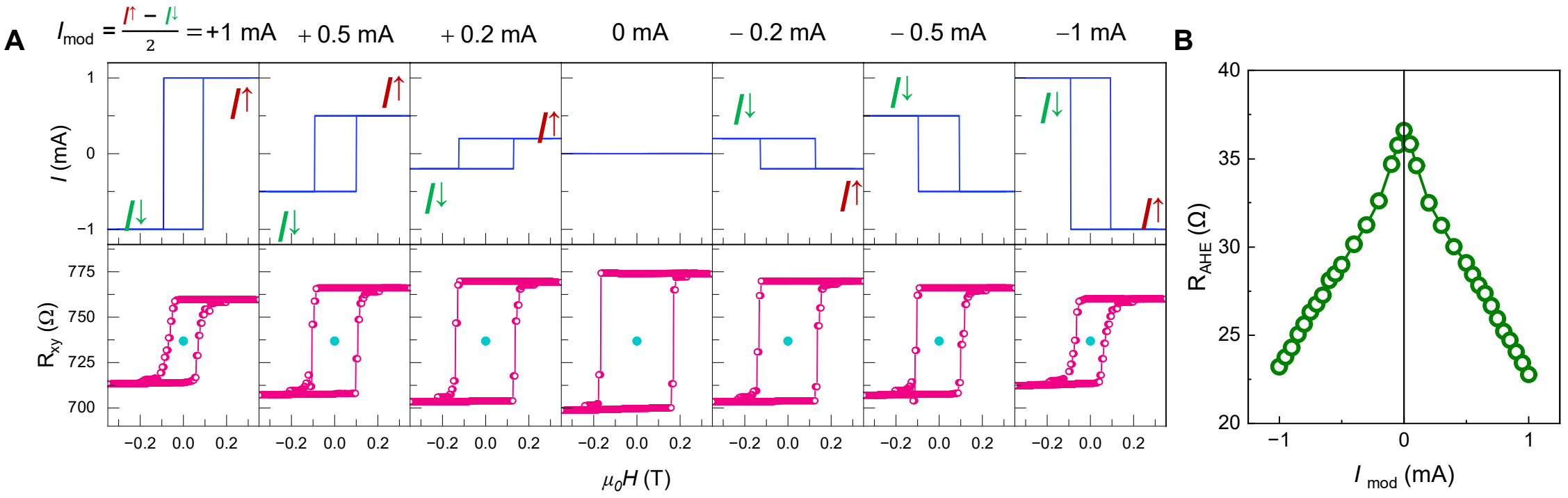

**Figure S26. Device DM8: modulation of the AHE for pure FGT.** (**A**): $R_{xy}$ loops at different $I_{mod}$. (**B**) $I_{mod}$ dependence of $R_{AHE}$.

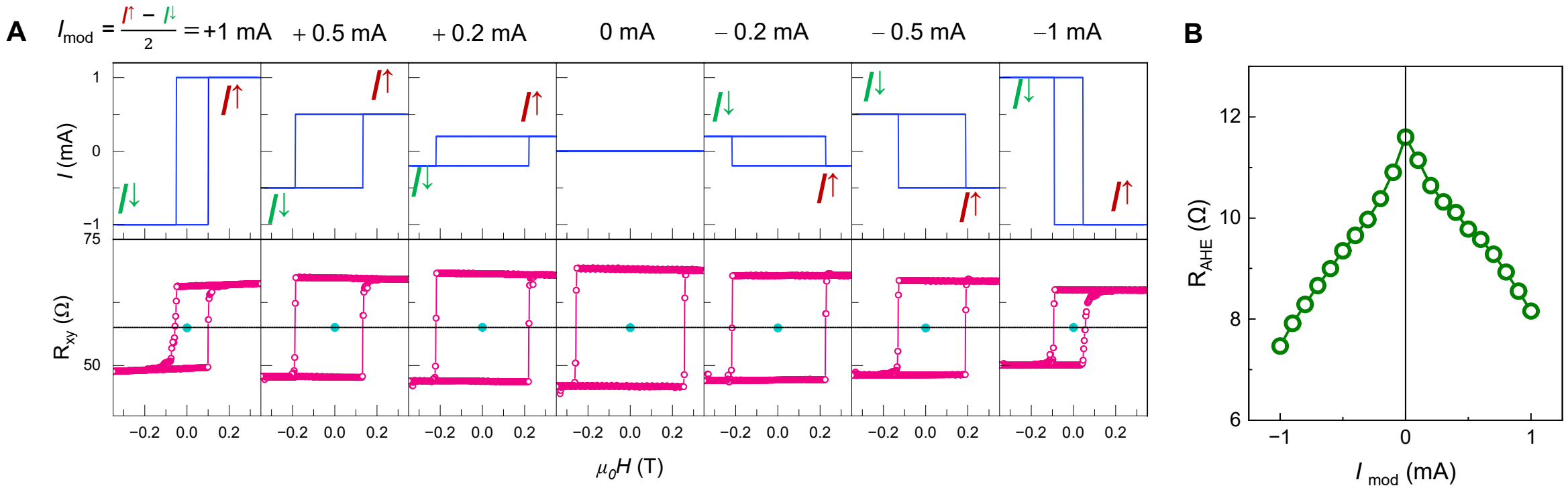

**Figure S27. Device DM9: modulation of the AHE for ∠ = 0º.** (**A**) $R_{xy}$ loops at different $I_{mod}$. (**B**) $I_{mod}$ dependence of $R_{AHE}$.

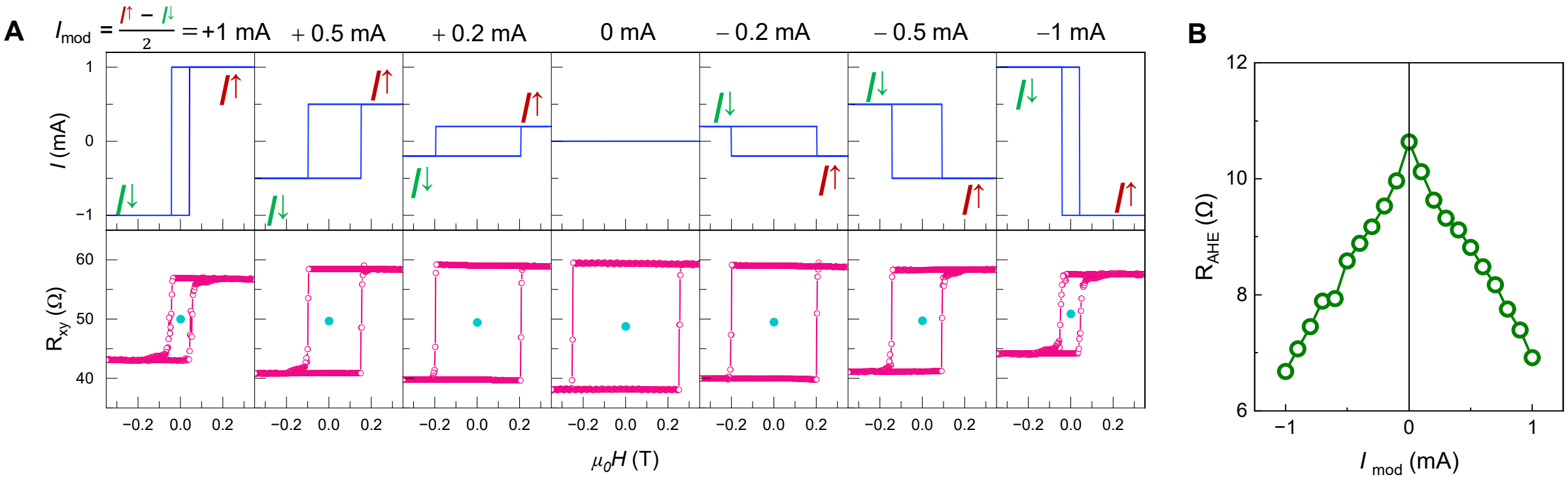

**Figure S28. Device DM9: modulation of the AHE for ∠ = 90º.** (**A**) $R_{xy}$ loops at different $I_{mod}$. (**B**) $I_{mod}$ dependence of $R_{AHE}$.

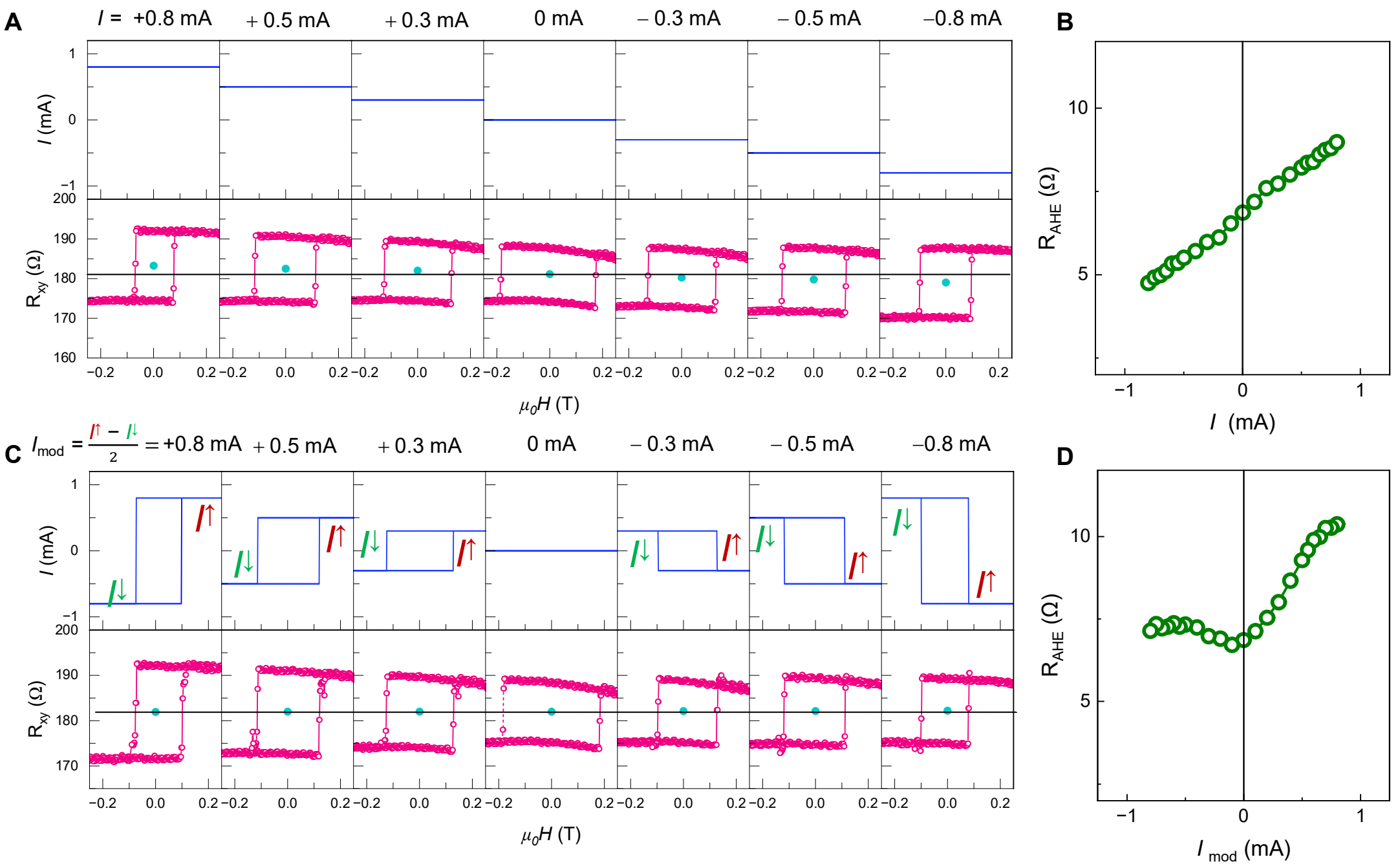

**Figure S29. Device DM10: modulation of the AHE for ∠ = 0º.** (**A**) and (**C**): $R_{xy}$ loops at different $I$ and $I_{mod}$. (**B**) and (**D**): $I$ and $I_{mod}$ dependence of $R_{AHE}$.

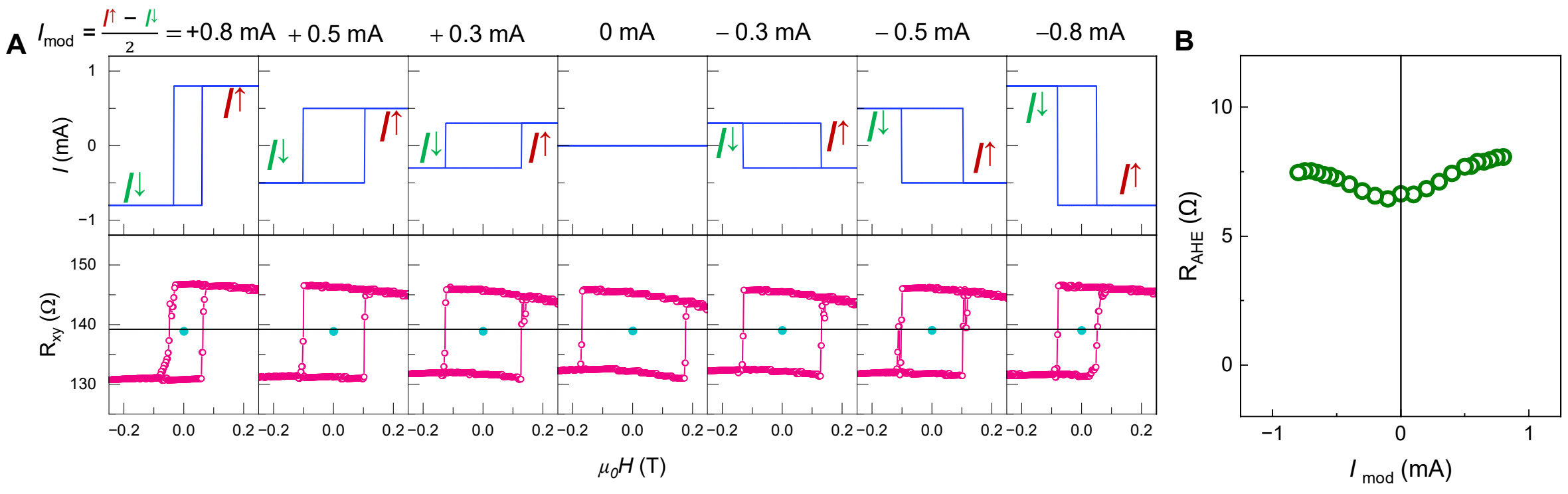

**Figure S30. Device DM10: modulation of the AHE for ∠ = 90º.** (**A**) $R_{xy}$ loops at different $I_{mod}$. (**B**) $I_{mod}$ dependence of $R_{AHE}$.

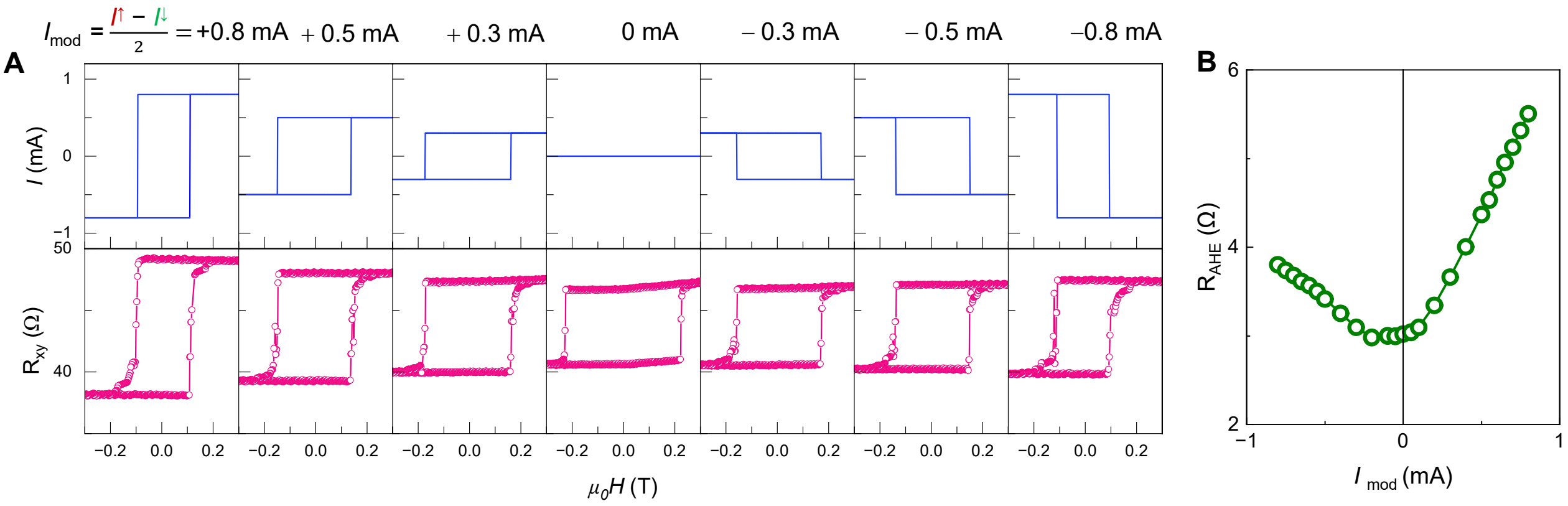

**Figure S31. Device DM11: modulation of the AHE for ∠ = 0º.** (**A**): $R_{xy}$ loops at different $I_{mod}$. (**B**) $I_{mod}$ dependence of $R_{AHE}$.

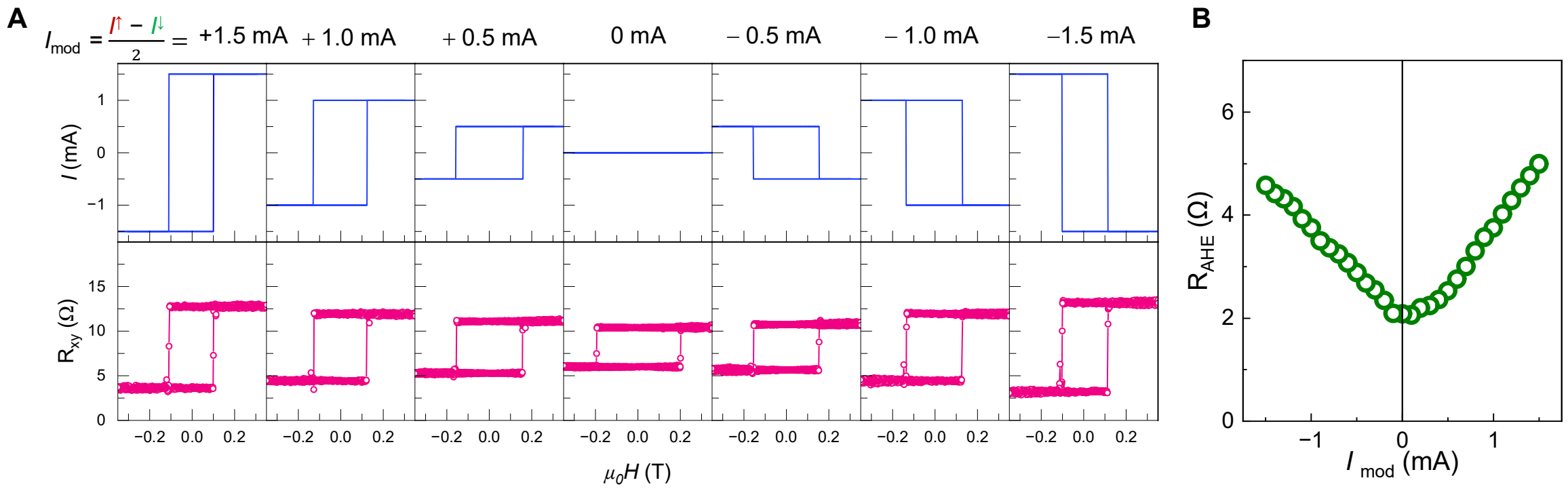

**Figure S32. Device DM12: modulation of the AHE for ∠ = 0º.** (**A**) and (**C**): $R_{xy}$ loops at different $I$ and $I_{mod}$. (**B**) and (**D**): $I$ and $I_{mod}$ dependence of $R_{AHE}$.

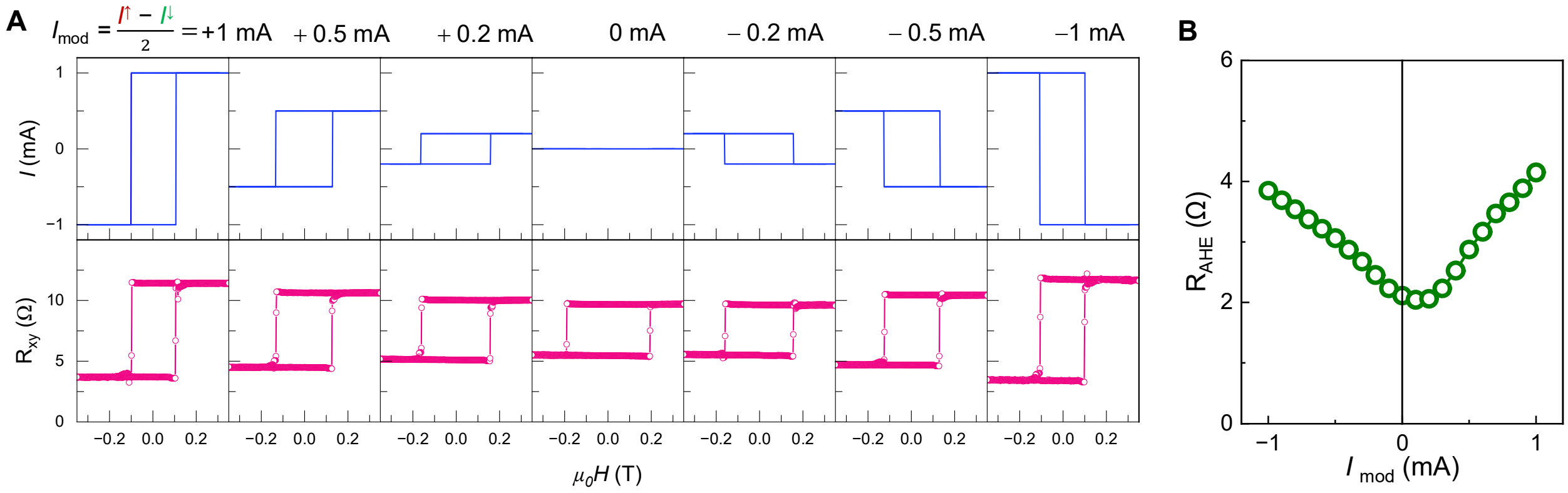

**Figure S33. Device DM12: modulation of the AHE for ∠ = 90º.** (**A**) and (**C**): $R_{xy}$ loops at different $I$ and $I_{mod}$. (**B**) and (**D**): $I$ and $I_{mod}$ dependence of $R_{AHE}$.

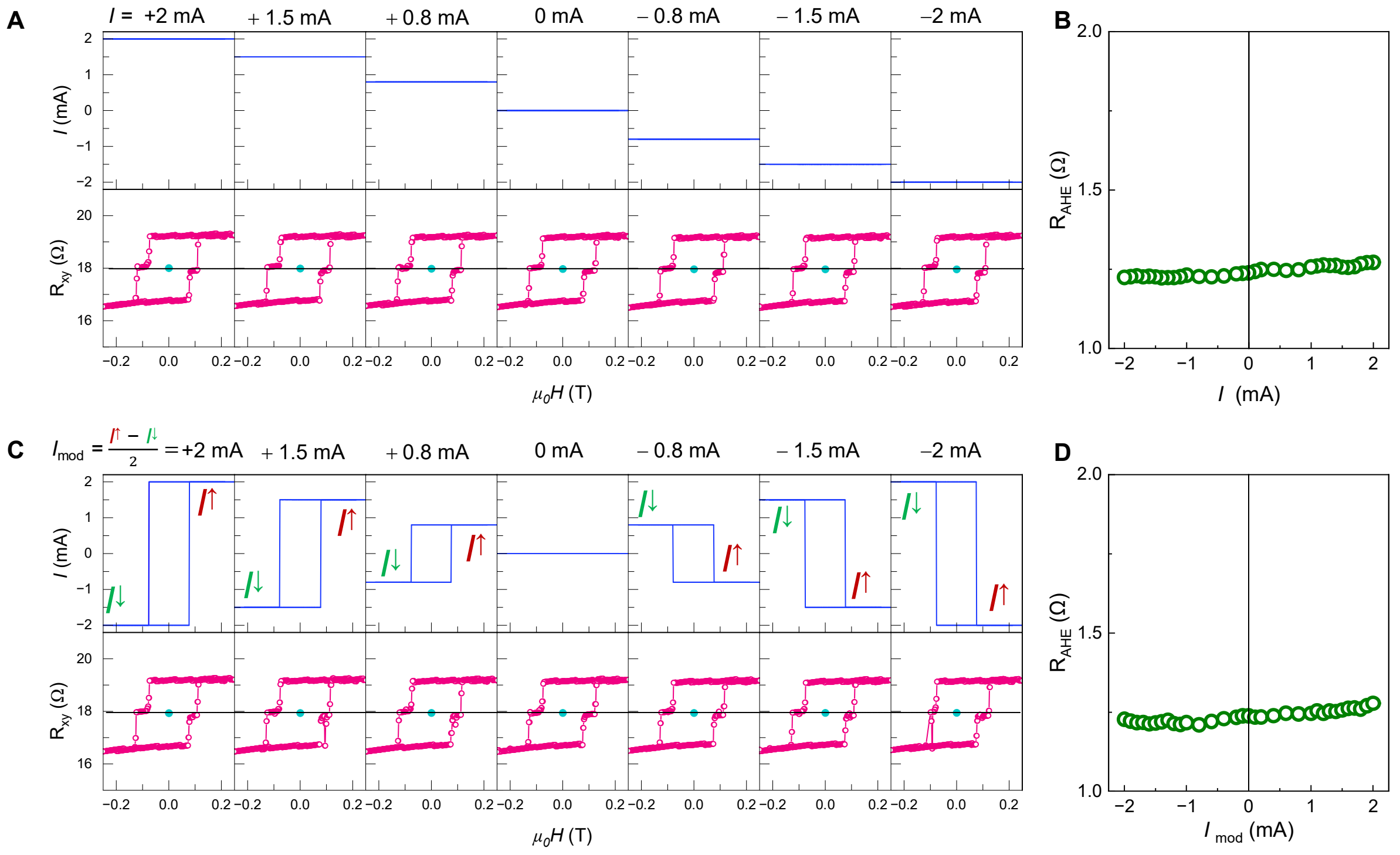

**Figure S34. Device DM13: modulation of the AHE for ∠ = 0º.** (**A**) and (**C**): $R_{xy}$ loops at different $I$ and $I_{mod}$. (**B**) and (**D**): $I$ and $I_{mod}$ dependence of $R_{AHE}$.

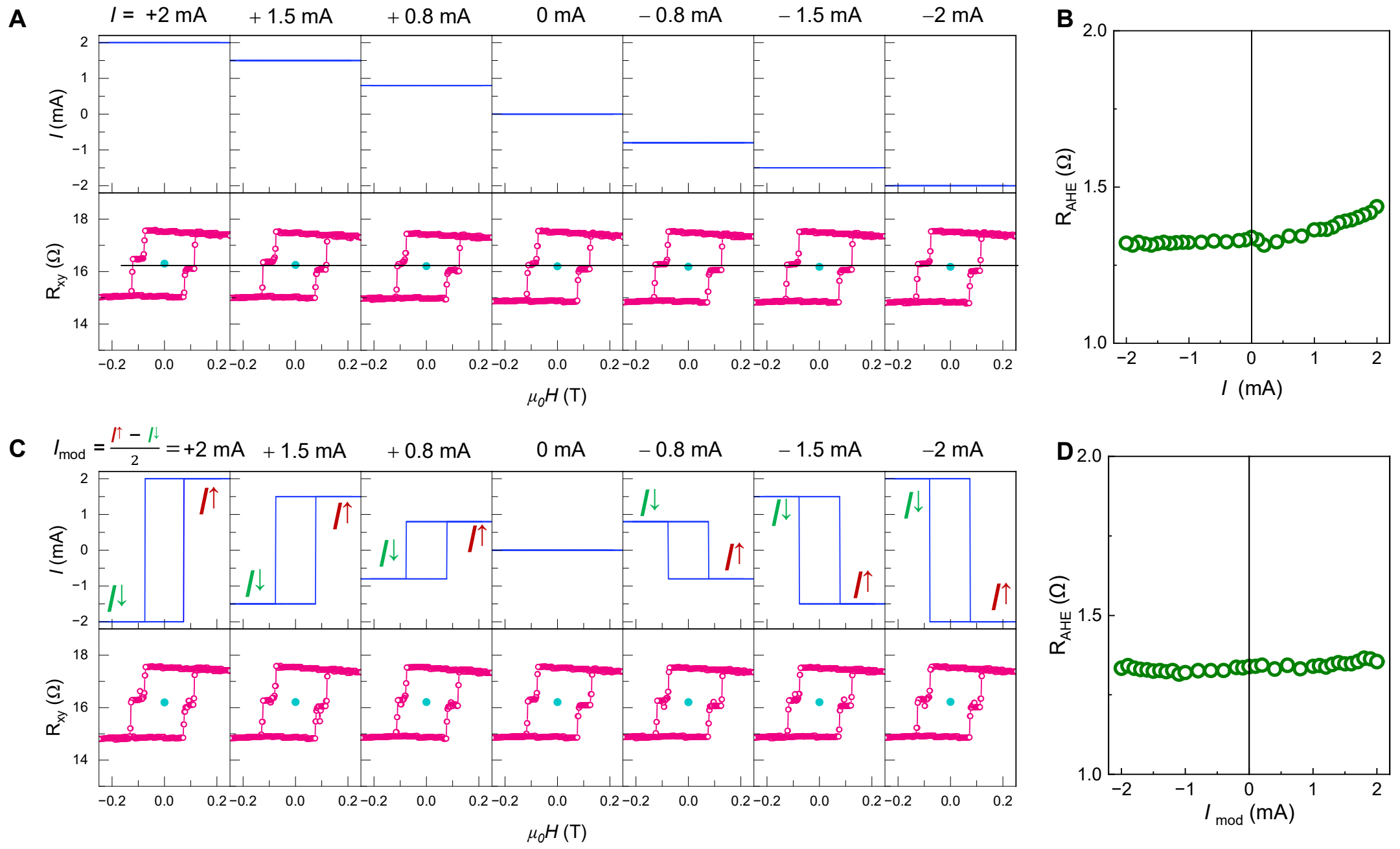

**Figure S35. Device DM13: modulation of the AHE for ∠ = 90º.** (**A**) and (**C**): $R_{xy}$ loops at different $I$ and $I_{mod}$. (**B**) and (**D**): $I$ and $I_{mod}$ dependence of $R_{AHE}$.

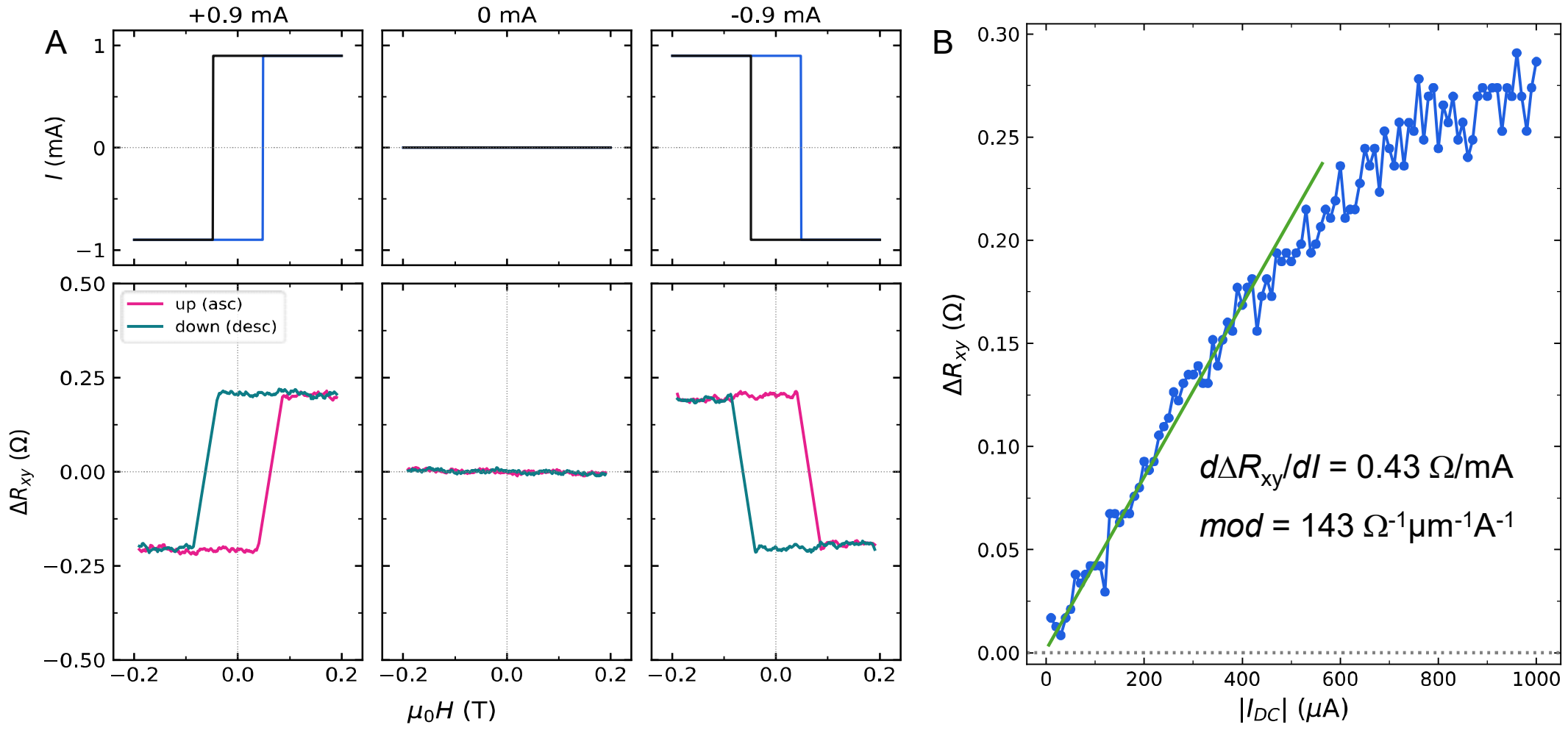

**Figure S36.** Current induced AHE in pure $WTe_2$ (width: 3 mm, thickness: 3.6 nm). The magnitude of signal is only about 0.25 Ω for $I = 1$ mA, corresponding to *mod* = 143 $\Omega^{-1}\mu m^{-1}A^{-1}$. This is about 13 times smaller than the AHE modulation in $WTe_2$/FGT stack.